\documentclass{aa}  
\usepackage{graphicx}
\usepackage{gensymb}
\usepackage{txfonts}
\usepackage[hidelinks]{hyperref}
\hypersetup{
    colorlinks=true,
    allcolors=blue,}

\usepackage{orcidlink}

\begin{document} 

\DeclareRobustCommand{\orcid}[1]{\orcidlink{#1}}

\defcitealias{lamarcaDustPowerUnravelling2024}{LM24}

\title{Galactic bars and active galactic nucleus fuelling in the second half of cosmic history}

\author{A.~La~Marca
\orcid{0000-0002-7217-5120}
\inst{1,2}\thanks{\email{antonio.la.marca.astro@gmail.com}}
\and
M.~T.~Nardone
\orcid{0009-0001-4102-9630}
\inst{2,3}
\and
L.~Wang\orcid{0000-0002-6736-9158}
\inst{1,2}
\and
B.~Margalef-Bentabol\orcid{0000-0001-8702-7019}
\inst{1}
\and
S.~Kruk
\orcid{0000-0001-8010-8879}
\inst{3}
\and S.~C.~Trager\orcid{0000-0001-6994-3566}
\inst{2}
}

\institute{SRON Netherlands Institute for Space Research, Landleven 12, 9747 AD Groningen, The Netherlands
\and
Kapteyn Astronomical Institute, University of Groningen, Postbus 800, 9700 AV Groningen, The Netherlands
\and
European Space Agency (ESA), European Space Astronomy Centre (ESAC), Camino Bajo del Castillo s/n, 28692, Villaneuva de la Cañada, Madrid, Spain
   }

\date{Received -; accepted -}

 
  \abstract
   {
   We investigate the role of galactic bars in fuelling and triggering Active Galactic Nucleus (AGN) in disc galaxies up to $z\sim 0.8$.
   We utilise a Deep Learning model, fine-tuned on Galaxy Zoo volunteer classifications, to identify (strongly and weakly) barred and unbarred disc galaxies in Hyper Suprime-Cam Subaru Strategic Program $i$-band images. We select AGN using three independent diagnostics: mid-infrared colours, X-ray detections, and spectral energy distribution (SED) fitting. The SED analysis, performed using CIGALE, quantifies the relative AGN contribution to the total galaxy luminosity ($f_{\rm AGN}$) and the AGN luminosity ($L_{\rm disc}$). We assess the impact of bars by comparing AGN incidence and properties in barred galaxies against carefully constructed redshift-, stellar mass-, and colour-matched unbarred control samples. 
   Our binary AGN classification experiment demonstrates that barred disc galaxies host a higher fraction of AGN compared to their unbarred counterparts, though the significance depends on the AGN selection method, with a more modest excess for SED AGN, and control sample size.
   This suggests a contributing role for bars in the global AGN budget.
   The contribution of bars to AGN fuelling appears confined to systems where the AGN has a lower relative contribution to the host galaxy's emission ($f_{\rm AGN} < 0.75$). Crucially, we find a significant dearth of barred disc galaxies hosting AGN with $f_{\rm AGN} > 0.75$, independent of bar strength. Consistent with this, the fraction of barred galaxies among AGN hosts decreases with increasing $L_{\rm disc}$. Combined with previous results, we suggest that bars may contribute to fuelling the population of low-to-moderate luminosity AGN, but major mergers are the principal mechanism for triggering the most powerful and dominant accretion events.
   }
   


   \maketitle
%

\section{Introduction}

Supermassive black holes (SMBHs) reside at the centres of the most massive galaxies and, when actively accreting matter, manifest as Active Galactic Nuclei (AGN). Understanding the mechanisms that fuel SMBHs and trigger AGN is key for comprehending galaxy-SMBH co-evolution \citep[see][for a review]{heckman_coevolution_2014}. While major galaxy mergers were initially considered the primary drivers of AGN activity \citep{di_matteo_energy_2005, hopkins_unified_2006, hopkins_cosmological_2008}, the role of secular processes, particularly in powering low-to-moderate-luminosity AGN, has gained significant attention \citep{Romeo2015DoubleMolecular, Romeo2016WhatPowers, martin_normal_2018, smethurst_evidence_2023}. Among these secular mechanisms, large-scale galactic bars are prominent candidates for funnelling gas towards the central SMBH, potentially igniting AGN in disc-dominated galaxies \citep{sakamoto_bar-driven_1999, athanassoula_what_2003, lin_hydrodynamical_2013}. This scenario is supported by observations showing correlations between bars and central molecular gas concentrations \citep[e.g.,][]{yu_edge-califa_2022}.
However, this framework is complicated by theoretical works showing that while such bars can efficiently channel gas to the inner hundred parsecs, their influence diminishes at smaller scales \citep[e.g., ][]{Shlosman1989BarsWithin, Hopkins2010HowDoMassive}.

Disc-dominated galaxies, given their likely merger-free histories \citep{somerville_physical_2015}, are ideal laboratories for studying secular AGN triggering. Bars are common in local disc galaxies, with observed fractions at optical wavelengths of $30-60\%$ \citep{barazza_bars_2008, aguerri_population_2009, masters_galaxy_2011, MendezAbreu2017TwoDimensional, geron_galaxy_2021, Euclid2025Q1BarFraction}, and even higher, up to $\sim 70\%$, in the infrared \citep{Eskridge2000FrequencyBarred, MenendezDelmestre2007NearInfrared}.
At higher redshifts, the observed bar fraction decreases, down to $15-20\%$ at $z\approx2$ \citep{LeConte2024JWSTInvestigation, Guo2025Abundance}.
However, the precise connection between bars and AGN activity remains debated. Some studies find that barred galaxies are more likely to host AGN \citep{galloway_galaxy_2015, alonso_impact_2018, silva-lima_revisiting_2022, garland_most_2023, garland_galaxy_2024, Marels2025RoleBars}, while others report no significant correlation or find that differences are reduced after controlling for host galaxy properties like stellar mass \citep{lee_bars_2012, cheung_galaxy_2013, cheung_galaxy_2015, cisternas_role_2015, zee_unraveling_2023}. 

Discrepancies in the literature may arise from several factors: diverse AGN selection techniques probing different AGN populations \citep{alexander_what_2012}; the strong link between AGN and host galaxy properties \citep[e.g.,][]{kauffmann_host_2003}; and the correlation of bar presence with both stellar mass and galaxy colour, with barred galaxies often being redder than their unbarred counterparts at fixed mass \citep[e.g.,][]{masters_galaxy_2011, skibba_galaxy_2012, Erwin2018Dependence, kruk_galaxy_2018}. Additionally, the differing timescales of AGN activity \citep[$\sim 10$--$100\,{\rm Myr}$;][]{marconi_local_2004} and bar persistence \citep[several Gyr;][]{sellwood_secular_2014, deSaFreitas2023NewMethod, deSaFreitas2025BarAges} can complicate observational connections. Furthermore, AGN feedback or the intense SMBH growth phase itself might disrupt or destroy bar structures \citep{zee_unraveling_2023}.

In recent work \citep[][hereafter \citetalias{lamarcaDustPowerUnravelling2024}]{lamarcaDustPowerUnravelling2024}, we found compelling evidence that major mergers are the primary, if not the only, triggers for the most powerful and dominant AGN. Our study introduced a novel approach, examining the AGN contribution to the total galaxy luminosity (AGN fraction, $f_{\rm AGN}$) and exploring the merger-AGN connection from a continuous perspective, finding that major mergers are the main trigger of AGN-dominated galaxies ($f_{\rm AGN}>0.80$), while they appear less significant for fuelling less dominant AGN. This led us to hypothesise that secular processes, particularly galactic bars, could be more prominent mechanisms for these AGN.

This paper investigates the galactic bar-AGN relationship in disc-dominated galaxies up to $z\sim 0.8$, using the same parent sample from \citetalias{lamarcaDustPowerUnravelling2024} but focusing on non-merging systems. We employ a Deep Learning (DL) model \citep[\texttt{Zoobot}:][]{walmsley_zoobot_2023} to identify barred and unbarred disc galaxies in Hyper-Suprime-Cam Subaru Strategic Program \citep[HSC-SSP;][]{aihara_hyper_2018} images. Using rich multi-wavelength data, we select AGN via mid-infrared (MIR) colours, X-ray detections, and spectral energy distribution (SED) fitting, quantifying $f_{\rm AGN}$ and AGN luminosity. We compare AGN incidence and properties in barred galaxies against carefully matched unbarred control samples to determine the role of bars in AGN fuelling across the second half of cosmic history.

This paper is organised as follows. Section~\ref{sect:data} presents a brief summary of the multi-wavelength observations and the characteristics of the galaxy sample used in this work. Section~\ref{sect:Methods} describes in detail our method employed for detecting galactic bars, how we selected AGN, and how we constructed control samples. Section~\ref{sect:Results} presents our results, Sect.~\ref{sect:Discussion} discusses the impact of major mergers on our analysis, and Sect.~\ref{sect:Conclusions} summarises our conclusions. Throughout the paper we assume a flat $\Lambda$CDM universe with $\Omega_M=0.2865$, $\Omega_{\Lambda}=0.7135$, and $H_0=69.32$ km s$^{-1}$ Mpc$^{-1}$ \citep{hinshaw_nine-year_2013}. Unless otherwise stated, all magnitudes are in the AB system.


\section{Data}\label{sect:data}

In this section, we first introduce the multi-wavelength dataset used to construct the galaxy sample.
Second, we briefly illustrate how we previously measured the galaxies' physical properties.
Then, we describe the Galaxy Zoo (GZ) projects and the HSC-SSP imaging data used for identifying galactic bars.

\subsection{Multi-wavelength catalogue}\label{sect:observ}

This investigation utilises the galaxy sample constructed in \citetalias{lamarcaDustPowerUnravelling2024}. This parent sample was drawn from a sky area of approximately $65\,{\rm deg^2}$ within the Kilo-Degree Survey \citep[KiDS;][]{de_jong_kilo-degree_2013} North-West 2 (N-W2) equatorial field \citep{kuijken_fourth_2019}. This field benefits from rich multi-wavelength coverage and was almost entirely observed by the Galaxy And Mass Assembly spectroscopic survey \citep[GAMA;][]{driver_galaxy_2011}, whose extensive spectroscopic redshifts facilitated the calibration of photometric redshifts for the broader galaxy sample.
The catalogue compiled in \citetalias{lamarcaDustPowerUnravelling2024} incorporated data from numerous surveys. These included: X-ray data ($0.2-2.3\,{\rm keV}$) from the extended ROentgen Survey with an Imaging Telescope Array \citep[eROSITA;][]{predehl_erosita_2021} as part of the eROSITA Final Equatorial Depth Survey \citep[eFEDS;][]{brunner_erosita_2022}; optical ($ugri$) and near-infrared (NIR; $ZYJHK_S$) photometry from the combined VISTA Kilo-degree INfrared Galaxy survey \citep[KiDS-VIKING, hereafter KV;][]{kuijken_fourth_2019, edge_vista_2013} and the HSC-SSP survey \citep[DR2, \emph{grizy};][]{aihara_hyper_2018}; mid-infrared (MIR) data (centred at 3.4, 4.6, 12, and 22 $\mu{\rm m}$) from the NASA Wide-field Infrared Survey Explorer \citep[WISE;][]{wright_wide-field_2010}; and far-IR and sub-millimetre data (centred at 100, 160, 250, 350, 500 $\mu{\rm m}$) from the Herschel Astrophysical Terahertz Large Area Survey \citep[H-ATLAS;][]{valiante_herschel-atlas_2016}.

To construct this parent galaxy sample, \citetalias{lamarcaDustPowerUnravelling2024} started by selecting all detected sources in the KiDS-N-W2 field from the final nine-band photometric KV catalogue, which had photometric redshifts \citep{kuijken_fourth_2019}. These sources were then cross-matched with detections from the other aforementioned surveys. For detailed information on the cross-matching procedure, we refer the reader to \citetalias{lamarcaDustPowerUnravelling2024}. Subsequent data cleaning involved selecting objects within the redshift range $0.1\leq z \leq 1.0$ and removing clearly identified stars, objects with unreliable photometry, and problematic detections \citep[see][for more details]{kuijken_fourth_2019}. After this cleaning, the sample comprised just over one million galaxies. The analysis in \citetalias{lamarcaDustPowerUnravelling2024}, and consequently this work, focuses on the redshift range $0.1\leq z \leq 0.76$.

\subsection{Galaxy properties}\label{sect:CIGALE}

The galaxy catalogue inherited from \citetalias{lamarcaDustPowerUnravelling2024} contains physical properties derived from SED decomposition. We employed the SED fitting and modelling tool Code Investigating GALaxy Emission \citep[CIGALE;][]{burgarella_star_2005, noll_analysis_2009, boquien_cigale_2019} to estimate key properties, including stellar mass ($M_{\star}$), and the AGN fraction ($f_{\rm AGN}$). The $f_{\rm AGN}$ parameter is defined as the AGN contribution to the total galaxy luminosity within a specific rest-frame wavelength range (in this case the MIR range $3$--$30\,\mu{\rm m}$, denoted as $L_{\rm AGN} / L_{\rm tot}$). In addition to these SED-derived properties, the \citetalias{lamarcaDustPowerUnravelling2024} catalogue also includes AGN selected using three distinct diagnostics (detailed in Sect.~\ref{sect:AGN selection}) and classifications for major mergers. The following paragraphs summarise the main aspects of the SED fitting procedure employed in \citetalias{lamarcaDustPowerUnravelling2024}.

We utilised \texttt{X-CIGALE} version 2022.1 \citep{yang_x-cigale_2020,yang_fitting_2022} for the derivation of physical properties. The star formation history was modelled using a delayed-$\tau$ component plus an optional recent (1-150 Myr) exponential starburst. 
The delayed component models the bulk of the stellar population (large $\tau$ values for late-type galaxies, small $\tau$ for early-types), while the exponential starburst provides some flexibility to the module, allowing it to represent the latest episode of star formation \citep[e.g.,][]{Malek2018HELPModelling}.
A \citet{bruzual_stellar_2003} single stellar population model was adopted, assuming a \citet{Chabrier2003} initial mass function and solar metallicity. For dust attenuation, we selected the \citet{calzetti_dust_2000} law, and for the dust emission component, we chose the \citet{draine_andromedas_2014} models. The SKIRTOR model \citep{stalevski_3d_2012, Stalevski2016} was used as the AGN template. This model assumes a clumpy two-phase dusty torus with a flared disk geometry, consisting of high-density clumps embedded in a lower-density medium. Crucially, \texttt{X-CIGALE} incorporates an X-ray module, enabling the simultaneous fitting of X-ray emission from both AGN and host galaxy components (such as hot gas and X-ray binaries). For a comprehensive list of all parameters used in the \texttt{X-CIGALE} configuration, we refer the reader to Table~2 of \citetalias{lamarcaDustPowerUnravelling2024}.

To ensure robust constraints on the $f_{\rm AGN}$ parameter, particularly in the MIR regime where AGN emission can dominate, the sample selection in \citetalias{lamarcaDustPowerUnravelling2024} was restricted to galaxies with significant MIR detections. Specifically, we selected objects with a signal-to-noise ratio (S/N) $>1$ in both the WISE W1 ($3.4\,\mu{\rm m}$) and W2 ($4.6\,\mu{\rm m}$) bands, and additionally required S/N $>3$ in either W1 or W2. Furthermore, to avoid saturation issues, only sources fainter than the saturation limits of W1 $>8\,{\rm mag}$ and W2 $>7\,{\rm mag}$ (Vega system) were included. This MIR pre-selection ensures that \texttt{CIGALE} has sufficient information to reliably model the AGN component. Furthermore, to maintain a sample of reliable SED fits, only galaxies with a reduced chi-squared ($\chi_{red}^2$) $<5$ from the fitting process were retained.

\begin{figure}
    \centering
    \includegraphics[width=0.49\textwidth]{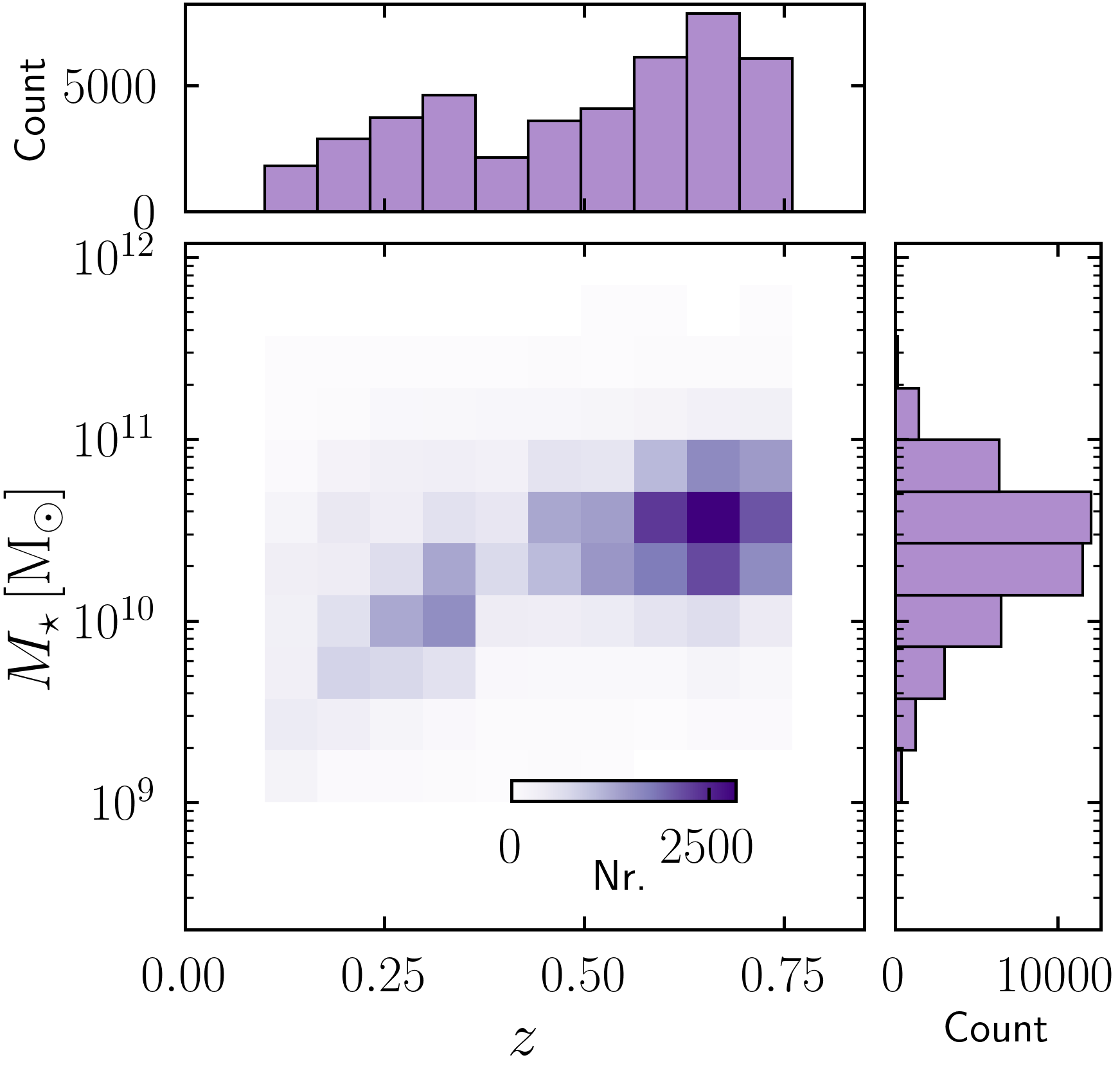}
    \caption{2D histogram (linear scaling) representation of the stellar mass--redshift distribution for the \citetalias{lamarcaDustPowerUnravelling2024} parent sample used in this work. The marginal histograms display the individual $z$ and $M_{\star}$ distributions.}
    \label{fig:Mstar_z}
\end{figure}

In \citetalias{lamarcaDustPowerUnravelling2024}, we constructed stellar mass--limited samples within three redshift bins:
\begin{itemize}
    \item $0.1\leq z < 0.31$, with $M_{\star}\geq10^9\,M_{\odot}$;
    \item $0.31\leq z < 0.52$, with $M_{\star}\geq10^9\,M_{\odot}$;
    \item $0.52\leq z < 0.76$, with $M_{\star}\geq2.5\cdot10^9\,M_{\odot}$.
\end{itemize}
These mass limits were adopted from \citet{wright_kidsviking-450_2019} to ensure mass completeness. After applying this mass selection, the final mass-complete sample comprises 9\,880, 10\,364, and 22\,593 galaxies in redshift bins 1, 2, and 3, respectively. Figure~\ref{fig:Mstar_z} shows the redshift and stellar mass distributions of the final mass-complete sample. The main panel displays stellar mass versus redshift, while the marginal histograms show the overall redshift distribution and stellar mass distribution for the sample used in this study. 

It is important to note that for the primary analysis presented in this paper (Sect.~\ref{sect:Results}), we explicitly exclude galaxies classified as major mergers in \citetalias{lamarcaDustPowerUnravelling2024}. This exclusion ensures that our investigation focuses on the influence of bars in relatively isolated, disc-dominated systems, reducing potential contamination from merger-induced AGN activity. The impact of including major mergers on our findings is explored separately in Sect.~\ref{sect:Discussion}.

\subsection{SC-SSP imaging and Galaxy Zoo classifications for bar identification}\label{sect:GZ}

To visually identify galactic bars, our work leverages citizen science classifications from multiple GZ projects applied to HSC-SSP survey images.

\subsubsection{HSC-SSP imaging data}\label{sect:HSC_images}

We utilised $i$-band coadded images (which include sky subtraction) from the HSC-SSP Public Data Release 3 \citep[PDR3;][]{aihara_third_2022}, selected for their generally superior seeing compared to other bands. The HSC-SSP {\it Wide} survey layer reaches a depth of approximately $i \sim 26\,{\rm mag}$ ($5\sigma$ for point sources) with an average $i$-band seeing of $0.61\arcsec$ \citep{aihara_third_2022}. For each galaxy in our parent sample (Sect.~\ref{sect:CIGALE}), we downloaded an image cutout with a semi-width and semi-height equal to four times the $i$-band Kron radius, as measured by the HSC-SSP PDR3 pipeline (\texttt{i kronflux radius}).

To ensure reliable morphological classification, we performed stringent image quality control. Each selected galaxy cutout was required not to have any of the following critical flags raised in the HSC-SSP PDR3 catalogue: 
\texttt{i pixelflags edge}(source is outside usable exposure area), \texttt{i pixelflags bad} (bad pixel in the source footprint), \texttt{i pixelflags crcenter} (cosmic ray in the source centre), \texttt{i pixelflags saturatedcenter} (saturated pixel in the source centre), \texttt{i pixelflags interpolatedcenter} (interpolated pixel in the source centre), \texttt{i kronflux flag} (general failure flag for Kron fit), \texttt{i kronradius flag bad radius} (bad Kron radius), or \texttt{i cmodel flag} (flag set if the final cmodel fit, or any previous fit, failed).
We also removed images if their associated mask images contained any of the flags: \texttt{MP BAD} (pixel is physically bad), \texttt{MP SAT} (pixel flux exceeded full-well), \texttt{MP NO DATA} (pixel has no input data), or \texttt{MP UNMASKEDNAN} (a NaN occurred in this pixel in instrument signature removal).

\subsubsection{Galaxy Zoo classifications}\label{sect:GZ_projects}

\begin{figure}
    \centering
    \includegraphics[width=0.49\textwidth]{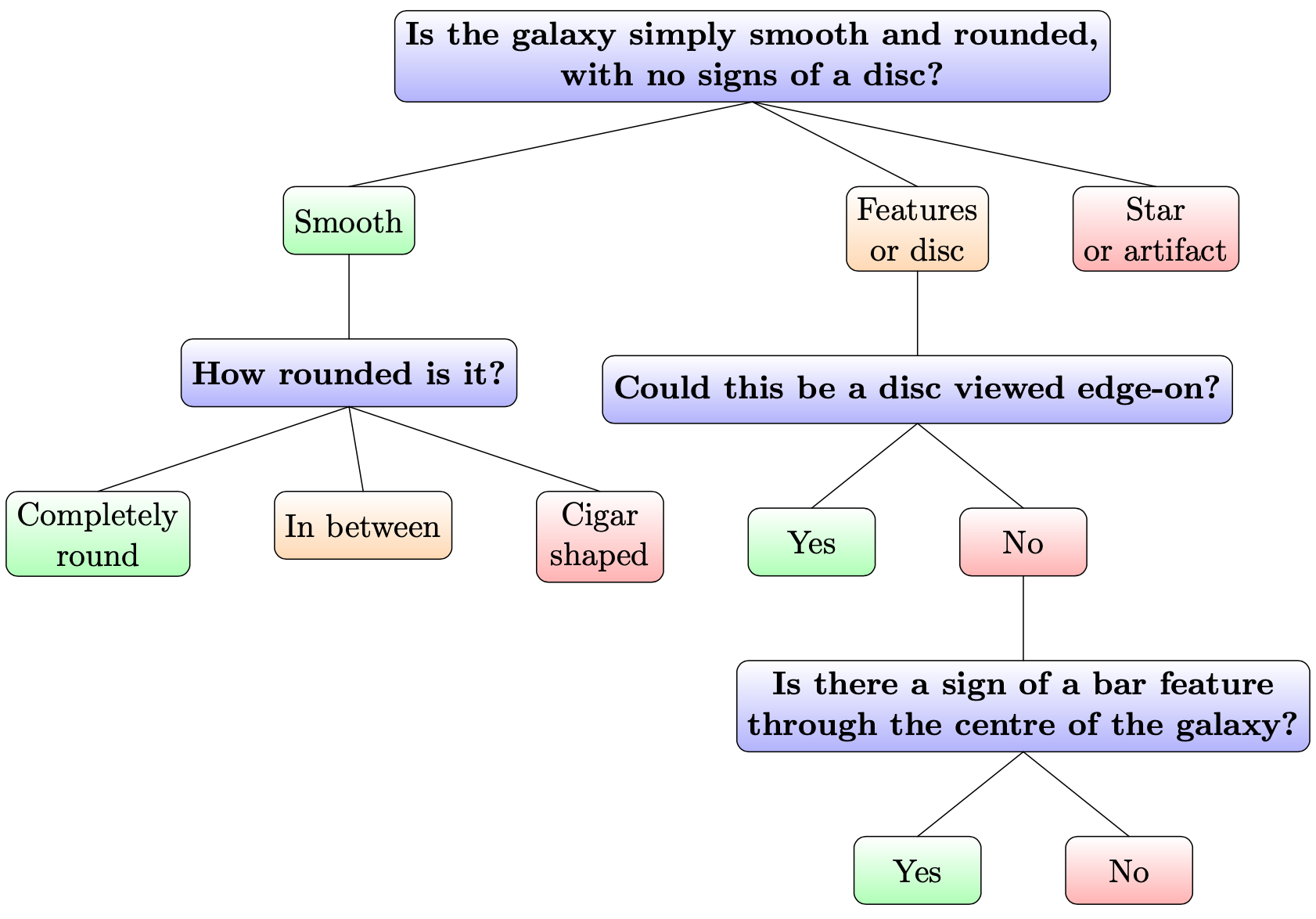}
    \caption{A simplified representation of the relevant questions from the Galaxy Zoo decision tree \citep[adapted from][]{willett_galaxy_2013} used to classify galaxies for the training set.
    }
    \label{fig:tree}
\end{figure}

Bar identifications were sourced from four GZ science projects: GZ Hubble \citep[GZH;][]{willett_galaxy_2017}, GZ2 \citep{hart_galaxy_2016}, GZ DECaLS \citep[Dark Energy Camera Legacy Survey;][]{walmsley_galaxy_2022}, and GZ GAMA-KiDS \citep{Holwerda2024GalaxyZoo}. In these projects, citizen scientists visually classified galaxies by answering a series of questions\footnote{More information available at \url{https://data.galaxyzoo.org/}} organised in a decision tree structure \citep[an example relevant to bar identification is shown in Fig.~\ref{fig:tree};][]{willett_galaxy_2013}. Most GZ projects provide a ``weighted'' and an ``unweighted'' vote fraction for each possible answer; we consistently utilised the ``unweighted'' vote fractions, as the impact of weighting is generally minimal \citep{willett_galaxy_2017}.
Given potential overlaps between the GZ survey footprints, we established the following priority for sourcing classifications for a given galaxy: GZ DECaLS, then GZ GAMA-KiDS, then GZH, and finally GZ2. GZ DECaLS was prioritised as it employed a modified decision tree specifically aimed at improving the identification of weak bars \citep{walmsley_galaxy_2022}, which is pertinent to this study. 

\subsubsection{Defining barred and unbarred galaxies }\label{sect:training-sample}

To construct the training dataset for our bar detection model, we focused on specific answers from the GZ decision trees. 
To select barred galaxies, we were primarily interested in responses to the question, ``Is there a sign of a bar feature through the centre of the galaxy?'' (Fig.~\ref{fig:tree}). Volunteers could only answer this question if they first identified the galaxy as disc-like or if it shows any type of feature (i.e. is not smooth), and is not edge-on. We considered only those galaxies for which at least 20 unique users answered the bar/no-bar question. If a galaxy received fewer than 20 votes in its highest-priority GZ project (as defined in Sect.~\ref{sect:GZ_projects}), we consulted the next GZ project in our priority list, ensuring a single, robust bar classification per galaxy. The bar likelihood ($p_{\rm bar}$) for each galaxy represents the fraction of positive ("yes, there is a bar") answers to this question\footnote{In the GZ DECaLS project, the possible answers to the bar question included ``strong bar'', ``weak bar'', and ``no bar''. For consistency with other GZ projects, we defined $p_{\rm bar}$ as the sum of the vote fractions for ``strong bar'' and ``weak bar'' in GZ DECaLS.}. 

Previous GZ studies \citep[e.g.,][]{masters_galaxy_2012, willett_galaxy_2013} established thresholds for $p_{\rm bar}$ based on comparisons with expert visual classifications \citep[e.g.,][]{nair_catalog_2010}. A $p_{\rm bar} \geq 0.5$ is generally considered reliable for identifying strongly barred galaxies. \citet{masters_galaxy_2012} suggested that a likelihood of $0.2 < p_{\rm bar} < 0.5$ often corresponds to weakly barred features, while \citet{willett_galaxy_2013} proposed $p_{\rm bar} = 0.3$ as an optimal threshold to include both strong and weak bars. To create our training classes, we defined:
\begin{itemize}
    \item Strongly barred discs: $p_{\rm bar} \geq 0.5$
    \item Weakly barred discs: $0.2< p_{\rm bar} < 0.5$
    \item Unbarred discs: $p_{\rm bar} \leq 0.2$
\end{itemize}

To effectively train a machine learning algorithm for identifying barred galaxies within a diverse galaxy population, it is essential to provide examples of various morphological types. Therefore, in addition to the unbarred disc galaxies defined by $p_{\rm bar} \leq 0.2$, we also included ``smooth'' and ``edge-on'' galaxies in our training dataset. 
\begin{itemize}
    \item Smooth galaxies: These were defined as galaxies for which volunteers answered ``smooth'' to the initial GZ question, ``Is the galaxy simply smooth and rounded, with no signs of a disc?'' (Fig.~\ref{fig:tree}). We further refined this class by excluding galaxies likely to be ``cigar-shaped'' (which can be confused with edge-on discs) by requiring $p_{\rm cigar} \leq 0.1$ in the follow-up ``How rounded is it?'' question. This ensures the smooth class predominantly contains elliptical and S0-type galaxies. 

    \item Edge-on disc galaxies: These were selected using a high threshold in the ``Could this be a disc viewed edge-on?'' question, specifically requiring $p_{\rm edge-on} \geq 0.8$, with at least 10 votes cast for this question.
\end{itemize}
Under these criteria, edge-on galaxies have a median vote count of 17, compared to 30 for the other classes. The inclusion of these distinct classes (strongly barred, weakly barred, unbarred discs, smooth, and edge-on galaxies) ensures our training set captures a comprehensive range of galaxy morphologies relevant to the bar identification task.

\begin{table}[]
    \centering
    \caption{Number of galaxies in the five classes in the final GZ dataset. }
    \begin{tabular}{lr}
        \hline\hline\\[-7pt]
        Class & $N$ \\
        \hline\\[-7pt]
         Strong bar & 4\,105 \\
         Weak bar & 4\,390 \\
         Unbarred disc & 4\,525 \\
         Smooth galaxy & 4\,591 \\
         Edge-on galaxy & 4\,236 \\
         \hline\\[-7pt]
         All samples & 21\,847 \\
    \end{tabular}
    \label{tab:GZ_num}
\end{table}

\begin{figure}
    \centering
    \includegraphics[width=0.49\textwidth]{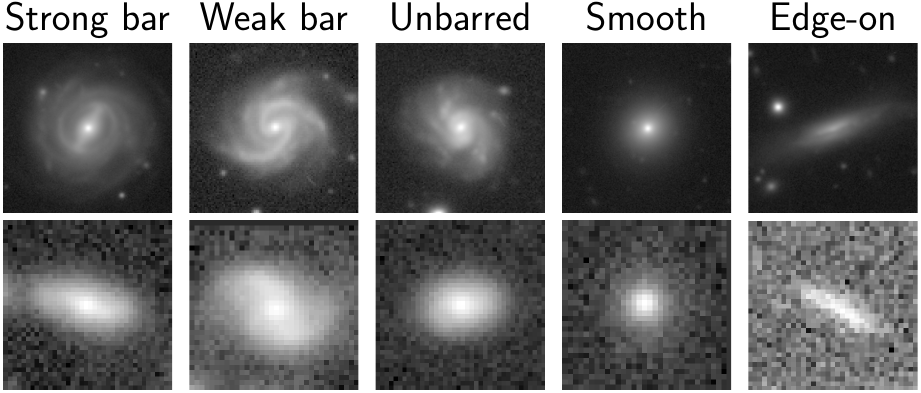}
    \caption{Examples of GZ training galaxies for each class. Top row shows galaxies with $z<0.5$, bottom row shows galaxies at $z\geq0.5$. Each cutout has a size of $8\times 8$ Kron radius (measured in the $i$-band). Images are displayed using a logarithmic scaling.}
    \label{fig:examples_zoobot}
\end{figure}

To create a balanced training sample, crucial for optimal machine learning model performance, we aimed for a similar number of objects in each of the five defined morphological classes. The ``strongly barred'' class contained the fewest galaxies, numbering approximately 5\,300. Consequently, we limited all other classes to this size by randomly selecting 5\,300 galaxies from each. During this process, we ensured that no galaxy appeared in more than one class, removing any duplicates. Finally, we downloaded the HSC-SSP $i$-band image cutouts for these selected galaxies and applied the image quality cuts detailed in Sect.~\ref{sect:HSC_images}.
The final GZ training dataset consists of 21\,847 galaxies with redshifts up to $z \sim 1$, almost equally divided among the five classes: strongly barred, weakly barred, unbarred disc, smooth, and edge-on. The final classifications were sourced from the GZ projects in the following proportions: 51\% GZH, 35\% GZ DECaLS, 12\% GZ GAMA-KiDS, and 2\% GZ2. Table~\ref{tab:GZ_num} reports the precise number of objects in each class. Figure~\ref{fig:examples_zoobot} shows examples of galaxies from these different classes at various redshifts, illustrating the visual characteristics captured in our training set.


\section{Methodology}\label{sect:Methods}

In this section, we first describe the DL approach used to identify galactic bars in the HSC-SSP survey images. 
Secondly, we outline the three different methods employed to select AGN within our sample. Finally, we explain the construction of control samples, essential for robustly assessing the bar-AGN connection.

\subsection{Detecting bars with \texttt{Zoobot}}\label{sect:CNN-perf}

To perform the multi-class morphological classification task defined by the five galaxy types established in Sect.~\ref{sect:training-sample} (strongly and weakly barred, unbarred discs, smooth, and edge-on galaxies), we employed Convolutional Neural Networks (CNNs). CNNs are a class of DL models particularly effective for image analysis, featuring multiple hidden layers capable of automatically learning and extracting hierarchical features from input images \citep{lecunGradientbasedLearningApplied1998, krizhevsky2012imagenet}. The output of a CNN is typically a set of scores, one for each predefined class, which are then used for classification. Training a CNN involves passing a large number of labelled images through its architecture and iteratively adjusting the network's weights to minimise the difference between its predictions and the true labels.

We utilised the \texttt{Zoobot} Python package \citep{walmsley_galaxy_2022}, which provides several DL models pre-trained on vast datasets of GZ volunteer responses. For this work, we fine-tuned the `zoobot-encoder-convnext\_nano' pre-trained model. This model architecture is based on the ConvNeXt family of CNNs \citep{liu2022convnet}, and was originally trained by \citet{walmsley_zoobot_2023} on approximately 820\,000 images and over 100 million volunteer responses from all major GZ campaigns, including GZ DECaLS, GZ2, and GZH. Fine-tuning allows us to adapt this powerful, general-purpose morphology model to our specific five-class problem using our curated GZ training dataset.

We randomly split our final GZ dataset of 21\,847 galaxies (Table~\ref{tab:GZ_num}) into three subsets: 70\% for training (15\,293 galaxies), 15\% for validation (3\,277 galaxies), and 15\% for testing (3\,277 galaxies). The training set is used to adjust the model weights, the validation set to monitor performance during training and optimise hyperparameters (preventing overfitting to the training data), and the test set for a final evaluation of the model's performance on unseen data. Input images were resized to $224 \times 224$ pixels. To increase the effective size of our training set and make the model more robust to variations, we implemented data augmentation during the training phase. This involved applying a random horizontal flip (with a 50\% probability) and a random rotation by multiples of 90 degrees (also with a 50\% probability) to each image in a batch before feeding it to the model.

We employed an early stopping criterion during training, with a patience of 10 epochs. This means training was halted if the validation loss did not improve for 10 consecutive epochs, preventing overfitting. Hyperparameter optimisation was performed via a grid search over different values for batch size, the number of unfrozen blocks in the pre-trained encoder, and the learning rate. The learning rate decay factor was kept fixed at 0.5. The best performance, as determined by the minimum validation loss, was achieved by unfreezing and training the last 4 (out of 5) blocks of the ConvNeXt encoder, using a batch size of 128, and an initial learning rate of $10^{-3}$.

\subsubsection{Model predictions and performance}

\begin{table}
\centering
\caption{Performance of the \texttt{Zoobot} model trained on the five classes task.}
\begin{tabular}{lccc}
\hline\hline\\[-7pt]
Class & Precision & Recall & $F_1$-score \\
 & (\%) & (\%) & (\%) \\
\hline\\[-7pt]
Strong bar & 81.41 & 60.77 & 69.59 \\
Weak bar & 46.12 & 50.24 & 48.09 \\
Unbarred disc & 52.76 & 63.96 & 57.82 \\
Smooth galaxy & 81.90 & 75.76 &  78.71 \\
Edge-on galaxy & 91.93 & 94.42 & 93.15 \\
\hline
\end{tabular}
\tablefoot{Precision, recall, and $F_1$-score for each class.}
\label{tab:performance}
\end{table}

\begin{figure}[h]
    \centering
    \includegraphics[width=0.42\textwidth]{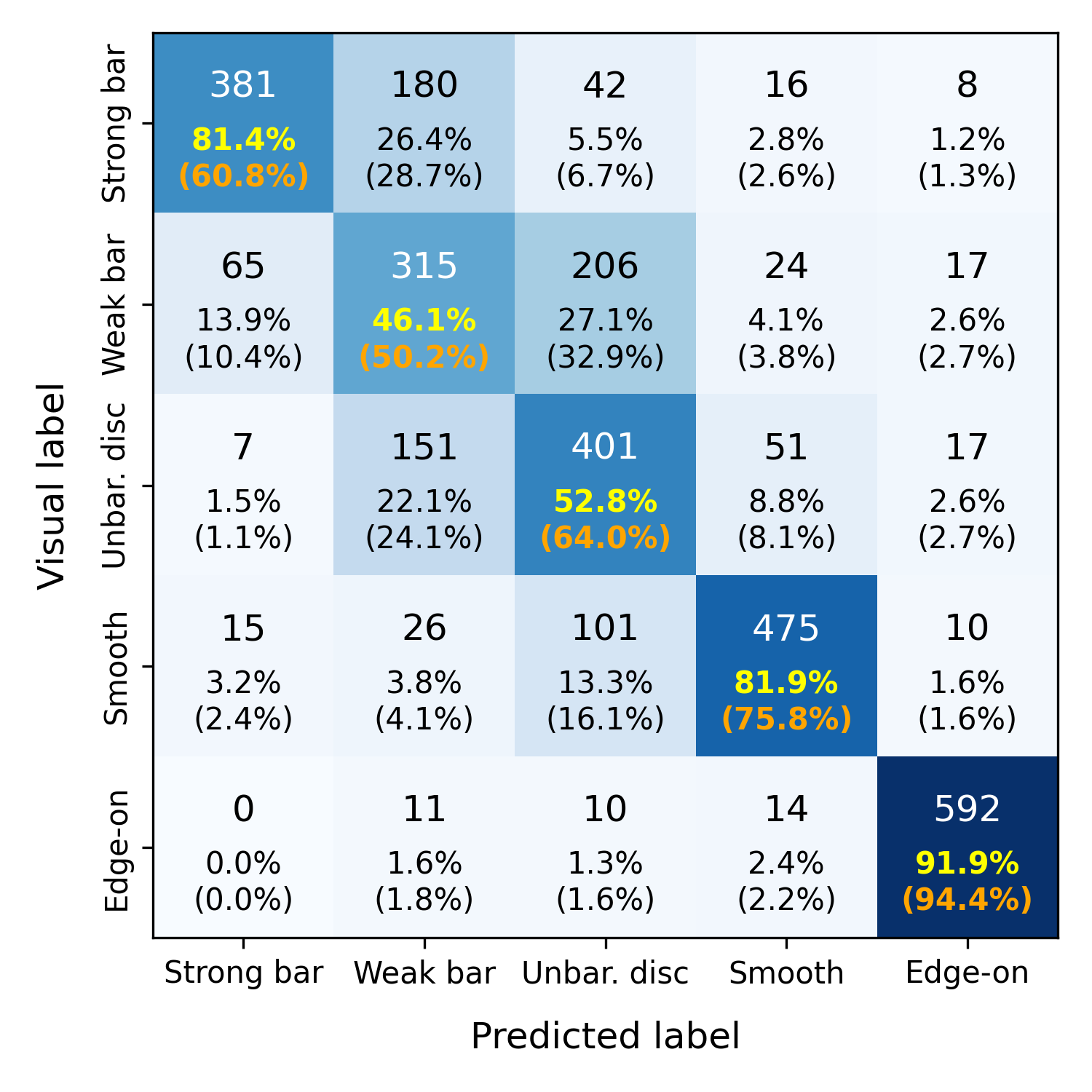}
    \caption{Confusion matrix for the fine-tuned \texttt{Zoobot} model, evaluated on the balanced test set. The matrix is colour-coded relative to the total number of galaxies in each true morphological class (rows). Each cell displays the raw count of galaxies, the column-wise percentage (precision, yellow boldface text on the diagonal), the row-wise percentage (recall, orange text in brackets on the diagonal).}
    \label{fig:confusion_all_classes}
\end{figure}

The fine-tuned \texttt{Zoobot} model outputs five ``prediction probabilities'' for each input galaxy, corresponding to the five morphological classes, which sum to unity. A galaxy was classified as belonging to the class with the highest prediction probability. The overall accuracy of our model -- defined as the ratio of correctly classified objects to the total number of objects -- is 69.0\%, calculated on a balanced test set. The balanced test set was created by randomly selecting an equal number of instances from each true class. To provide a more detailed evaluation of the classifier's performance for each class, we calculated precision, recall, and the $F_1$-score on the balanced test set. Precision for a given class is the ratio of correctly identified galaxies among all galaxies predicted to belong to that class. Recall (or sensitivity) measures the fraction of correctly identified galaxies among all true instances of that class. The $F_1$-score is the harmonic mean of precision and recall, providing a single metric that balances both. These metrics for each class, evaluated on the test set, are reported in Table~\ref{tab:performance}.
The model's performance is also visualised as a confusion matrix in Fig.~\ref{fig:confusion_all_classes}. Each cell in the confusion matrix shows the raw count of galaxies, the precision (as a percentage of predictions for that class), and the recall (as a percentage of true labels for that class). The diagonal elements highlight how well the model predicts each specific class.

Overall, the model demonstrates reasonable performance. As seen in Table~\ref{tab:performance} and Fig.~\ref{fig:confusion_all_classes}, the `Edge-on galaxy' class achieves the highest $F_1$-score (93.15\%), with high precision (91.93\%) and recall (94.42\%), indicating these are relatively easy to distinguish. `Strong bar' and `Smooth galaxy' classes also show good $F_1$-scores (69.59\% and 78.71\%, respectively). The `Unbarred disc' class has an $F_1$-score of 57.82\%. The `Weak bar' class proves to be the most challenging, with an $F_1$-score of 48.09\%. This is likely due to the inherent subtlety of weak bar features, which can be difficult to distinguish from other central structures like prominent bulges or even some features in unbarred discs, leading to a somewhat fuzzy distinction and consequently lower classification performance. The confusion matrix (Fig.~\ref{fig:confusion_all_classes}) further illustrates these trends, showing, for example, that a notable fraction of true `Weak bar' galaxies are misclassified as `Unbarred disc' or vice-versa.

\subsubsection{Selecting barred and unbarred discs}\label{sect:bar-selection}

After training and evaluating the \texttt{Zoobot} model, we applied it to the full mass-complete galaxy sample (described in Sect.~\ref{sect:CIGALE}) to identify barred and unbarred disc galaxies for our scientific analysis of the bar-AGN connection. The criteria for this selection were optimised to ensure high purity for the defined classes, minimising contamination which could affect our statistical conclusions. We classified galaxies in our sample based on the \texttt{Zoobot} output probabilities ($P_{\rm Strong\,bar}$, $P_{\rm Weak\, bar}$, $P_{\rm Unbar.\,disc}$, $P_{\rm Smooth}$, $P_{\rm Edge-on}$) as follows. A galaxy was classified as barred ($A_{\rm bar}$) if the sum of probabilities for being either strongly or weakly barred was higher than 0.65, $P_{\rm Bar}=P_{\rm Strong\, bar}+P_{\rm Weak\, bar}\geq 0.65$. Within the $A_{\rm bar}$ sample, a galaxy was further classified as strongly barred ($S_{\rm bar}$) if $P_{\rm Strong\, bar} \geq P_{\rm Weak\, bar}$. Conversely, a barred galaxy was classified as weakly barred ($W_{\rm bar}$) if $P_{\rm Strong\, bar} < P_{\rm Weak\, bar}$. 

A galaxy was classified as an unbarred disc ($U_{\rm bar}$) if its probability of being an unbarred disc was $P_{\rm Unbar.\, disc}>0.45$ {\it and} this probability was significantly higher than the next highest probability among the other four classes. Specifically, we required $\Delta P = P_{\rm Unbar.\, disc} - P_{\rm second\,highest}> 0.1$, where $P_{\rm second\,highest}$ is the probability of the class with the second-highest score from the \texttt{Zoobot} output. This additional criterion was imposed to create a sample of unbarred discs with high purity, minimising contamination from galaxies that might have ambiguous classifications (e.g., potentially weak bars or smooth galaxies with disc-like features). Examples of galaxies classified as strongly barred, weakly barred, and unbarred discs using these criteria are shown in Appendix~\ref{app:example_img} (Fig.~\ref{fig:examples}).

\begin{figure}
    \centering
     \includegraphics[width=0.25\textwidth]{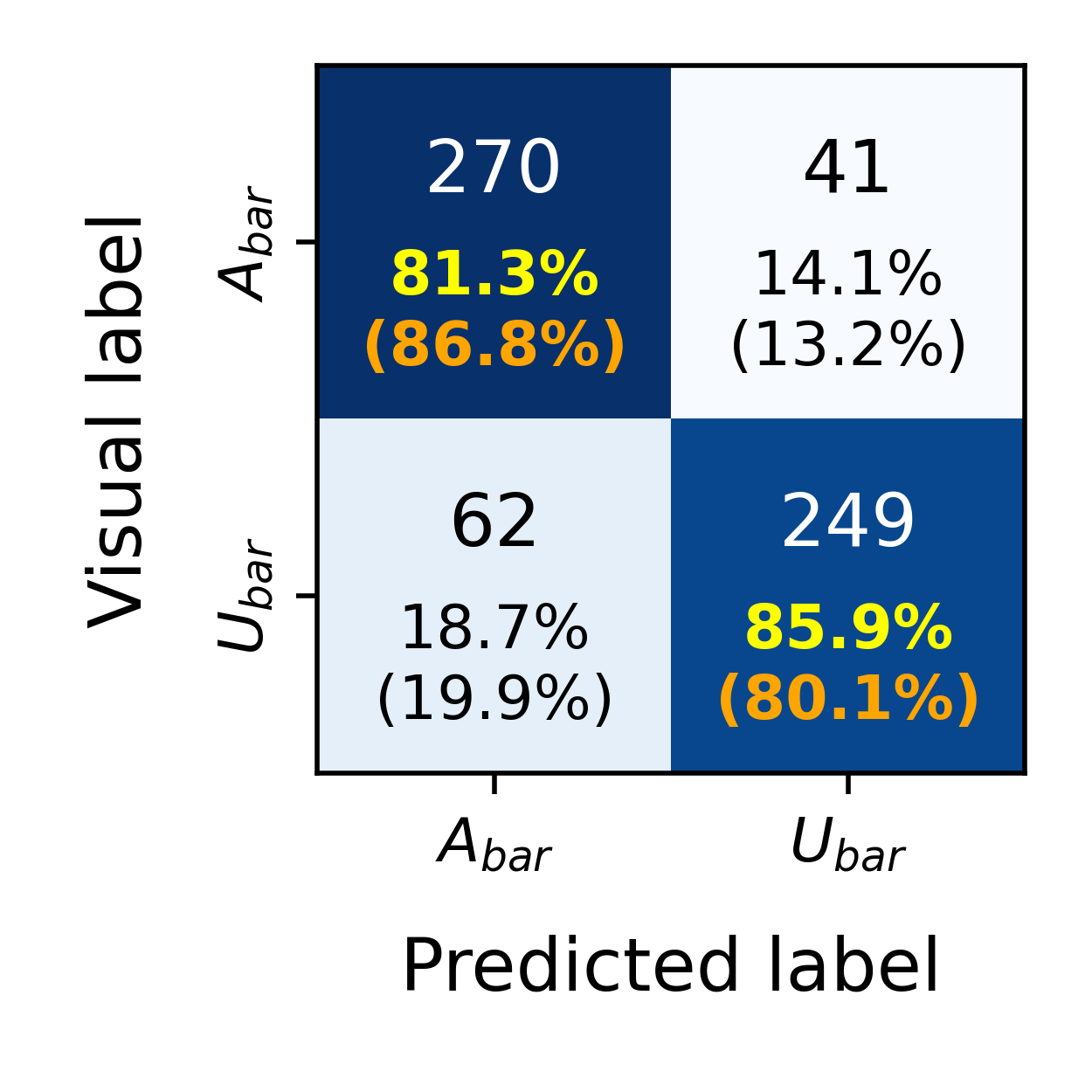}
    \includegraphics[width=0.32\textwidth]{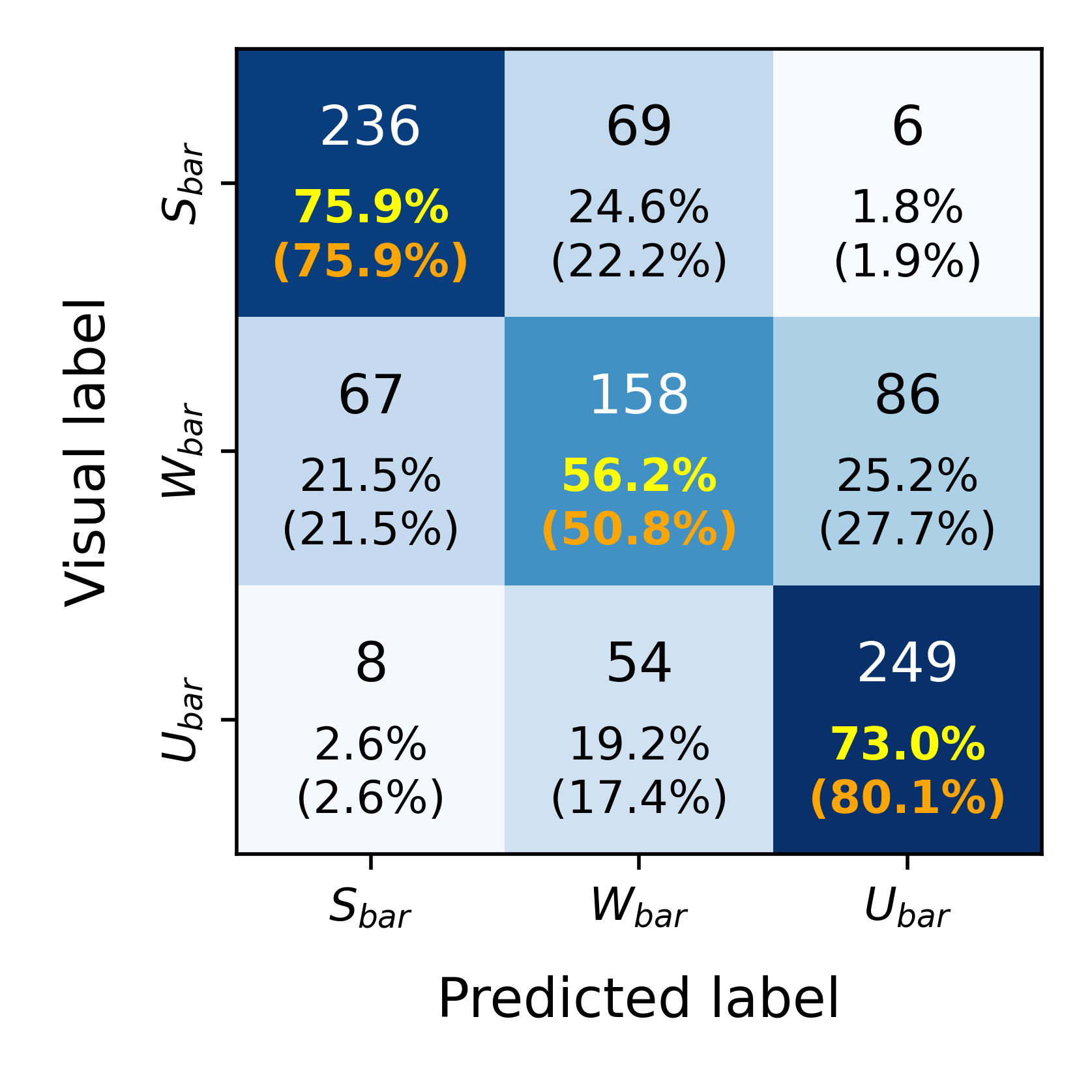}
    \caption{Confusion matrices for the sample selected using the criteria described in Sect.~\ref{sect:bar-selection}, colour-coded according to the total number of galaxies in each row. The content of each cell is the same as in Fig.~\ref{fig:confusion_all_classes}. 
    \textit{Top}: Confusion matrix considering all bars as a single class ($A_{\rm bar}$). We selected the same number of $A_{\rm bar}$ and $U_{\rm bar}$ examples.
    \textit{Lower}: Confusion matrix dividing the bars into $S_{\rm bar}$ and $W_{\rm bar}$. 
    }
    \label{fig:confusion_strong_weak_nobar}
\end{figure}

To assess the impact of these selection criteria on classification performance, we evaluated them on the same balanced test set used in Sect.~\ref{sect:CNN-perf}. The results are presented as confusion matrices in Fig.~\ref{fig:confusion_strong_weak_nobar}. The top panel of Fig.~\ref{fig:confusion_strong_weak_nobar} shows the confusion matrix when considering all barred galaxies as a single class ($A_{\rm bar}$) versus unbarred discs ($U_{\rm bar}$). Note that we randomly select $A_{\rm bar}$ galaxies to have the same number as $U_{\rm bar}$ galaxies. For the $A_{\rm bar}$ class, the model achieves a precision of 81.3\% and a recall of 86.8\%. For the $U_{\rm bar}$ class, the precision is 85.9\% and the recall is 80.1\%. These values indicate a good ability to distinguish barred from unbarred discs with relatively high purity for both classes. The stricter criteria for $U_{\rm bar}$ (especially the $\Delta P$ cut) contribute to its high precision, minimising the fraction of contaminants.

The lower panel of Fig.~\ref{fig:confusion_strong_weak_nobar} further divides the barred galaxies into $S_{\rm bar}$ and $W_{\rm bar}$. For $S_{\rm bar}$, the precision is 75.9\%, with a recall of 75.9\% from the visually labelled strong bars in the test set. Most confusion for $S_{\rm bar}$ arises from $W_{\rm bar}$ galaxies, which is expected. For $W_{\rm bar}$, the precision is 56.2\% and the recall is 50.8\%. This class remains the most challenging, with significant confusion with both $S_{\rm bar}$ and $U_{\rm bar}$ galaxies. This is expected given the continuous nature of bar strengths and reflects the inherent difficulty in distinguishing subtle weak bar features. For $U_{\rm bar}$, the precision remains high at 73.0\% (slightly lower than when $A_{\rm bar}$ was a single class), and the recall is 80.1\%. A higher contamination of visually labelled $W_{\rm bar}$ galaxies emerges. 

Overall, these criteria provide a robust separation, particularly between the general `barred' ($A_{\rm bar}$) and `unbarred' ($U_{\rm bar}$) categories, and a reasonably pure sample of `strongly barred' ($S_{\rm bar}$) galaxies. While the `weakly barred' ($W_{\rm bar}$) class has lower purity, its inclusion allows for a more nuanced investigation of bar strength effects. In Appendix~\ref{app:example_img}, we further demonstrate how the adopted definitions, especially for $U_{\rm bar}$, effectively mitigate confusion with the `smooth' galaxy class (Fig.~\ref{fig:confusion_bar_nobar_smooth}).
With this adopted classification scheme, the final mass-complete galaxy sample used for the subsequent analysis (from Sect.~\ref{sect:CIGALE}) contains 3\,174 $U_{\rm bar}$ galaxies and 2\,405 $A_{\rm bar}$ galaxies. Of the barred galaxies, 1\,261 are classified as $S_{\rm bar}$ and 1\,144 as $W_{\rm bar}$.

\subsection{AGN selections}\label{sect:AGN selection}

To investigate the bar-AGN connection, we identified AGN within our mass-complete disc galaxy sample using three different diagnostics (MIR colours, X-ray detections, and SED fitting), leveraging the multi-wavelength data available. 
We selected AGN based on their MIR colours using the criterion $W1-W2 > 0.8\,{\rm mag}$ (Vega system), as proposed by \citet{stern_mid-infrared_2012}. This selection was applied to all galaxies in our sample with WISE $W2\leq15\,{\rm mag}$ (Vega) and a $S/N \geq 5$ in both the $W1$ and $W2$ bands. This method effectively identifies AGN whose hot dust emission dominates the MIR spectrum, yielding a sample of 15 MIR AGN within our final disc galaxy sample. 
We identified X-ray AGN by cross-matching our galaxy sample with the eFEDS main catalogue \citep{salvato_erosita_2022}. Galaxies coincident with an eFEDS X-ray source classified as an AGN in that catalogue were flagged as X-ray AGN. This selection resulted in 143 X-ray AGN.
Following the methodology in \citetalias{lamarcaDustPowerUnravelling2024}, we identified AGN based on the results of our \texttt{CIGALE} SED fitting. In this work, however, we adopted a more conservative selection criterion, classifying a galaxy as an SED AGN if its AGN fraction in the $3$--$30,\mu{\rm m}$ range satisfies $f_{\rm AGN} \geq 0.1$. This threshold was chosen to ensure a more robust identification of AGN-dominated sources and is motivated by dedicated tests showing that galaxies with $f_{\rm AGN} \geq 0.1$ exhibit a significantly worse SED fit when the AGN component is omitted (see Appendix~\ref{app:agn_thresh}). Using this criterion, we identified 852 SED AGN.
 
\begin{figure}
    \centering
    \includegraphics[width=0.33\textwidth]{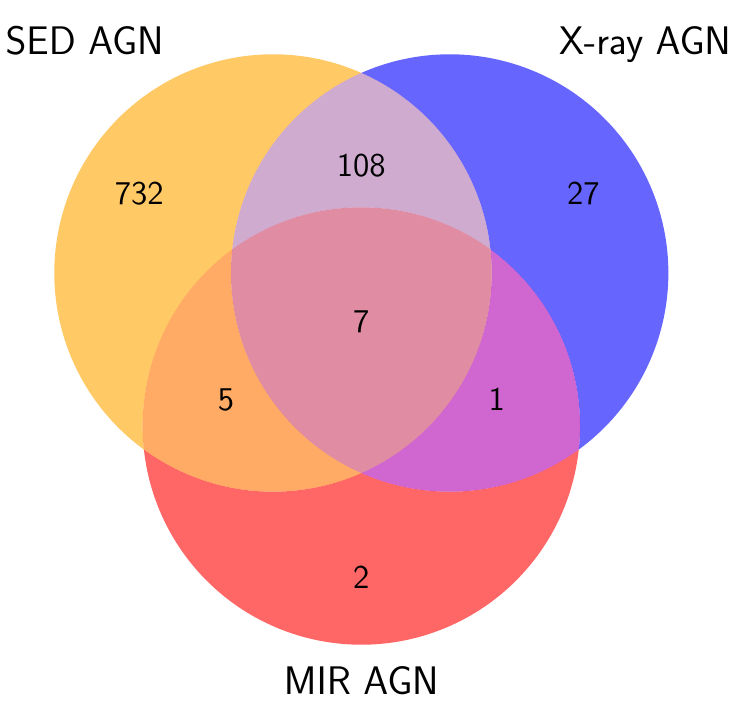}
    \caption{Venn diagram showing the number of unique and overlapping AGN identified by the three selection methods within the combined sample of barred and unbarred disc galaxies analysed in this work.}
    \label{fig:AGN_counts}
\end{figure}

Figure~\ref{fig:AGN_counts} presents a Venn diagram illustrating the overlap between these three AGN selection methods for the total sample of barred and unbarred disc galaxies used in this work. While there is some overlap, with certain AGN being identified by multiple methods (e.g., 7 AGN are identified by all three methods, and 108 are common to SED and X-ray selections), each technique also uniquely identifies a distinct subset of AGN. This highlights the complementary nature of these selection methods in capturing different facets of the AGN population.

\begin{figure}
    \centering
    \includegraphics[width=0.43\textwidth]{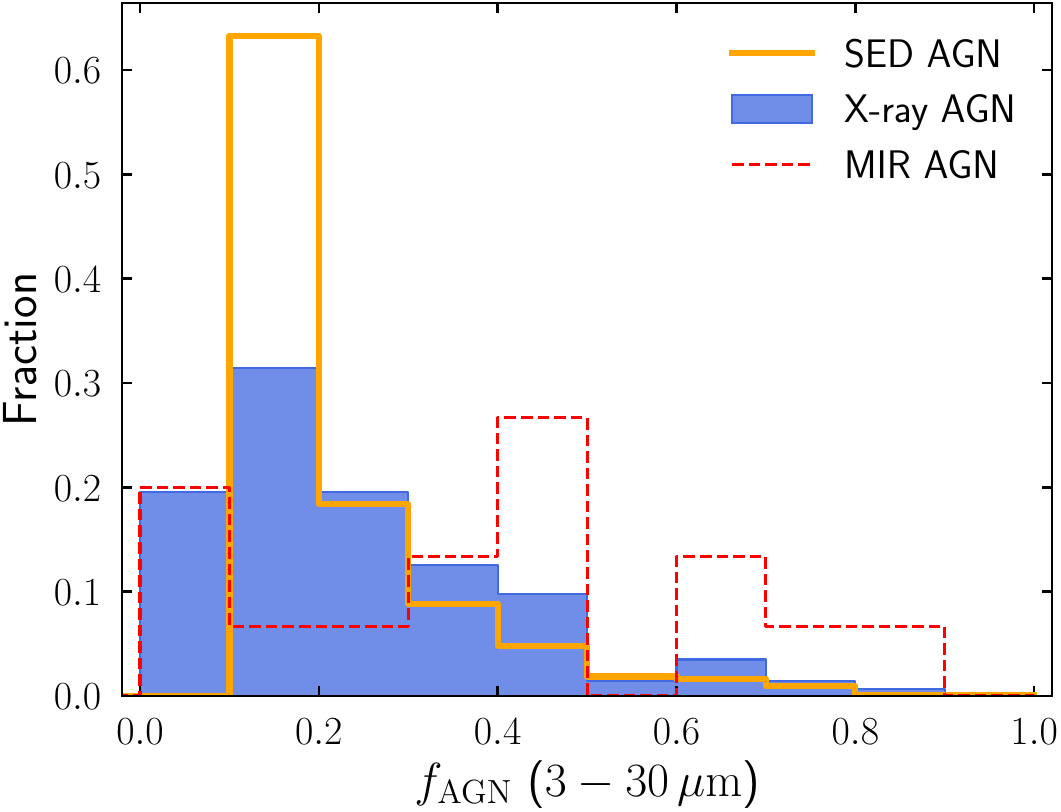}
    \caption{Distributions of AGN fraction, $f_{\rm AGN}(3$--$30\mu {\rm m})$, for the MIR-selected (red dashed line), X-ray-selected (blue filled histogram), and SED-selected (orange solid line) AGN samples. 
    }
    \label{fig:fAGN_dist}
\end{figure}

\begin{figure}
    \centering
    \includegraphics[width=0.49\textwidth]{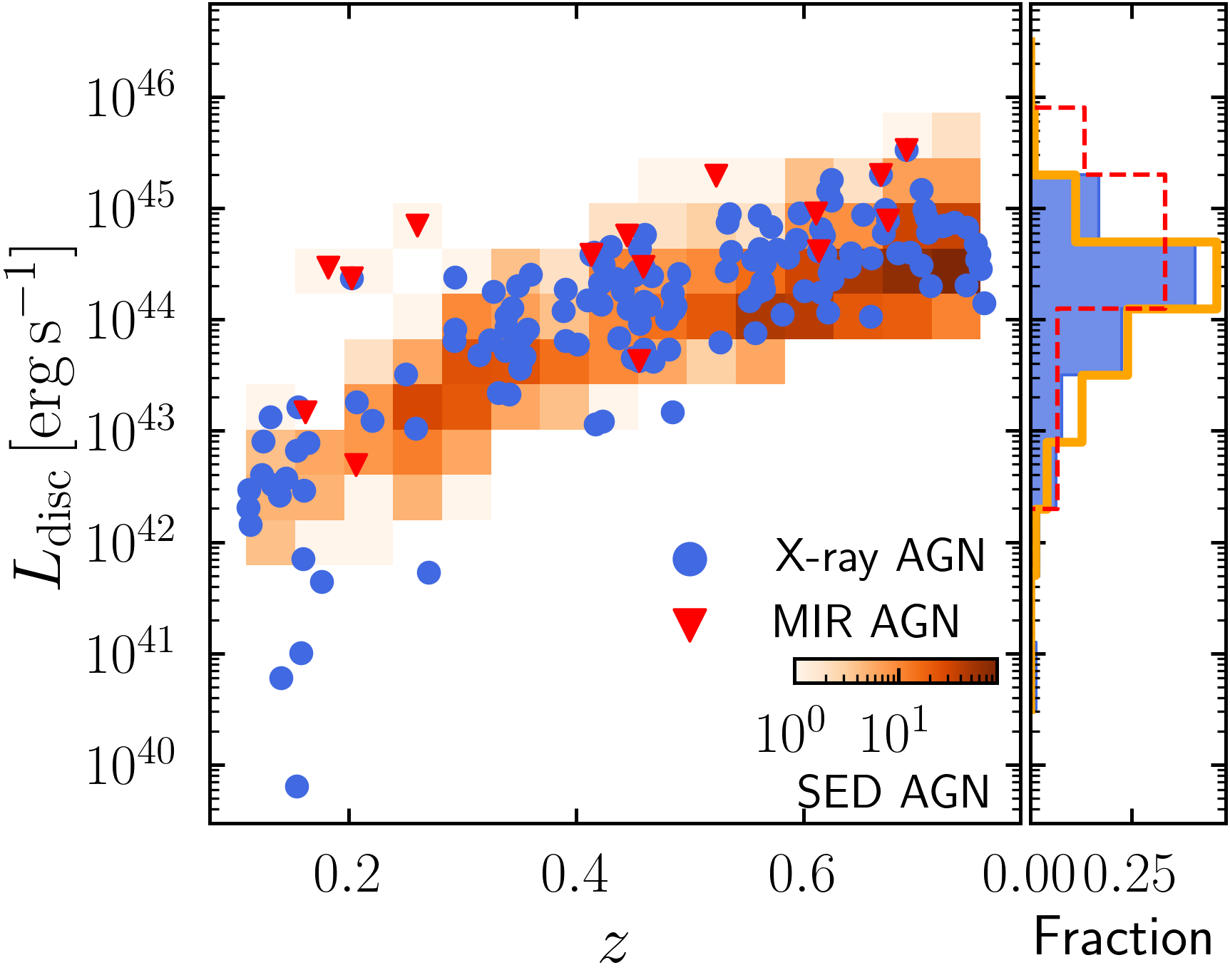}
    \caption{
    Accretion disc luminosity ($L_{\rm disc}$) as a function of redshift for MIR AGN (red triangles), X-ray AGN (blue circles), and SED AGN (2D histogram, colour-coded by the number of sources). The marginal histogram on the right shows the $L_{\rm disc}$ distributions for MIR-selected (red dashed line), X-ray-selected (blue filled histogram), and SED-selected (orange solid line) AGN.
    }
    \label{fig:AGN_lum_z}
\end{figure}

The primary AGN properties derived from the CIGALE SED fitting that we explore in this paper are the AGN fraction, $f_{\rm AGN}\,(3$--$30\mu{\rm m})$, and the AGN accretion disc luminosity, $L_{\rm disc}$, which is computed as the viewing angle-averaged accretion disc luminosity. The accretion disc luminosity considered is the intrinsic luminosity, i.e. before reprocessing by the torus \citep[][]{yang_linking_2018}. Figure~\ref{fig:fAGN_dist} displays the distribution of $f_{\rm AGN}\,(3$--$30\mu{\rm m})$ for each of the three AGN types.
SED AGN, by selection, have $f_{\rm AGN}\geq 0.1$ and show a distribution that generally decreases towards higher $f_{\rm AGN}$ values, with most having $f_{\rm AGN} < 0.3$. X-ray AGN exhibit a similar $f_{\rm AGN}$ distribution with a peak around $f_{\rm AGN}\approx 0.2$, but include objects with $f_{\rm AGN}<0.1$. Although the MIR AGN sample is small, they typically display high $f_{\rm AGN}$ values, consistent with the selection criterion targeting MIR-dominant sources (see \citetalias{lamarcaDustPowerUnravelling2024}).

Figure~\ref{fig:AGN_lum_z} shows $L_{\rm disc}$ as a function of redshift for the three AGN types. The main panel illustrates that for all types, a higher number of AGN, particularly the more luminous ones, are found at relatively higher redshifts ($z > 0.4$) within our sample range (up to $z = 0.76$). The marginal histogram on the right of Fig.~\ref{fig:AGN_lum_z} displays the overall $L_{\rm disc}$ distributions for each AGN type. Both SED and X-ray AGN span a broad range from approximately $10^{42}$ to $\sim 10^{45}\,{\rm erg\,s^{-1}}$. SED AGN present a slightly higher fraction of fainter sources compared to X-ray AGN. The bulk of SED and X-ray AGN have $L_{\rm disc}$ between $10^{44}$ and $10^{45}\,{\rm erg\,s^{-1}}$.
MIR AGN are consistently the most luminous in our sample, with approximately 85\% of them having $L_{\rm disc} > 10^{44}\,{\rm erg\,s^{-1}}$.

\subsection{Control pools}\label{sect:control_pools}

To robustly assess the impact of bars on AGN activity, it is crucial to construct appropriate control samples of unbarred galaxies. This is because AGN occurrence is known to correlate with host galaxy properties such as stellar mass \citep[e.g.,][]{kauffmann_host_2003}, and bar incidence itself can depend on stellar mass and colour \citep[e.g.,][]{masters_galaxy_2011}. Therefore, for each barred galaxy ($S_{\rm bar}$ or $W_{\rm bar}$), we identified a pool of unbarred ($U_{\rm bar}$) control galaxies. These control galaxies were required to simultaneously match the barred galaxy in redshift ($z$), stellar mass ($M_{\star}$), and rest-frame ($g-r$) colour, according to the following criteria:
\begin{align}\label{control_eq}
    &|z_{\rm control}-z_{\rm sample}| \leq \Delta z \times (1+z_{\rm sample})\, ,\\
    &|{\rm log} \; \text{M}_{*,\rm control} - {\rm log} \; \text{M}_{*,\rm  sample} | \leq \Delta \text{M}_* \, ,\\
    &|(g-r)_{\rm control} - (g-r)_{\rm sample}| \leq \Delta(g-r) \, ,
\end{align}
where the initial matching tolerances were set to $\Delta z = 0.05$, $\Delta M_{\star} = 0.1\,{\rm dex}$, and $\Delta (g-r) =0.1$. These values were chosen based on the typical photo-$z$ precision, the median uncertainties in our \texttt{CIGALE}-derived $M_{\star}$, and the median photometric uncertainties in ($g-r$) colour, respectively. For each barred galaxy in the sample, we required at least 10 unique $U_{\rm bar}$ control counterparts satisfying these criteria. If fewer than 10 controls were found, the tolerances were iteratively increased by a factor of 1.5 (up to a maximum of three iterations). If, after these iterations, fewer than 10 controls were still found, the original barred galaxy was excluded from analyses requiring matched controls. When more than 10 controls were available, 10 were randomly selected to form the control pool for that specific barred galaxy. 
This larger population of unbarred galaxies compared to the barred sample minimises potential biases related to host galaxy properties. 

\begin{figure}
    \centering
    \includegraphics[width=.49\textwidth]{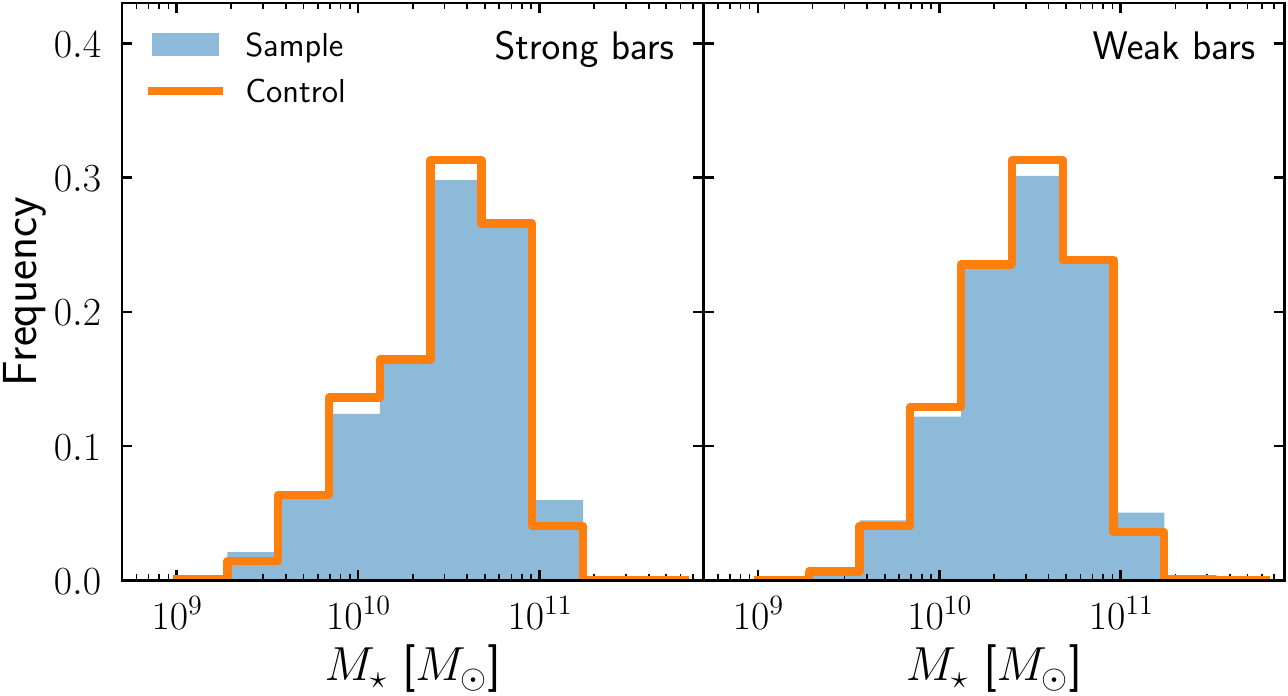}
    \includegraphics[width=.49\textwidth]{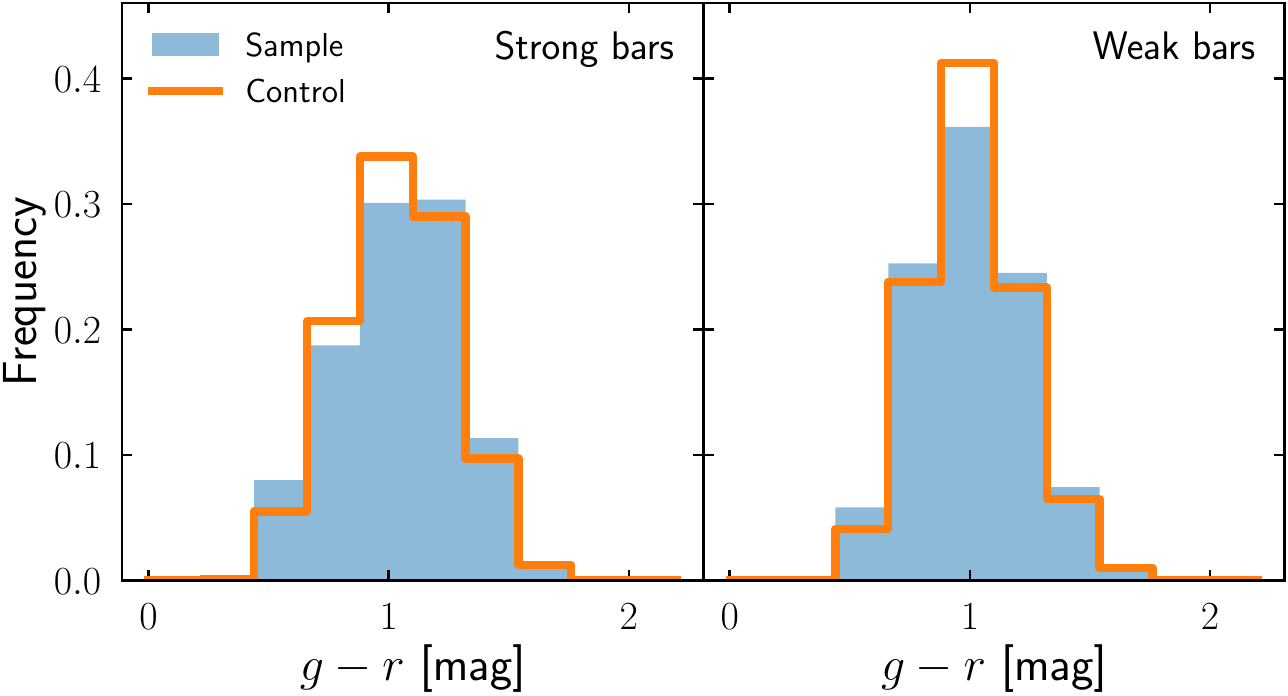}
    \caption{
    $M_{\star}$ (\textit{top panels}) and $(g-r)$ colour (\textit{bottom panels}) distributions for strongly and weakly barred galaxies (blue filled histograms), compared to their corresponding unbarred control samples (thick orange lines). }
    \label{fig:control_mstar}
\end{figure}

We compare the $M_{\star}$ and ($g-r$) distributions for barred galaxies and relative unbarred controls to verify the effectiveness of our control sample matching procedure. Figure~\ref{fig:control_mstar} displays the stellar mass ($M_{\star}$, top panels) and observed ($g - r$) colour (bottom panels) distributions for the $S_{\rm bar}$ and $W_{\rm bar}$ samples (filled blue histograms) compared to their respective $U_{\rm bar}$ control samples (thick orange lines). 
In both cases, the distributions demonstrate excellent agreement between the samples and their controls after the matching process. This confirms that our methodology successfully removes biases related to stellar mass, redshift, and colour when comparing barred and unbarred populations.

\section{Results}\label{sect:Results}

This section presents our findings on the bar-AGN connection. We first analyse this relationship by comparing AGN frequency in barred versus unbarred disc galaxies (binary classification). Subsequently, we explore how relative AGN power ($f_{\rm AGN}$) and absolute AGN luminosity ($L_{\rm disc}$) vary with bar presence and type.

\subsection{Bar-AGN connection using a binary AGN classification}\label{sect:binary_agn_res}

\begin{table*}
\small
\centering
\caption{MIR, X-ray, and SED AGN frequency in $S_{\rm bar}$, $W_{\rm bar}$, and $A_{\rm bar}$ disc galaxies, compared to their respective matched $U_{\rm bar}$ control samples.}
\label{tab:agn_freq}

\begin{tabular}{lccc|ccc|ccc}
\hline \hline\\[-7pt]
Type & AGN freq. & AGN freq. & Excess & AGN freq. & AGN freq. & Excess & AGN freq. & AGN freq. & Excess \\
 & in $S_{\rm bar}$ & in $U_{\rm bar}$ contr. & & in $W_{\rm bar}$ & in $U_{\rm bar}$ contr. & & in $A_{\rm bar}$ & in $U_{\rm bar}$ contr. & \\
\hline\\[-7pt]
MIR & $0.4\pm0.2\%$ & $0.05\pm 0.02\%$ & $8\pm5$ & $0.2\pm 0.1\%$ & $0.05\pm 0.02 \%$ & $4\pm3$ & $0.3\pm0.1\%$ & $0.05\pm0.2\%$ & $6\pm3$ \\
 & (5/1137) & (6/11\,370) & & (2/1049) & (5/10\,490) & & (7/2186) & (11/21\,860) & \\
X-ray & $2.0\pm0.4\%$ & $0.7 \pm 0.1\%$ & $2.9\pm0.7$ & $1.6\pm0.4\%$ & $0.7\pm 0.1\%$ & $2.5\pm0.7$ & $1.8\pm0.3\%$ & $0.7\pm0.1\%$ & $2.7\pm0.5$\\
& (23/1137) & (79/11\,370) & & (17/1049) & (69/10\,490) & & (40/2186) & (148/21\,860) & \\

SED & $11\pm1\%$ & $9.6\pm 0.3\%$ & $1.2\pm0.1$ & $15\pm1\%$ & $14.1\pm0.3\%$ & $1.07\pm0.08$ & $13.2\pm0.7\%$ & $11.8\pm0.2\%$ & $1.12\pm0.06$ \\
& (130/1137) & (1097/11\,370) & & (158/1049) & (1481/10\,490) & &  (288/2186) & (2578/21\,860) & \\
\hline
\end{tabular}

\tablefoot{
The ``Excess'' column reports the ratio of AGN frequency in the barred sample to that in its control sample. Errors are calculated using binomial statistics. Numbers in brackets indicate the raw count of AGN over the total number of galaxies in each specific sample.
}
\end{table*}

\begin{figure}
    \centering
    \includegraphics[width=.49\textwidth]{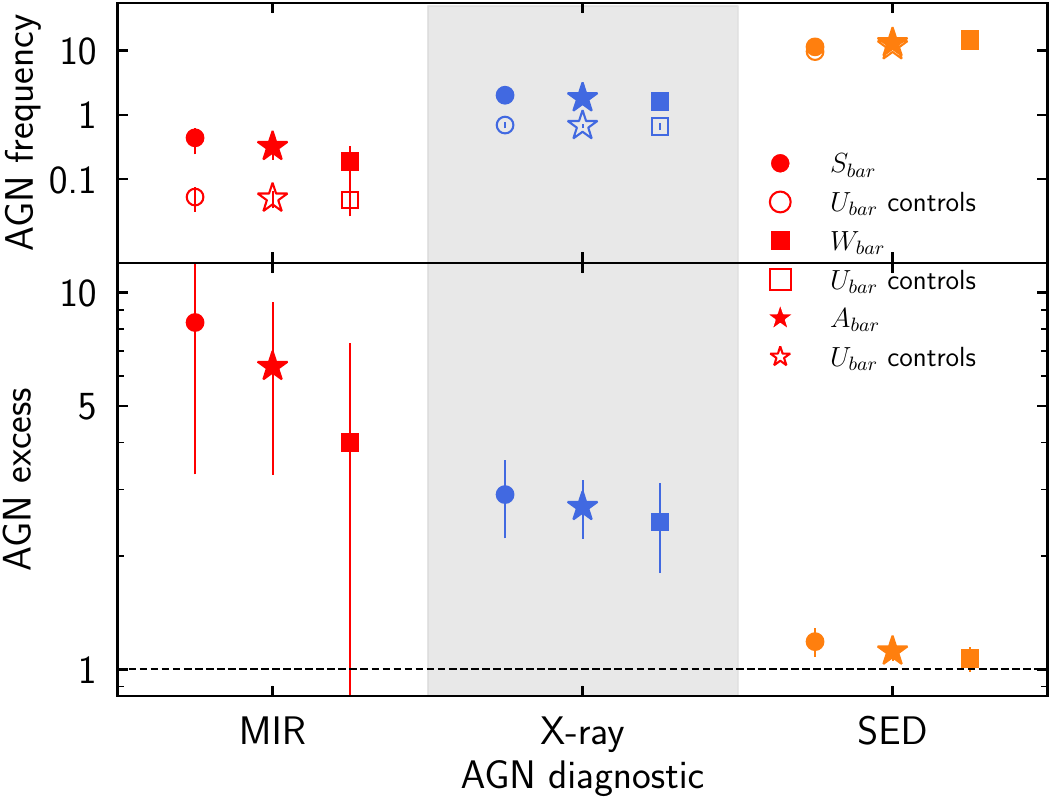}
    \caption{
    AGN frequency and relative AGN excess in barred versus unbarred disc galaxies. \emph{Top panel}: Frequency of AGN (MIR: red, X-ray: blue, SED: orange) in $S_{\rm bar}$ (circles), $W_{\rm bar}$ (squares), $A_{\rm bar}$ (stars), and the corresponding $U_{\rm bar}$ control samples (empty symbols of the same type). Errors are calculated using binomial statistics. 
    \emph{Bottom panel}: Ratio of the AGN frequency in the barred samples to that in their corresponding $U_{\rm bar}$ controls (AGN excess).} 
    \label{fig:agn_excess}
\end{figure}

To assess if galactic bars contribute to AGN fuelling, we compared AGN frequencies (i.e., the fraction of galaxies hosting an AGN) in $S_{\rm bar}$, $W_{\rm bar}$, and $A_{\rm bar}$ samples against their respective $U_{\rm bar}$ control samples, which were carefully matched in redshift, stellar mass, and ($g-r$) colour (Sect.~\ref{sect:control_pools}). The ratio of the AGN frequency in a barred sample to that in its corresponding control sample gives the ``AGN excess'', where a value greater than 1 indicates an enhancement of AGN activity in the presence of a bar. Results are detailed in Table~\ref{tab:agn_freq} and Fig.~\ref{fig:agn_excess}.

Our analysis consistently reveals an enhanced AGN fraction in barred galaxies across all bar types and AGN selection methods. MIR-selected AGN show a notable, though statistically limited due to small numbers, enhancement in barred galaxies, with an AGN excess of approximately 6 for the three bar samples ($\approx 0.3 \pm 0.1\%$ vs. $\approx 0.05\pm0.02\%$ in controls). 
For the X-ray AGN, this enhancement is statistically more robust: the $A_{\rm bar}$ sample hosts X-ray AGN at a frequency of $1.8\pm0.3\%$, significantly higher than the $0.7\pm0.1\%$ in controls, yielding an AGN excess ratio of $2.7\pm 0.5$. Similar excesses are observed for $S_{\rm bar}$ ($2.9\pm0.6$) and $W_{\rm bar}$ ($2.5\pm 0.6$).
For SED AGN, the most prevalent type, the enhancement in barred galaxies is only modest. The $S_{\rm bar}$ sample has an SED AGN frequency of $11\pm1\%$, compared to $9.6\pm 0.3\%$ in the matched control sample (excess ratio of $1.2\pm0.1$). The absolute difference in frequency is 1.4 percentage points. Propagating the binomial errors in quadrature yields an error on this difference of $\approx 1.04\%$, corresponding to a significance of approximately $1.3\sigma$. While this is below the conventional threshold for high statistical significance, a Pearson's chi-squared test on the raw galaxy counts (see Table~\ref{tab:agn_freq}) yields a $p$-value of 0.04, indicating a marginal statistical rejection of the null hypothesis that AGN fraction is independent of the presence of a strong bar. We therefore consider this a tentative detection.

The $W_{\rm bar}$ sample shows a higher frequency of $15 \pm 1\%$ versus $14.1 \pm 0.3\%$ in controls (excess of $1.07\pm0.08$). However, this difference is not statistically significant.
For $A_{\rm bar}$, the frequency is $13 \pm 1\%$ compared to $11.8 \pm 0.2\%$ in their $U_{\rm bar}$ controls, yielding an overall excess of $1.12 \pm 0.06$. The absolute difference in frequency is 1.4 percentage points, with an error of $\approx1.02\%$, corresponding to a significance of $\approx1.4\sigma$. Similarly to the $S_{\rm bar}$ sample, a chi-squared test on the raw counts yields a $p$-value of 0.027. This strengthens the argument that, while the absolute effect is modest, there is a detectable enhancement of SED-selected AGN in the general population of barred galaxies within your sample.

To ensure that the imbalance in sample size (with control samples being $10 \times$ larger than the barred samples) does not introduce a bias, we performed a resampling test. We generated 1000 random subsets of the $U_{\rm bar}$ control samples, each with a size matching the corresponding barred sample ($S_{\rm bar}$, $W_{\rm bar}$, or $A_{\rm bar}$). We found that the median SED AGN fraction of these subsets remained identical, confirming that the full control sample provides an unbiased reference. However, due to the reduced statistical power (and larger error bars) of the smaller subsets, individual $p$-values frequently exceed the 0.05 threshold. For $S_{\rm bar}$ and $A_{\rm bar}$, we observed a $p$-value$<0.05$ in 16.7\% and 19.3\% of the cases, respectively. This confirms that while the excess is present, the statistical significance of the SED-selected AGN result is marginal and relies on the precision provided by the large control sample.

Our finding of an enhanced AGN fraction in barred galaxies is broadly consistent with several previous observational studies. For instance, \citet{galloway_galaxy_2015} found that barred galaxies are more likely to host an AGN compared to unbarred ones. Similarly, \citet{alonso_impact_2018} reported a higher incidence of AGN in local barred galaxies. More recently, \citet{garland_most_2023, garland_galaxy_2024}, using a variety of AGN selection techniques, also found that barred galaxies are more likely to host AGN. 
Some studies, such as \citet{lee_bars_2012} and \citet{cheung_galaxy_2015}, found a higher AGN frequency in barred than in unbarred galaxies, but this difference disappeared after controlling for host galaxy properties.
Indeed, \citet{cisternas_role_2015} showed that differences in bar fractions between active and inactive galaxies can be suppressed after controlling for $M_{\star}$, suggesting that direct comparisons without careful matching can be misleading. Our results, based on meticulously matched control samples, reinforce the idea that bars do play a role in AGN triggering, even if the effect varies with AGN type and luminosity. 

The varying strength of the AGN excess across our three selection methods warrants further discussion. For the MIR-selected sample, the very small number of objects (only 7 AGN in $A_{\rm bar}$ galaxies) precludes a robust physical interpretation of its large excess; while tantalising, this result requires confirmation with much larger samples. For the more numerous X-ray-selected sample, the observed excess could be influenced by selection effects inherent to our methodology. Our analysis excludes, by necessity, edge-on galaxies where bars cannot be identified, which favours face-on systems. Since X-ray emission can be significantly obscured by the high column densities of the host galaxy's interstellar medium (in addition to the AGN torus), our sample may be systematically biased towards less obscured systems, a selection effect that may not be perfectly accounted for in our control matching. More broadly, it is well-established that different selection techniques are sensitive to distinct AGN populations and physical processes, which evolve on different timescales \citep{alexander_what_2012}. A full deconvolution of these effects is beyond the scope of this work, but represents a crucial avenue for future investigation.


\subsection{Bar$-$AGN connection using the relative and absolute AGN power}\label{sect:continuous_exp}

In this Section, we present the results on the bar-AGN connection using the continuous $f_{\rm AGN}$ and $L_{\rm disc}$ parameters, derived from SED fitting. Since the MIR AGN sample is limited to 15 instances, we focus only on X-ray and SED AGN.

\subsubsection{The relative AGN power in barred and unbarred galaxies}\label{sect:fbar_fAGN}

\begin{figure}[]
    \centering
    \includegraphics[width=.49\textwidth]{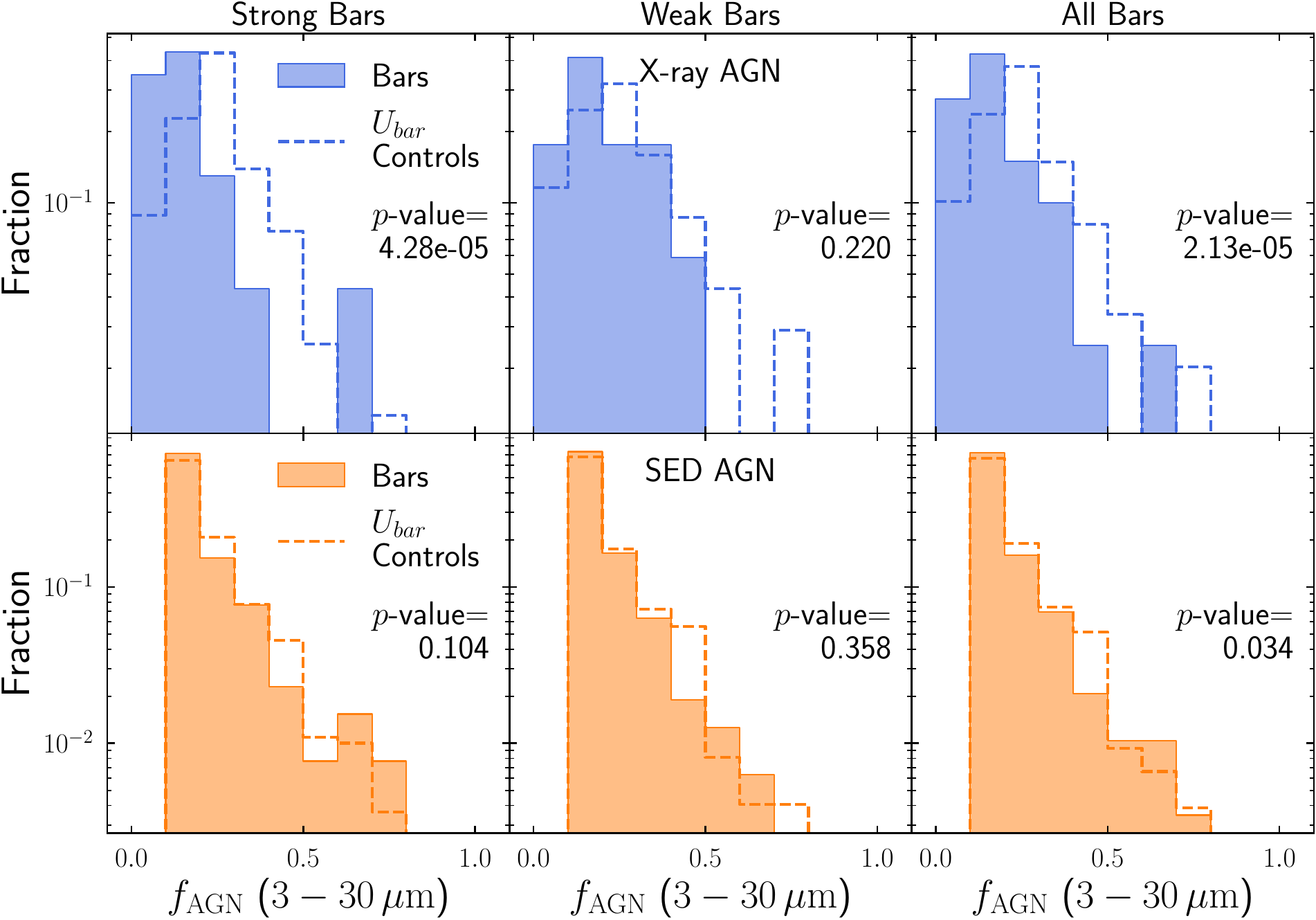}
    \caption{
    Normalised distributions of $f_{\rm AGN}$ for X-ray (top panels) and SED (bottom panels) AGN. Distributions are shown for $S_{\rm bar}$, $W_{\rm bar}$, and $A_{\rm bar}$ galaxies (filled histograms), compared to their respective $U_{\rm bar}$ control samples (dashed lines). The $p$-value from a KS test comparing the barred sample with its control is reported in each panel. 
    }
    \label{fig:fagn_hist}
\end{figure}

We now investigate the bar-AGN connection by examining the distribution of relative AGN power, quantified by the $f_{\rm AGN}\,(3-30\,\mu{\rm m})$ parameter. Figure~\ref{fig:fagn_hist} displays the normalised $f_{\rm AGN}$ distributions for X-ray and SED AGN hosted in $S_{\rm bar}$, $W_{\rm bar}$, and $A_{\rm bar}$ galaxies, compared to their respective $U_{\rm bar}$ control samples. 
We used the two-sample Kolmogorov–Smirnov (KS) test \citep{hodges_significance_1958} to evaluate whether the $f_{\rm AGN}$ distributions of barred AGN differ significantly from those of their unbarred controls (Fig.~\ref{fig:fagn_hist}). The KS test compares the cumulative distributions of two samples; a small $p$-value (typically below a chosen significance level, here we take it to be 0.05) suggests that the two distributions are significantly different. 

For the X-ray AGN, the $f_{\rm AGN}$ distributions in $S_{\rm bar}$ and $A_{\rm bar}$ samples differ from their $U_{\rm bar}$ controls ($p$-value $<0.05$). Specifically, the $U_{\rm bar}$ controls show a higher fraction of galaxies in the $f_{\rm AGN}$ range of $0.2-0.5$. No significant difference is observed between the $W_{\rm bar}$ X-ray AGN and their controls ($p$-value $=0.22$), suggesting their $f_{\rm AGN}$ distributions are consistent with being drawn from the same parent population. For SED-selected AGN, there is no significant differences in the $f_{\rm AGN}$ distributions between the strongly and weakly barred galaxy types ($S_{\rm bar}$, $W_{\rm bar}$) and their corresponding $U_{\rm bar}$ controls ($p$-value$>0.05$). On the contrary, in the $A_{\rm bar}$ case, we cannot rule out the hypothesis that the two distributions are drawn from the same parent distribution, since we obtained a $p$-value $<0.05$.

\begin{figure}
    \centering
    \includegraphics[width=0.49\textwidth]{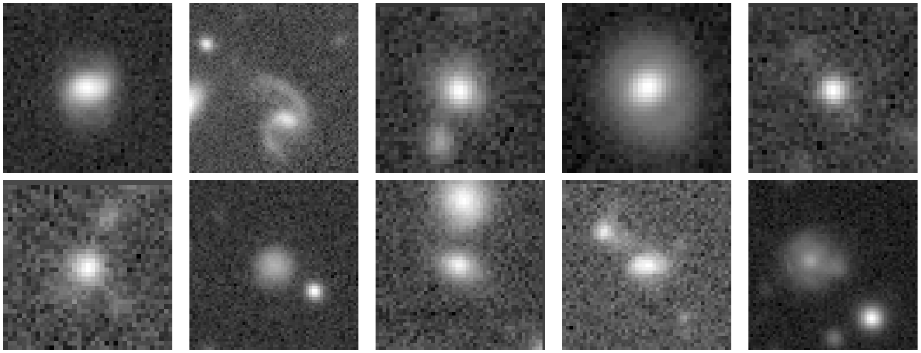}
    \caption{
     HSC-SSP $i$-band image cutouts of the ten disc galaxies from the parent sample with the highest $f_{\rm AGN}$ values (all $f_{\rm AGN} > 0.7$). Cutout size and scaling as Fig.~\ref{fig:examples_zoobot}. On top of each panel is indicated the predicted class.
    }
    \label{fig:AGN_dominant}
\end{figure}

A striking feature across all bar samples and AGN types is the near absence of galaxies with $f_{\rm AGN} > 0.8$. The opposite trend is observed in studies focusing on major mergers \citepalias[e.g.,][]{lamarcaDustPowerUnravelling2024}, which often host such AGN-dominated systems. This scarcity suggests that bar-driven secular processes are unlikely to be the primary drivers of the most AGN-dominated phases. To further explore this, Fig.~\ref{fig:AGN_dominant} shows image cutouts of the ten galaxies in our entire disc galaxy sample (barred and unbarred) with the highest $f_{\rm AGN}$ values (all $f_{\rm AGN} > 0.7$). None of these galaxies exhibit clear bar features, while approximately half show signs of ongoing interaction or disturbed morphologies (i.e., merger features). This observation lends further support to the idea that major mergers, rather than bars, are predominantly responsible for triggering AGN with very high $f_{\rm AGN}$.

\begin{figure*}[h]
    \centering
    \includegraphics[width=0.85\textwidth]{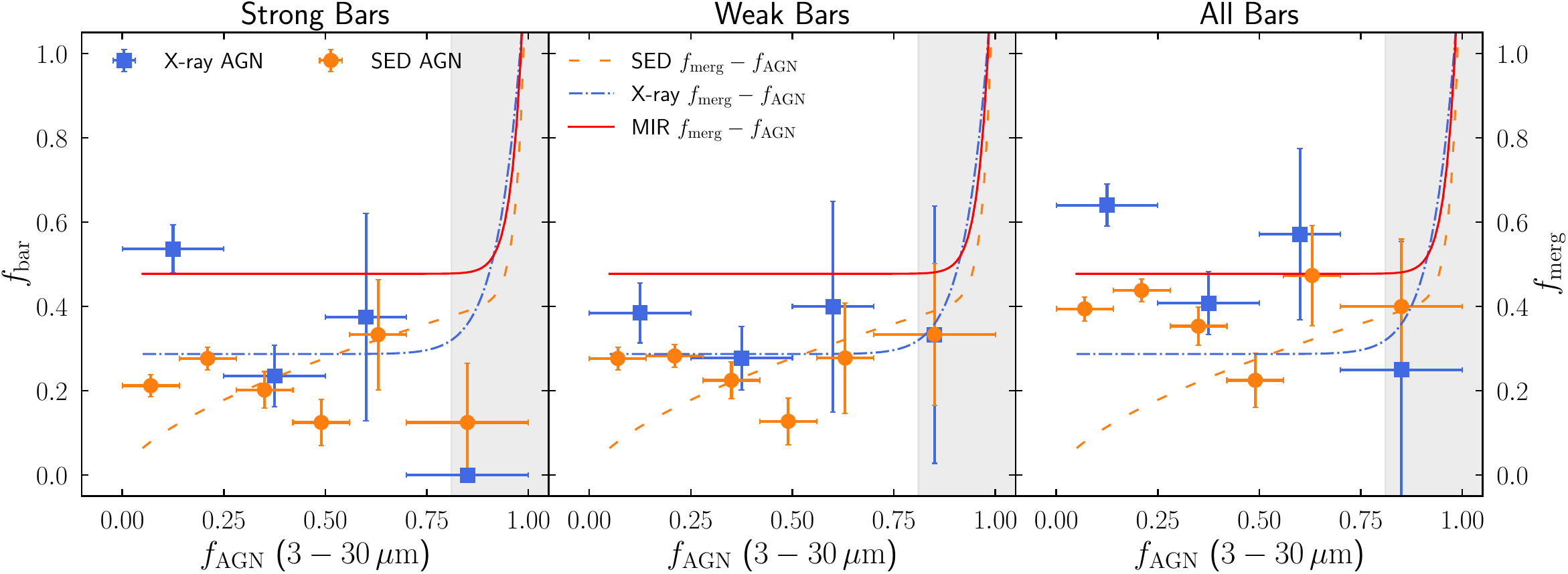}
    \caption{
    Bar fraction ($f_{\rm bar}$) as a function of $f_{\rm AGN}$ for X-ray AGN (blue squares) and SED AGN (orange circles). Results are shown for $S_{\rm bar}$, $W_{\rm bar}$, and $A_{\rm bar}$ separately. The $f_{\rm bar}$ is calculated in $f_{\rm AGN}$ bins using bootstrapping with resampling. The values reported are the median $f_{\rm bar}$, with $1\sigma$ standard deviation as uncertainty, and $x$-axis error bars indicate bin width. 
    Overlaid are the best-fit merger fraction ($f_{\rm merg}$)--$f_{\rm AGN}$ relations from \citetalias{lamarcaDustPowerUnravelling2024} for MIR (red solid line), X-ray (blue dotted-dashed line), and SED (dashed orange line) AGN, parametrised as described in Appendix~\ref{app:fmerg_fagn}. In each subplot, the grey shaded area highlights the region where $f_{
    \rm merg}$ sharply rises as a function of $f_{\rm AGN}$.
    }
    \label{fig:fbar_fagn}
\end{figure*}

To quantify the relationship between bar presence and $f_{\rm AGN}$, we calculated the bar fraction ($f_{\rm bar}$) as a function of $f_{\rm AGN}$. This was done by dividing the number of barred AGN ($S_{\rm bar}$, $W_{\rm bar}$, or $A_{\rm bar}$) by the total number of AGN (barred + unbarred) in bins of $f_{\rm AGN}$. For the X-ray AGN, we used four equally spaced bins between $f_{\rm AGN} = 0$ and $1$. For the SED AGN, given the larger number statistics, we adopted six bins: five equally spaced between $f_{\rm AGN} = 0$ and $0.7$, and one final bin spanning $0.7 < f_{\rm AGN} \leq 1$\footnote{We varied the number and width of the bins, but the overall trends remained unchanged.}. 
The bar fraction is defined as $f_{\rm bar} = N_{\rm bar, \,class} / (N_{\rm bar, \,class} + N_{\rm unbar})$\footnote{$S_{\rm bar}$ and $W_{\rm bar}$ are treated independently; when computing one, the other is excluded from the sample.}.
We used bootstrapping with resampling (1000 samples for each population) for calculating $f_{\rm bar}$ in each $f_{\rm AGN}$ bin. 

Figure~\ref{fig:fbar_fagn} presents the median $f_{\rm bar}$--$f_{\rm AGN}$ relations for X-ray and SED AGN, for strong, weak, and all bars.
For both AGN selections, $f_{\rm bar}$ remains largely constant as a function of $f_{\rm AGN}$, fluctuating between $\sim 20-50\%$ depending on the bar type. Only for the $S_{\rm bar}$ sample, we notice a mild declining trend in $f_{\rm bar}$ with $f_{\rm AGN}$, which is likely due to the poor number statistics.
For the most dominant AGN ($f_{\rm AGN}>0.75$), we observe larger statistical uncertainties due to the smaller numbers of galaxies. As shown already in Fig.~\ref{fig:fAGN_dist}, this near-absence of bars at $f_{\rm AGN}>0.75$ is consistent across all bar classes and both AGN types.

This behaviour contrasts with the merger fraction--AGN fraction ($f_{\rm merg}$--$f_{\rm AGN}$) relation presented for the first time in our previous work \citepalias[][]{lamarcaDustPowerUnravelling2024}. In Fig.~\ref{fig:fbar_fagn}, we overlay the best-fit parametrisations of the $f_{\rm merg}$--$f_{\rm AGN}$ relation from \citetalias{lamarcaDustPowerUnravelling2024} for MIR, X-ray, and SED AGN. The details of how these relations were parametrised and fit are provided in Appendix~\ref{app:fmerg_fagn}. These overlaid lines show that $f_{\rm merg}$ is flat up to $f_{\rm AGN}\approx 0.8$, followed by a rapid increase for $f_{\rm AGN}>0.8$. This opposing trend between $f_{\rm bar}$ and $f_{\rm merg}$ at high $f_{\rm AGN}$ suggests that while both bars and mergers can fuel AGN with low to intermediate $f_{\rm AGN}$, major mergers become the dominant (if not the sole) mechanism for triggering the most AGN-dominated systems ($f_{\rm AGN}>0.8$), where bar-driven fuelling appears inefficient or the bars themselves may be disrupted.

\subsubsection{The absolute AGN power in barred and unbarred galaxies}

\begin{figure}[h]
    \centering
    \includegraphics[width=.49\textwidth]{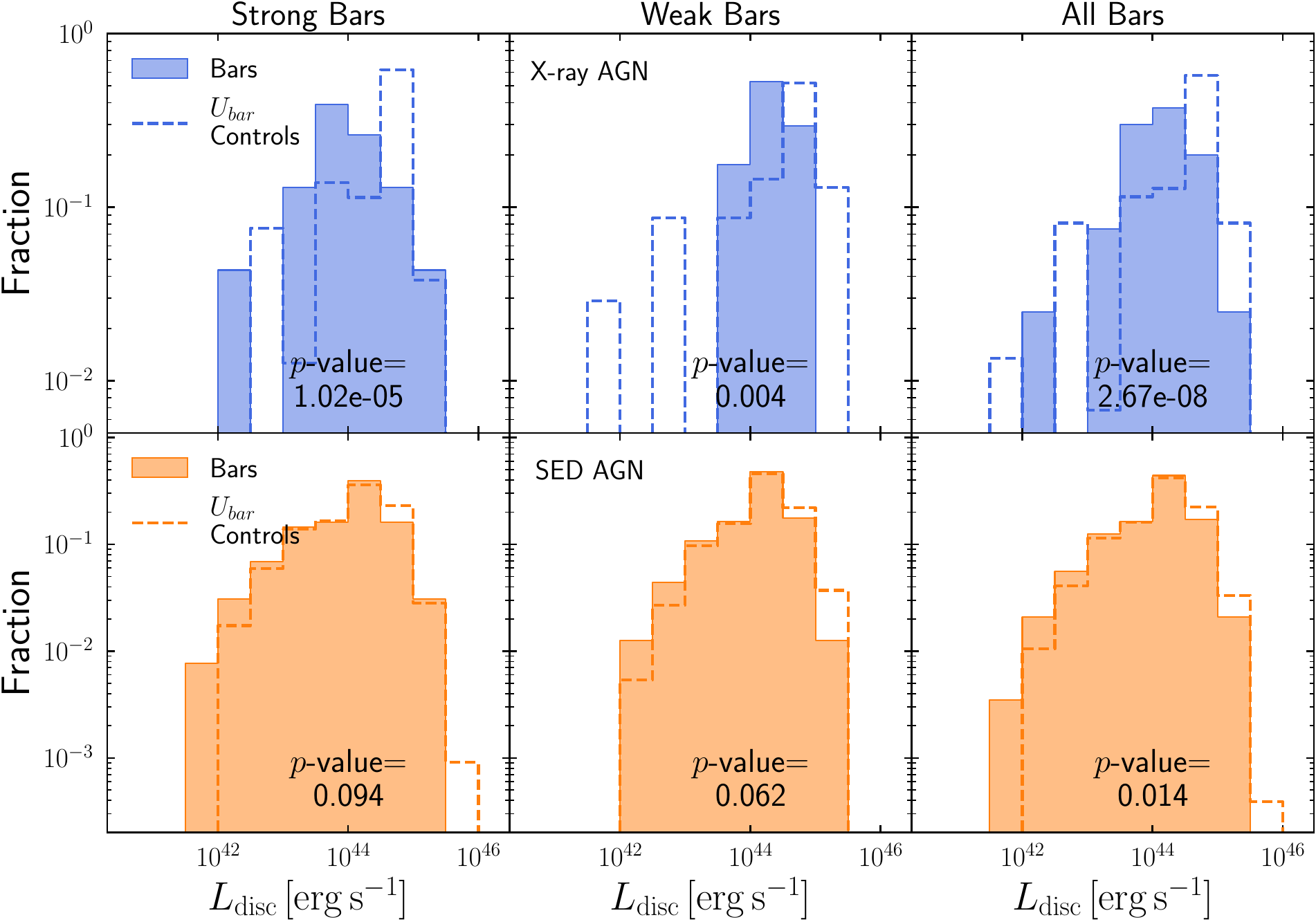}
    \caption{Normalised distributions of AGN accretion disc luminosity, $L_{\rm disc}$, in X-ray (top panels) and SED (bottom panels) AGN. Distributions are shown for $S_{\rm bar}$, $W_{\rm bar}$, and $A_{\rm bar}$ (filled histograms), compared to their respective $U_{\rm bar}$ controls (dashed lines). The $p$-value from a KS test comparing the barred sample with its control is reported in each panel. 
    }
    \label{fig:Ldisc_hist}
\end{figure}

In addition to relative AGN power, we analysed the distributions of absolute AGN accretion disc luminosity ($L_{\rm disc}$) for barred and unbarred galaxies. Figure~\ref{fig:Ldisc_hist} presents the normalised $L_{\rm disc}$ distributions for X-ray and SED-selected AGN hosted in $S_{\rm bar}$, $W_{\rm bar}$, and $A_{\rm bar}$ galaxies, compared to their $U_{\rm bar}$ controls. We used the two-sample KS test to assess the statistical significance of any observed differences. 
For X-ray AGN, the $L_{\rm disc}$ distributions in barred galaxies ($S_{\rm bar}$, $W_{\rm bar}$, and $A_{\rm bar}$) are all statistically different from their respective $U_{\rm bar}$ controls ($p$-values $<0.05$ in all cases). Figure~\ref{fig:Ldisc_hist} shows that the $U_{\rm bar}$ controls tend to host a larger fraction of high luminosity X-ray AGN ($L_{\rm disc}>10^{44.5}\,{\rm erg\,s^{-1}}$) compared to barred galaxies. Conversely, barred galaxies show a relative excess of X-ray AGN at intermediate luminosities ($L_{\rm disc}\approx 10^{43}$--$10^{44.5}\,{\rm erg\,s^{-1}}$).
For SED-selected AGN, however, there are no significant differences in the $L_{\rm disc}$ distributions between $S_{\rm bar}$, $W_{\rm bar}$ and their $U_{\rm bar}$ controls ($p$-values $>0.05$), suggesting that bar presence does not strongly influence the typical luminosities of SED-selected AGN in our sample. When the two samples are combined, $A_{\rm bar}$, a mild difference with the unbarred controls ($p$-value$=0.014$) emerges.

\begin{figure*}
    \centering
    \includegraphics[width=0.85\textwidth]{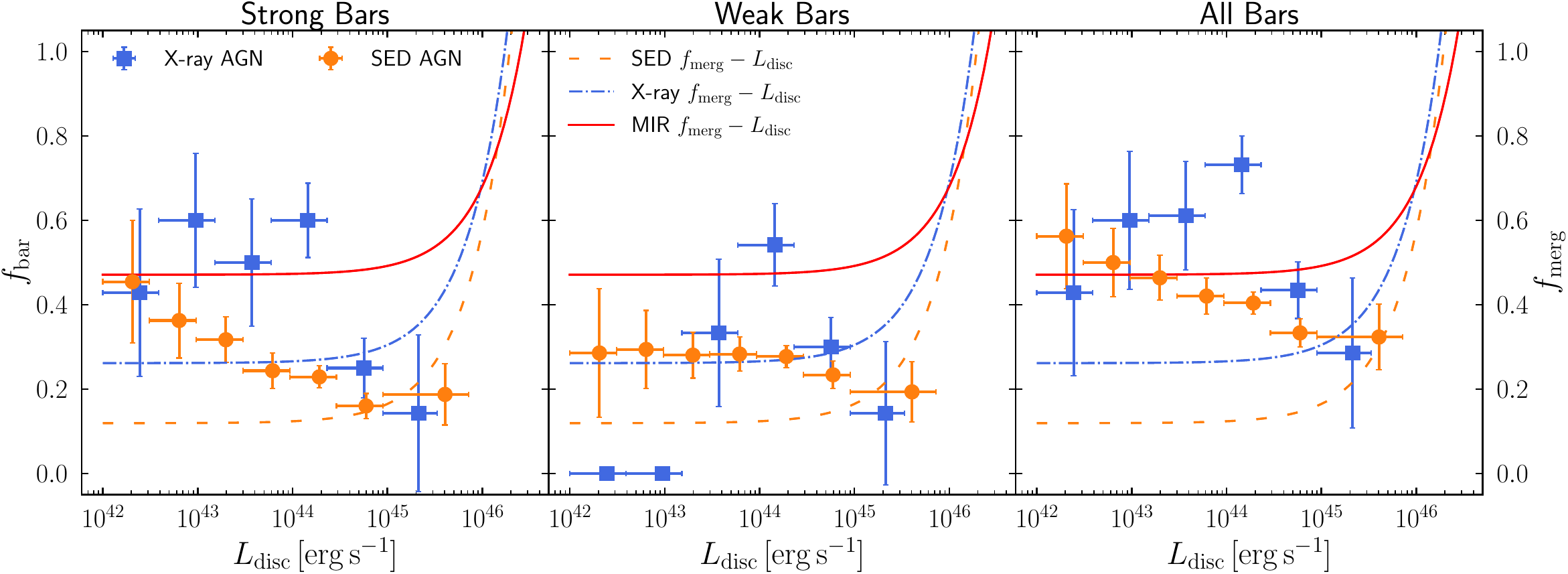}
    \caption{Bar fraction as a function of the AGN total disc accretion luminosity, $L_{disc}$, for X-ray (blue lines) and SED (orange circles) AGN. The $f_{\rm bar}$ is calculated for the $S_{\rm bar}$ and the $W_{\rm bar}$ individually (left and middle panels, respectively), and for the total $A_{\rm bar}$ sample (right panel). Error bars as Fig.~\ref{fig:fbar_fagn}.
    Overlaid are the best-fit $f_{\rm merg}$--$L_{\rm disc}$ relations from \citetalias{lamarcaDustPowerUnravelling2024} for MIR (red solid line), X-ray (blue dotted-dashed line), and SED (dashed orange line) AGN, parametrised as described in Appendix~\ref{app:fmerg_fagn}.
    }
    \label{fig:fbar_Ldisc}
\end{figure*}

To further explore how bar presence relates to AGN luminosity, we calculated $f_{\rm bar}$ as a function of $L_{\rm disc}$. The results are shown in Fig.~\ref{fig:fbar_Ldisc} for X-ray and SED AGN, separately for $S_{\rm bar}$, $W_{\rm bar}$, and $A_{\rm bar}$ classifications. The $f_{\rm bar}$ was calculated in $L_{\rm disc}$ bins (6 for X-ray AGN and 7 for SED AGN), logarithmically spaced between $10^{42}$ and $10^{45}\,{\rm erg\, s^{-1}}$ and an additional bin for AGN brighter than $10^{45}\,{\rm erg\, s^{-1}}$. For SED AGN, for $S_{\rm bar}$ and $A_{\rm bar}$, a clear trend emerges: $f_{\rm bar}$ generally decreases with increasing $L_{\rm disc}$: from $\approx 40-50\%$ at $L_{\rm disc}=10^{42}\,{\rm erg\,s^{-1}}$, to $\approx20-30\%$ at $L_{\rm disc}=10^{45.5}\,{\rm erg\,s^{-1}}$. Similarly, X-ray AGN in the same bar types show a $f_{\rm bar}\approx 50\%$ at $L_{\rm disc}\leq 10^{44}\,{\rm erg\,s^{-1}}$, dropping to $f_{\rm bar}\approx30\%$ for brighter AGN. For $W_{\rm bar}$ SED AGN, $f_{\rm bar}$ roughly stays constant at $30\%$, while X-ray AGN show a larger scatter due to the low number statistics. 
Notably, very few or no barred galaxies are found hosting the most luminous AGN ($L_{\rm disc}\approx 10^{46}\,{\rm erg\,s^{-1}}$) in our sample.

This observed decrease in bar fraction at high AGN luminosities is opposite to the behaviour of the merger-AGN relation, where the merger fraction typically increases with AGN luminosity \citep[e.g.,][\citetalias{lamarcaDustPowerUnravelling2024}]{treisterMajorGalaxyMergers2012, glikmanMajorMergersHost2015}. 
We overlay the $f_{\rm merg}$--$L_{\rm disc}$ parametrised relation in Fig.~\ref{fig:fbar_Ldisc} (we report the details of the parametrisation in Appendix~\ref{app:fmerg_fagn}).
Our findings suggest that while bars may effectively trigger and fuel AGN up to intermediate luminosities, they are not efficient mechanisms for sustaining the most powerful AGN. Several studies support the idea that bars are more commonly associated with lower-to-moderate luminosity AGN activity. For instance, \citet{KnapenSubarcsecond2000} and \citet{LaineNested2002} found connections between bars and Seyfert nuclei, often indicative of moderate accretion. Similarly, in their sample of X-ray AGN, \citet{KossHostGalaxy2011} found that mergers are prevalent among the most luminous AGN, while bars (and other secular processes) might predominantly power moderate luminosity AGN. \citet{Marels2025RoleBars} showed that the fraction of barred luminous AGN is at most $25\%$.
Furthermore, \citet{CisternasXray2013} found no significant correlation between bar strength and SMBH fuelling, suggesting that the extent of the bar-driven inflow is not directly connected with the degree of ongoing accretion (i.e. AGN luminosity). 
The decline we observe at high $L_{\rm disc}$ might imply that either bars cannot channel material at sufficiently high rates to power the brightest AGN, or that the energetic feedback from such luminous AGN could disrupt or destroy the bar structure over time \citep[e.g.,][]{zee_unraveling_2023}. 

The observed $f_{\rm bar}$--$L_{\rm disc}$ trend is in excellent agreement with the emerging paradigm of a dichotomy in AGN fuelling. In this scenario, the most luminous AGN, like quasars, are predominantly triggered by violent galaxy major mergers \citep[e.g.,][]{RamosAlmeida2011OpticalMorphologies, RamosAlmeida2012AreLuminous, treisterMajorGalaxyMergers2012, EuclidCollaboration2025Q1Mergers}. Conversely, less luminous, Seyfert-like activity has been linked to internal secular processes, among which bars are a viable path \citep[e.g.,][]{Cisternas2011SecularEvolution}.

\section{The effect of mergers on the continuous analysis}\label{sect:Discussion}

In our analysis in Sect.~\ref{sect:Results} above, we deliberately excluded galaxies classified as major mergers in \citetalias{lamarcaDustPowerUnravelling2024} to isolate the influence of bars on AGN fuelling in relatively undisturbed disc galaxies. However, it is important to consider how this exclusion might affect our conclusions, particularly regarding the trends observed in the $f_{\rm bar}$ - $f_{\rm AGN}$ and $f_{\rm bar}$ - $L_{\rm disc}$ relations at high AGN dominance/power, where mergers are known to be significant. In this section, we explore two key aspects: first, whether any of the most dominant or luminous AGN, previously excluded as mergers, might also host bars; and second, how the continuous relations change when mergers are re-included in the sample.

\begin{figure}
    \centering
    \includegraphics[width=0.45\textwidth]{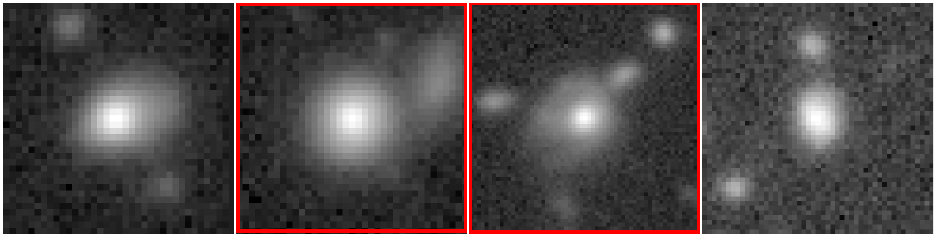}
    \includegraphics[width=0.45\textwidth]{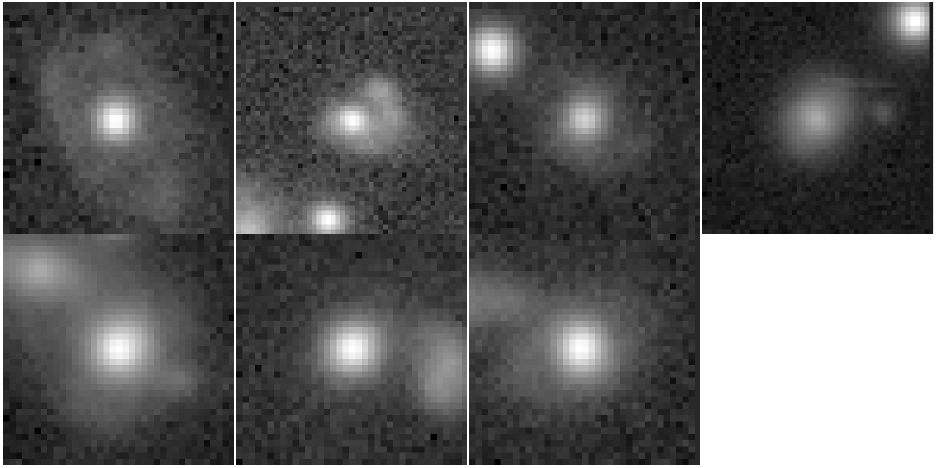}
    \caption{HSC-SSP $i$-band cutouts of the most dominant ($f_{\rm AGN}>0.7$, top row) and luminous AGN ($L_{\rm disc}>10^{44.5}\,{\rm erg\,s^{-1}}$, mid and low rows) in our sample, previously classified as major mergers in \citetalias{lamarcaDustPowerUnravelling2024}. Red boxes highlight the galaxies in both samples. Same scaling as Fig.~\ref{fig:examples_zoobot}.}
    \label{fig:dominant_luminous_examples}
\end{figure}

First, we quantified the overlap between our \texttt{Zoobot} bar classifications and the merger classifications from \citetalias{lamarcaDustPowerUnravelling2024}. Within the parent sample matched between both catalogues, we found that 226 $A_{\rm bar}$ galaxies (102 $S_{\rm bar}$, 124 $W_{\rm bar}$) and 224 $U_{\rm bar}$ discs were classified as major mergers. These figures correspond to a $10\%$ increase in both samples. We then examined the AGN properties of these ``barred mergers''. We found that none of the 226 barred mergers host AGN with $f_{\rm AGN}>0.7$. Similarly, nearly no mergers hosting an AGN with $L_{\rm disc} > 10^{44.5}\,{\rm erg\, s^{-1}}$ were labelled as barred. This indicates that even when considering galaxies that are both barred and undergoing a major merger, such systems do not contribute to the population of extremely dominant or luminous AGN in our sample. Figure~\ref{fig:dominant_luminous_examples}, which displays the $i$-band images of all merger-classified galaxies (barred or unbarred by \texttt{Zoobot}) hosting $f_{\rm AGN}>0.7$ or $L_{\rm disc}> 10^{44.5}\,{\rm erg\,s^{-1}}$, further supports this: none of these dominant/luminous AGN hosts exhibit clear bar structures, while most of them show merging features. This confirms that excluding major mergers did not inadvertently remove a significant population of barred galaxies hosting extremely powerful AGN; such systems appear to be genuinely rare or absent.

Next, we re-evaluated the continuous relations -- $f_{\rm bar}$ as a function of $f_{\rm AGN}$ and $L_{\rm disc}$ -- by including the previously excluded mergers in our analysis. For this test, we recalculated $f_{\rm bar}$ in $f_{\rm AGN}$ and $L_{\rm disc}$ bins, as done in Sect.~\ref{sect:continuous_exp}. The revised $f_{\rm bar}$--$f_{\rm AGN}$ and $f_{\rm bar}$--$L_{\rm disc}$ relations are presented in Appendix~\ref{app:revised_relations}, as Fig.~\ref{fig:fbar_relations_merg}. This experiment revealed that the overall trends presented in Figures~\ref{fig:fbar_fagn} and \ref{fig:fbar_Ldisc} remain qualitatively unchanged. To conclude, these tests robustly indicate that our primary conclusions regarding the inefficiency of bars in fuelling the most dominant (high $f_{\rm AGN}$) and most luminous (high $L_{\rm disc}$) AGN are not significantly biased by the initial exclusion of mergers. The dearth of bars in these extreme AGN regimes appears to be a genuine feature, with major mergers being the more likely drivers of such activity.

\section{Summary and Conclusions}\label{sect:Conclusions}

We investigated the role of galactic bars in fuelling AGN in disc-dominated galaxies up to $z\approx0.8$. Our analysis utilised a large, multi-wavelength catalogue from which we selected galaxies based on HSC-SSP $i$-band imaging \citepalias[][]{lamarcaDustPowerUnravelling2024}. We employed a Deep Learning model \citep[\texttt{Zoobot},][]{walmsley_zoobot_2023}, fine-tuned on Galaxy Zoo classifications, to identify strongly barred, weakly barred, and unbarred disc galaxies. AGN were selected via three independent methods (MIR colours, X-ray detections, and SED fitting with \texttt{CIGALE}), and their properties, specifically the AGN fraction ($f_{\rm AGN}(3$--$30\,\mu{\rm m})$) and accretion disc luminosity ($L_{\rm disc}$), were quantified. The impact of bars was assessed by comparing AGN incidence and properties in barred galaxies against carefully constructed redshift, stellar mass, and colour-matched unbarred control samples. We explicitly excluded major mergers from our primary analysis to isolate bar-driven secular processes, with their impact assessed separately.
Our main findings can be summarised as follows: 

\begin{enumerate}
    \renewcommand{\labelenumi}{\it \roman{enumi})}
    
    \item 
    Barred discs exhibit a higher fraction of AGN compared to their unbarred control counterparts across all three AGN selections. The AGN excess is more pronounced and robust for MIR AGN (factor of $\sim  6$) and X-ray AGN (factor of $\sim 2.7$). For the most numerous SED AGN population, the excess is more modest (factor of $\sim 1.1$), and its statistical significance depends on the control sample size adopted. 
    This consistent trend across multiple independent tracers suggests that galactic bars might contribute to the triggering of AGN activity.
    
    \item The fraction of AGN hosted in barred galaxies ($f_{\rm bar}$) is largely consistent with a flat trend as a function of relative AGN power ($f_{\rm AGN}$), fluctuating between $\sim 20-50\%$. We find a significant scarcity of barred galaxies, whether strongly or weakly, hosting AGN with $f_{\rm AGN}>0.75$. 

    \item Similarly, $f_{\rm bar}$ is flat or mildly decreases with increasing $L_{\rm disc}$. While bars are associated with AGN up to intermediate luminosities (peaking around $L_{\rm disc} \approx 10^{44}\,{\rm erg\, s^{-1}}$ for X-ray AGN), there is a sharp decline in $f_{\rm bar}$ at higher luminosities, with very few or no barred galaxies found hosting the most luminous AGN ($L_{\rm disc}>10^{45}\,{\rm erg\,s^{-1}}$). 
    
\end{enumerate}

Therefore, our work helps to resolve the long-standing debate on the role of bars by defining their operational boundaries: they appear to be a viable, though not dominant, mechanism for sustaining low AGN activity, but are inefficient at fuelling the moderately and most luminous and dominant phases of black hole growth.
Our analysis, including the reintroduction of major mergers, confirms that these extreme AGN are preferentially hosted in unbarred merging galaxies, and rarely, if ever, in clearly barred disc galaxies in our sample. 
In conclusion, our findings align with the previously proposed dual-mode picture of AGN fuelling: internal processes, such as bars, sustain a population of low-luminosity AGN, while major mergers are the primary drivers of the most powerful, high-luminosity AGN.

Future work should incorporate detailed bar morphologies (e.g., bar length) and host galaxy star formation to clarify bar-driven AGN fuelling conditions. Larger, deeper galaxy and AGN samples are crucial for robustly probing these relationships across wider parameter ranges, especially at high AGN power, and for studying their evolution with redshift, a task for which our current sample lacks sufficient statistical power. The forthcoming ESA \textit{Euclid} mission data release \citep{Euclid2025Euclid} will provide vast, high-quality imaging, identifying hundreds of thousands of barred systems \citep{Euclid2025Q1BarFraction}. This unprecedented dataset will enable a more detailed and statistically powerful investigation of bar-driven AGN fuelling and its evolution across cosmic time, decisively advancing our understanding of secular processes in galaxy evolution, particularly at higher redshifts.

\section*{Data Availability}

Catalogues containing the galaxy morphological classification are only available in electronic form at the CDS via anonymous ftp to cdsarc.u-strasbg.fr (130.79.128.5) or via \url{http://cdsweb.u-strasbg.fr/cgi-bin/qcat?J/A+A/}, and from the Zenodo repository \url{https://doi.org/10.5281/zenodo.18299075}


\begin{acknowledgements}
We thank the anonymous referee for their valuable feedback.

The Hyper Suprime-Cam (HSC) collaboration includes the astronomical communities of Japan and Taiwan, and Princeton University. The HSC instrumentation and software were developed by the National Astronomical Observatory of Japan (NAOJ), the Kavli Institute for the Physics and Mathematics of the Universe (Kavli IPMU), the University of Tokyo, the High Energy Accelerator Research Organization (KEK), the Academia Sinica Institute for Astronomy and Astrophysics in Taiwan (ASIAA), and Princeton University. Funding was contributed by the FIRST program from Japanese Cabinet Office, the Ministry of Education, Culture, Sports, Science and Technology (MEXT), the Japan Society for the Promotion of Science (JSPS), Japan Science and Technology Agency (JST), the Toray Science Foundation, NAOJ, Kavli IPMU, KEK, ASIAA, and Princeton University. 

Based on data collected at the Subaru Telescope and retrieved from the HSC data archive system, which is operated by Subaru Telescope and Astronomy Data Center at National Astronomical Observatory of Japan.

Based on data products from observations made with ESO Telescopes at the La Silla Paranal Observatory under programme IDs 177.A-3016, 177.A-3017 and 177.A-3018, and on data products produced by Target/OmegaCEN, INAF-OACN, INAF-OAPD and the KiDS production team, on behalf of the KiDS consortium. OmegaCEN and the KiDS production team acknowledge support by NOVA and NWO-M grants. Members of INAF-OAPD and INAF-OACN also acknowledge the support from the Department of Physics \& Astronomy of the University of Padova, and of the Department of Physics of Univ. Federico II (Naples).

This publication makes use of data products from the Wide-field Infrared Survey Explorer, which is a joint project of the University of California, Los Angeles, and the Jet Propulsion Laboratory/California Institute of Technology, funded by the National Aeronautics and Space Administration.

This work is based on data from eROSITA, the soft X-ray instrument aboard SRG, a joint Russian-German science mission supported by the Russian Space Agency (Roskosmos), in the interests of the Russian Academy of Sciences represented by its Space Research Institute (IKI), and the Deutsches Zentrum für Luft- und Raumfahrt (DLR). 

This publication is part of the project `Clash of the titans:
deciphering the enigmatic role of cosmic collisions' (with project number VI.Vidi.193.113) of the research programme Vidi which is (partly) financed by the Dutch Research Council (NWO).

The training and testing of the CNN were carried out on the Dutch National Supercomputer (Snellius).
We thank SURF (www.surf.nl) for the support in using the National Supercomputer Snellius.

We thank the Center for Information Technology of the University of Groningen for support and access to the H\'abr\'ok high performance computing cluster.

\end{acknowledgements}

\bibliography{Bar-AGN_references}

@article{de_jong_kilo-degree_2013,
	title = {The {Kilo}-{Degree} {Survey}},
	volume = {154},
	issn = {0722-6691},
	url = {https://ui.adsabs.harvard.edu/abs/2013Msngr.154...44D},
	journal = {The Messenger},
	author = {de Jong, J. T. A. and Kuijken, K. and Applegate, D. and Begeman, K. and Belikov, A. and Blake, C. and Bout, J. and Boxhoorn, D. and Buddelmeijer, H. and Buddendiek, A. and Cacciato, M. and Capaccioli, M. and Choi, A. and Cordes, O. and Covone, G. and Dall'Ora, M. and Edge, A. and Erben, T. and Franse, J. and Getman, F. and Grado, A. and Harnois-Deraps, J. and Helmich, E. and Herbonnet, R. and Heymans, C. and Hildebrandt, H. and Hoekstra, H. and Huang, Z. and Irisarri, N. and Joachimi, B. and Köhlinger, F. and Kitching, T. and La Barbera, F. and Lacerda, P. and McFarland, J. and Miller, L. and Nakajima, R. and Napolitano, N. R. and Paolillo, M. and Peacock, J. and Pila-Diez, B. and Puddu, E. and Radovich, M. and Rifatto, A. and Schneider, P. and Schrabback, T. and Sifon, C. and Sikkema, G. and Simon, P. and Sutherland, W. and Tudorica, A. and Valentijn, E. and van der Burg, R. and van Uitert, E. and van Waerbeke, L. and Velander, M. and Verdoes Kleijn, G. and Viola, M. and Vriend, W. -J.},
	month = dec,
	year = {2013},
	pages = {44--46},
}

@article{kuijken_fourth_2019,
	title = {The fourth data release of the {Kilo}-{Degree} {Survey}: \textit{ugri} imaging and nine-band optical-{IR} photometry over 1000 square degrees},
	volume = {625},
	issn = {0004-6361, 1432-0746},
	shorttitle = {The fourth data release of the {Kilo}-{Degree} {Survey}},
	url = {https://www.aanda.org/10.1051/0004-6361/201834918},
	doi = {10.1051/0004-6361/201834918},
	journal = {\aap},
	author = {Kuijken, K. and Heymans, C. and Dvornik, A. and Hildebrandt, H. and de Jong, J. T. A. and Wright, A. H. and Erben, T. and Bilicki, M. and Giblin, B. and Shan, H.-Y. and Getman, F. and Grado, A. and Hoekstra, H. and Miller, L. and Napolitano, N. and Paolilo, M. and Radovich, M. and Schneider, P. and Sutherland, W. and Tewes, M. and Tortora, C. and Valentijn, E. A. and Verdoes Kleijn, G. A.},
	month = may,
	year = {2019},
	pages = {A2},
}

@article{predehl_erosita_2021,
	title = {The {eROSITA} {X}-ray telescope on {SRG}},
	volume = {647},
	issn = {0004-6361, 1432-0746},
	url = {https://www.aanda.org/10.1051/0004-6361/202039313},
	doi = {10.1051/0004-6361/202039313},
	journal = {\aap},
	author = {Predehl, P. and Andritschke, R. and Arefiev, V. and Babyshkin, V. and Batanov, O. and Becker, W. and Böhringer, H. and Bogomolov, A. and Boller, T. and Borm, K. and Bornemann, W. and Bräuninger, H. and Brüggen, M. and Brunner, H. and Brusa, M. and Bulbul, E. and Buntov, M. and Burwitz, V. and Burkert, W. and Clerc, N. and Churazov, E. and Coutinho, D. and Dauser, T. and Dennerl, K. and Doroshenko, V. and Eder, J. and Emberger, V. and Eraerds, T. and Finoguenov, A. and Freyberg, M. and Friedrich, P. and Friedrich, S. and Fürmetz, M. and Georgakakis, A. and Gilfanov, M. and Granato, S. and Grossberger, C. and Gueguen, A. and Gureev, P. and Haberl, F. and Hälker, O. and Hartner, G. and Hasinger, G. and Huber, H. and Ji, L. and Kienlin, A. v. and Kink, W. and Korotkov, F. and Kreykenbohm, I. and Lamer, G. and Lomakin, I. and Lapshov, I. and Liu, T. and Maitra, C. and Meidinger, N. and Menz, B. and Merloni, A. and Mernik, T. and Mican, B. and Mohr, J. and Müller, S. and Nandra, K. and Nazarov, V. and Pacaud, F. and Pavlinsky, M. and Perinati, E. and Pfeffermann, E. and Pietschner, D. and Ramos-Ceja, M. E. and Rau, A. and Reiffers, J. and Reiprich, T. H. and Robrade, J. and Salvato, M. and Sanders, J. and Santangelo, A. and Sasaki, M. and Scheuerle, H. and Schmid, C. and Schmitt, J. and Schwope, A. and Shirshakov, A. and Steinmetz, M. and Stewart, I. and Strüder, L. and Sunyaev, R. and Tenzer, C. and Tiedemann, L. and Trümper, J. and Voron, V. and Weber, P. and Wilms, J. and Yaroshenko, V.},
	month = mar,
	year = {2021},
	pages = {A1},
}

@article{brunner_erosita_2022,
	title = {The {eROSITA} {Final} {Equatorial} {Depth} {Survey} ({eFEDS}): {X}-ray catalogue},
	volume = {661},
	issn = {0004-6361, 1432-0746},
	shorttitle = {The {eROSITA} {Final} {Equatorial} {Depth} {Survey} ({eFEDS})},
	url = {https://www.aanda.org/10.1051/0004-6361/202141266},
	doi = {10.1051/0004-6361/202141266},
	journal = {\aap},
	author = {Brunner, H. and Liu, T. and Lamer, G. and Georgakakis, A. and Merloni, A. and Brusa, M. and Bulbul, E. and Dennerl, K. and Friedrich, S. and Liu, A. and Maitra, C. and Nandra, K. and Ramos-Ceja, M. E. and Sanders, J. S. and Stewart, I. M. and Boller, T. and Buchner, J. and Clerc, N. and Comparat, J. and Dwelly, T. and Eckert, D. and Finoguenov, A. and Freyberg, M. and Ghirardini, V. and Gueguen, A. and Haberl, F. and Kreykenbohm, I. and Krumpe, M. and Osterhage, S. and Pacaud, F. and Predehl, P. and Reiprich, T. H. and Robrade, J. and Salvato, M. and Santangelo, A. and Schrabback, T. and Schwope, A. and Wilms, J.},
	month = may,
	year = {2022},
	pages = {A1},
}

@article{edge_vista_2013,
	title = {The {VISTA} {Kilo}-degree {Infrared} {Galaxy} ({VIKING}) {Survey}: {Bridging} the {Gap} between {Low} and {High} {Redshift}},
	volume = {154},
	issn = {0722-6691},
	shorttitle = {The {VISTA} {Kilo}-degree {Infrared} {Galaxy} ({VIKING}) {Survey}},
	url = {https://ui.adsabs.harvard.edu/abs/2013Msngr.154...32E},
	journal = {The Messenger},
	author = {Edge, A. and Sutherland, W. and Kuijken, K. and Driver, S. and McMahon, R. and Eales, S. and Emerson, J. P.},
	month = dec,
	year = {2013},
	pages = {32--34},
}

@article{aihara_hyper_2018,
	title = {The {Hyper} {Suprime}-{Cam} {SSP} {Survey}: {Overview} and survey design},
	volume = {70},
	issn = {0004-6264},
	shorttitle = {The {Hyper} {Suprime}-{Cam} {SSP} {Survey}},
	url = {https://ui.adsabs.harvard.edu/abs/2018PASJ...70S...4A},
	doi = {10.1093/pasj/psx066},
	journal = {\pasj},
	author = {Aihara, Hiroaki and Arimoto, Nobuo and Armstrong, Robert and Arnouts, Stéphane and Bahcall, Neta A. and Bickerton, Steven and Bosch, James and Bundy, Kevin and Capak, Peter L. and Chan, James H. H. and Chiba, Masashi and Coupon, Jean and Egami, Eiichi and Enoki, Motohiro and Finet, Francois and Fujimori, Hiroki and Fujimoto, Seiji and Furusawa, Hisanori and Furusawa, Junko and Goto, Tomotsugu and Goulding, Andy and Greco, Johnny P. and Greene, Jenny E. and Gunn, James E. and Hamana, Takashi and Harikane, Yuichi and Hashimoto, Yasuhiro and Hattori, Takashi and Hayashi, Masao and Hayashi, Yusuke and Hełminiak, Krzysztof G. and Higuchi, Ryo and Hikage, Chiaki and Ho, Paul T. P. and Hsieh, Bau-Ching and Huang, Kuiyun and Huang, Song and Ikeda, Hiroyuki and Imanishi, Masatoshi and Inoue, Akio K. and Iwasawa, Kazushi and Iwata, Ikuru and Jaelani, Anton T. and Jian, Hung-Yu and Kamata, Yukiko and Karoji, Hiroshi and Kashikawa, Nobunari and Katayama, Nobuhiko and Kawanomoto, Satoshi and Kayo, Issha and Koda, Jin and Koike, Michitaro and Kojima, Takashi and Komiyama, Yutaka and Konno, Akira and Koshida, Shintaro and Koyama, Yusei and Kusakabe, Haruka and Leauthaud, Alexie and Lee, Chien-Hsiu and Lin, Lihwai and Lin, Yen-Ting and Lupton, Robert H. and Mandelbaum, Rachel and Matsuoka, Yoshiki and Medezinski, Elinor and Mineo, Sogo and Miyama, Shoken and Miyatake, Hironao and Miyazaki, Satoshi and Momose, Rieko and More, Anupreeta and More, Surhud and Moritani, Yuki and Moriya, Takashi J. and Morokuma, Tomoki and Mukae, Shiro and Murata, Ryoma and Murayama, Hitoshi and Nagao, Tohru and Nakata, Fumiaki and Niida, Mana and Niikura, Hiroko and Nishizawa, Atsushi J. and Obuchi, Yoshiyuki and Oguri, Masamune and Oishi, Yukie and Okabe, Nobuhiro and Okamoto, Sakurako and Okura, Yuki and Ono, Yoshiaki and Onodera, Masato and Onoue, Masafusa and Osato, Ken and Ouchi, Masami and Price, Paul A. and Pyo, Tae-Soo and Sako, Masao and Sawicki, Marcin and Shibuya, Takatoshi and Shimasaku, Kazuhiro and Shimono, Atsushi and Shirasaki, Masato and Silverman, John D. and Simet, Melanie and Speagle, Joshua and Spergel, David N. and Strauss, Michael A. and Sugahara, Yuma and Sugiyama, Naoshi and Suto, Yasushi and Suyu, Sherry H. and Suzuki, Nao and Tait, Philip J. and Takada, Masahiro and Takata, Tadafumi and Tamura, Naoyuki and Tanaka, Manobu M. and Tanaka, Masaomi and Tanaka, Masayuki and Tanaka, Yoko and Terai, Tsuyoshi and Terashima, Yuichi and Toba, Yoshiki and Tominaga, Nozomu and Toshikawa, Jun and Turner, Edwin L. and Uchida, Tomohisa and Uchiyama, Hisakazu and Umetsu, Keiichi and Uraguchi, Fumihiro and Urata, Yuji and Usuda, Tomonori and Utsumi, Yousuke and Wang, Shiang-Yu and Wang, Wei-Hao and Wong, Kenneth C. and Yabe, Kiyoto and Yamada, Yoshihiko and Yamanoi, Hitomi and Yasuda, Naoki and Yeh, Sherry and Yonehara, Atsunori and Yuma, Suraphong},
	month = jan,
	year = {2018},
	pages = {S4},
}

@article{wright_wide-field_2010,
	title = {The {Wide}-field {Infrared} {Survey} {Explorer} ({WISE}): {Mission} {Description} and {Initial} {On}-orbit {Performance}},
	volume = {140},
	issn = {0004-6256},
	shorttitle = {The {Wide}-field {Infrared} {Survey} {Explorer} ({WISE})},
	url = {https://ui.adsabs.harvard.edu/abs/2010AJ....140.1868W},
	doi = {10.1088/0004-6256/140/6/1868},
	journal = {\apj},
	author = {Wright, Edward L. and Eisenhardt, Peter R. M. and Mainzer, Amy K. and Ressler, Michael E. and Cutri, Roc M. and Jarrett, Thomas and Kirkpatrick, J. Davy and Padgett, Deborah and McMillan, Robert S. and Skrutskie, Michael and Stanford, S. A. and Cohen, Martin and Walker, Russell G. and Mather, John C. and Leisawitz, David and Gautier, III, Thomas N. and McLean, Ian and Benford, Dominic and Lonsdale, Carol J. and Blain, Andrew and Mendez, Bryan and Irace, William R. and Duval, Valerie and Liu, Fengchuan and Royer, Don and Heinrichsen, Ingolf and Howard, Joan and Shannon, Mark and Kendall, Martha and Walsh, Amy L. and Larsen, Mark and Cardon, Joel G. and Schick, Scott and Schwalm, Mark and Abid, Mohamed and Fabinsky, Beth and Naes, Larry and Tsai, Chao-Wei},
	month = dec,
	year = {2010},
	pages = {1868--1881},
}

@article{valiante_herschel-atlas_2016,
	title = {The {Herschel}-{ATLAS} data release 1 - {I}. {Maps}, catalogues and number counts},
	volume = {462},
	issn = {0035-8711},
	url = {https://ui.adsabs.harvard.edu/abs/2016MNRAS.462.3146V},
	doi = {10.1093/mnras/stw1806},
	journal = {\mnras},
	author = {Valiante, E. and Smith, M. W. L. and Eales, S. and Maddox, S. J. and Ibar, E. and Hopwood, R. and Dunne, L. and Cigan, P. J. and Dye, S. and Pascale, E. and Rigby, E. E. and Bourne, N. and Furlanetto, C. and Ivison, R. J.},
	month = nov,
	year = {2016},
	pages = {3146--3179},
}

@article{burgarella_star_2005,
	title = {Star formation and dust attenuation properties in galaxies from a statistical ultraviolet-to-far-infrared analysis},
	volume = {360},
	issn = {0035-8711},
	url = {https://ui.adsabs.harvard.edu/abs/2005MNRAS.360.1413B},
	doi = {10.1111/j.1365-2966.2005.09131.x},
	journal = {\mnras},
	author = {Burgarella, D. and Buat, V. and Iglesias-Páramo, J.},
	month = jul,
	year = {2005},
	pages = {1413--1425},
}

@article{noll_analysis_2009,
	title = {Analysis of galaxy spectral energy distributions from far-{UV} to far-{IR} with {CIGALE}: studying a {SINGS} test sample},
	volume = {507},
	issn = {0004-6361, 1432-0746},
	shorttitle = {Analysis of galaxy spectral energy distributions from far-{UV} to far-{IR} with {CIGALE}},
	url = {http://www.aanda.org/10.1051/0004-6361/200912497},
	doi = {10.1051/0004-6361/200912497},
	number = {3},
	journal = {\aap},
	author = {Noll, S. and Burgarella, D. and Giovannoli, E. and Buat, V. and Marcillac, D. and Muñoz-Mateos, J. C.},
	month = dec,
	year = {2009},
	pages = {1793--1813},
}

@article{boquien_cigale_2019,
	title = {{CIGALE}: a python {Code} {Investigating} {GALaxy} {Emission}},
	volume = {622},
	issn = {0004-6361, 1432-0746},
	shorttitle = {{CIGALE}},
	url = {https://www.aanda.org/10.1051/0004-6361/201834156},
	doi = {10.1051/0004-6361/201834156},
	journal = {\aap},
	author = {Boquien, M. and Burgarella, D. and Roehlly, Y. and Buat, V. and Ciesla, L. and Corre, D. and Inoue, A. K. and Salas, H.},
	month = feb,
	year = {2019},
	pages = {A103},
}

@article{yang_fitting_2022,
	title = {Fitting {AGN}/{Galaxy} {X}-{Ray}-to-radio {SEDs} with {CIGALE} and {Improvement} of the {Code}},
	volume = {927},
	issn = {0004-637X},
	url = {https://ui.adsabs.harvard.edu/abs/2022ApJ...927..192Y},
	doi = {10.3847/1538-4357/ac4971},
	journal = {\apj},
	author = {Yang, Guang and Boquien, Médéric and Brandt, W. N. and Buat, Véronique and Burgarella, Denis and Ciesla, Laure and Lehmer, Bret D. and Małek, Katarzyna and Mountrichas, George and Papovich, Casey and Pons, Estelle and Stalevski, Marko and Theulé, Patrice and Zhu, Shifu},
	month = mar,
	year = {2022},
	pages = {192},
}

@article{yang_x-cigale_2020,
	title = {X-{CIGALE}: {Fitting} {AGN}/galaxy {SEDs} from {X}-ray to infrared},
	volume = {491},
	issn = {0035-8711},
	shorttitle = {X-{CIGALE}},
	url = {https://ui.adsabs.harvard.edu/abs/2020MNRAS.491..740Y},
	doi = {10.1093/mnras/stz3001},
	journal = {\mnras},
	author = {Yang, G. and Boquien, M. and Buat, V. and Burgarella, D. and Ciesla, L. and Duras, F. and Stalevski, M. and Brandt, W. N. and Papovich, C.},
	month = jan,
	year = {2020},
	pages = {740--757},
}

@article{stern_mid-infrared_2012,
	title = {Mid-infrared {Selection} of {Active} {Galactic} {Nuclei} with the {Wide}-{Field} {Infrared} {Survey} {Explorer}. {I}. {Characterizing} {WISE}-selected {Active} {Galactic} {Nuclei} in {COSMOS}},
	volume = {753},
	issn = {0004-637X},
	url = {https://ui.adsabs.harvard.edu/abs/2012ApJ...753...30S},
	doi = {10.1088/0004-637X/753/1/30},
	journal = {\apj},
	author = {Stern, Daniel and Assef, Roberto J. and Benford, Dominic J. and Blain, Andrew and Cutri, Roc and Dey, Arjun and Eisenhardt, Peter and Griffith, Roger L. and Jarrett, T. H. and Lake, Sean and Masci, Frank and Petty, Sara and Stanford, S. A. and Tsai, Chao-Wei and Wright, E. L. and Yan, Lin and Harrison, Fiona and Madsen, Kristin},
	month = jul,
	year = {2012},
	pages = {30},
}

@article{salvato_erosita_2022,
	title = {The {eROSITA} {Final} {Equatorial}-{Depth} {Survey} ({eFEDS}): {Identification} and characterization of the counterparts to point-like sources},
	volume = {661},
	issn = {0004-6361, 1432-0746},
	shorttitle = {The {eROSITA} {Final} {Equatorial}-{Depth} {Survey} ({eFEDS})},
	url = {https://www.aanda.org/10.1051/0004-6361/202141631},
	doi = {10.1051/0004-6361/202141631},
	journal = {\aap},
	author = {Salvato, M. and Wolf, J. and Dwelly, T. and Georgakakis, A. and Brusa, M. and Merloni, A. and Liu, T. and Toba, Y. and Nandra, K. and Lamer, G. and Buchner, J. and Schneider, C. and Freund, S. and Rau, A. and Schwope, A. and Nishizawa, A. and Klein, M. and Arcodia, R. and Comparat, J. and Musiimenta, B. and Nagao, T. and Brunner, H. and Malyali, A. and Finoguenov, A. and Anderson, S. and Shen, Y. and Ibarra-Medel, H. and Trump, J. and Brandt, W. N. and Urry, C. M. and Rivera, C. and Krumpe, M. and Urrutia, T. and Miyaji, T. and Ichikawa, K. and Schneider, D. P. and Fresco, A. and Boller, T. and Haase, J. and Brownstein, J. and Lane, R. R. and Bizyaev, D. and Nitschelm, C.},
	month = may,
	year = {2022},
	pages = {A3},
}

@article{yang_linking_2018,
	title = {Linking black hole growth with host galaxies: the accretion-stellar mass relation and its cosmic evolution},
	volume = {475},
	issn = {0035-8711},
	shorttitle = {Linking black hole growth with host galaxies},
	url = {https://ui.adsabs.harvard.edu/abs/2018MNRAS.475.1887Y},
	doi = {10.1093/mnras/stx2805},
	journal = {\mnras},
	author = {Yang, G. and Brandt, W. N. and Vito, F. and Chen, C. -T. J. and Trump, J. R. and Luo, B. and Sun, M. Y. and Xue, Y. Q. and Koekemoer, A. M. and Schneider, D. P. and Vignali, C. and Wang, J. -X.},
	month = apr,
	year = {2018},
	pages = {1887--1911},
}

@article{draine_andromedas_2014,
	title = {Andromeda's {Dust}},
	volume = {780},
	issn = {0004-637X},
	url = {https://ui.adsabs.harvard.edu/abs/2014ApJ...780..172D},
	doi = {10.1088/0004-637X/780/2/172},
	journal = {\apj},
	author = {Draine, B. T. and Aniano, G. and Krause, Oliver and Groves, Brent and Sandstrom, Karin and Braun, Robert and Leroy, Adam and Klaas, Ulrich and Linz, Hendrik and Rix, Hans-Walter and Schinnerer, Eva and Schmiedeke, Anika and Walter, Fabian},
	month = jan,
	year = {2014},
	pages = {172},
}

@article{calzetti_dust_2000,
	title = {The {Dust} {Content} and {Opacity} of {Actively} {Star}-forming {Galaxies}},
	volume = {533},
	issn = {0004-637X},
	url = {https://ui.adsabs.harvard.edu/abs/2000ApJ...533..682C},
	doi = {10.1086/308692},
	journal = {\apj},
	author = {Calzetti, Daniela and Armus, Lee and Bohlin, Ralph C. and Kinney, Anne L. and Koornneef, Jan and Storchi-Bergmann, Thaisa},
	month = apr,
	year = {2000},
	pages = {682--695},
}

@article{bruzual_stellar_2003,
	title = {Stellar population synthesis at the resolution of 2003},
	volume = {344},
	issn = {0035-8711},
	url = {https://ui.adsabs.harvard.edu/abs/2003MNRAS.344.1000B},
	doi = {10.1046/j.1365-8711.2003.06897.x},
	journal = {\mnras},
	author = {Bruzual, G. and Charlot, S.},
	month = oct,
	year = {2003},
	pages = {1000--1028},
}

@article{stalevski_3d_2012,
	title = {{3D} radiative transfer modelling of the dusty tori around active galactic nuclei as a clumpy two-phase medium},
	volume = {420},
	issn = {0035-8711},
	url = {https://ui.adsabs.harvard.edu/abs/2012MNRAS.420.2756S},
	doi = {10.1111/j.1365-2966.2011.19775.x},
	journal = {\mnras},
	author = {Stalevski, Marko and Fritz, Jacopo and Baes, Maarten and Nakos, Theodoros and Popović, Luka C.},
	month = mar,
	year = {2012},
	pages = {2756--2772},
}

@article{wright_kidsviking-450_2019,
	title = {{KiDS}+{VIKING}-450: {A} new combined optical and near-infrared dataset for cosmology and astrophysics},
	volume = {632},
	issn = {0004-6361, 1432-0746},
	shorttitle = {{KiDS}+{VIKING}-450},
	url = {https://www.aanda.org/10.1051/0004-6361/201834879},
	doi = {10.1051/0004-6361/201834879},
	journal = {\aap},
	author = {Wright, Angus H. and Hildebrandt, Hendrik and Kuijken, Konrad and Erben, Thomas and Blake, Robert and Buddelmeijer, Hugo and Choi, Ami and Cross, Nicholas and de Jong, Jelte T. A. and Edge, Alastair and Gonzalez-Fernandez, Carlos and González Solares, Eduardo and Grado, Aniello and Heymans, Catherine and Irwin, Mike and Kupcu Yoldas, Aybuke and Lewis, James R. and Mann, Robert G. and Napolitano, Nicola and Radovich, Mario and Schneider, Peter and Sifón, Cristóbal and Sutherland, William and Sutorius, Eckhard and Verdoes Kleijn, Gijs A.},
	month = dec,
	year = {2019},
	pages = {A34},
}

@article{willett_galaxy_2017,
	title = {Galaxy {Zoo}: morphological classifications for 120 000 galaxies in {HST} legacy imaging},
	volume = {464},
	issn = {0035-8711},
	shorttitle = {Galaxy {Zoo}},
	url = {https://ui.adsabs.harvard.edu/abs/2017MNRAS.464.4176W},
	doi = {10.1093/mnras/stw2568},
	journal = {\mnras},
	author = {Willett, Kyle W. and Galloway, Melanie A. and Bamford, Steven P. and Lintott, Chris J. and Masters, Karen L. and Scarlata, Claudia and Simmons, B. D. and Beck, Melanie and Cardamone, Carolin N. and Cheung, Edmond and Edmondson, Edward M. and Fortson, Lucy F. and Griffith, Roger L. and Häußler, Boris and Han, Anna and Hart, Ross and Melvin, Thomas and Parrish, Michael and Schawinski, Kevin and Smethurst, R. J. and Smith, Arfon M.},
	month = feb,
	year = {2017},
	pages = {4176--4203},
}

@article{willett_galaxy_2013,
	title = {Galaxy {Zoo} 2: detailed morphological classifications for 304 122 galaxies from the {Sloan} {Digital} {Sky} {Survey}},
	volume = {435},
	issn = {0035-8711},
	shorttitle = {Galaxy {Zoo} 2},
	url = {https://doi.org/10.1093/mnras/stt1458},
	doi = {10.1093/mnras/stt1458},
	number = {4},
	journal = {\mnras},
	author = {Willett, Kyle W. and Lintott, Chris J. and Bamford, Steven P. and Masters, Karen L. and Simmons, Brooke D. and Casteels, Kevin R. V. and Edmondson, Edward M. and Fortson, Lucy F. and Kaviraj, Sugata and Keel, William C. and Melvin, Thomas and Nichol, Robert C. and Raddick, M. Jordan and Schawinski, Kevin and Simpson, Robert J. and Skibba, Ramin A. and Smith, Arfon M. and Thomas, Daniel},
	month = nov,
	year = {2013},
	pages = {2835--2860},
}

@article{walmsley_galaxy_2022,
	title = {Galaxy {Zoo} {DECaLS}: {Detailed} visual morphology measurements from volunteers and deep learning for 314 000 galaxies},
	volume = {509},
	issn = {0035-8711},
	shorttitle = {Galaxy {Zoo} {DECaLS}},
	url = {https://ui.adsabs.harvard.edu/abs/2022MNRAS.509.3966W},
	doi = {10.1093/mnras/stab2093},
	journal = {\mnras},
	author = {Walmsley, Mike and Lintott, Chris and Géron, Tobias and Kruk, Sandor and Krawczyk, Coleman and Willett, Kyle W. and Bamford, Steven and Kelvin, Lee S. and Fortson, Lucy and Gal, Yarin and Keel, William and Masters, Karen L. and Mehta, Vihang and Simmons, Brooke D. and Smethurst, Rebecca and Smith, Lewis and Baeten, Elisabeth M. and Macmillan, Christine},
	month = jan,
	year = {2022},
	pages = {3966--3988},
}

@article{aihara_third_2022,
	title = {Third data release of the {Hyper} {Suprime}-{Cam} {Subaru} {Strategic} {Program}},
	volume = {74},
	issn = {0004-6264},
	url = {https://ui.adsabs.harvard.edu/abs/2022PASJ...74..247A},
	doi = {10.1093/pasj/psab122},
	journal = {\pasj},
	author = {Aihara, Hiroaki and AlSayyad, Yusra and Ando, Makoto and Armstrong, Robert and Bosch, James and Egami, Eiichi and Furusawa, Hisanori and Furusawa, Junko and Harasawa, Sumiko and Harikane, Yuichi and Hsieh, Bau-Ching and Ikeda, Hiroyuki and Ito, Kei and Iwata, Ikuru and Kodama, Tadayuki and Koike, Michitaro and Kokubo, Mitsuru and Komiyama, Yutaka and Li, Xiangchong and Liang, Yongming and Lin, Yen-Ting and Lupton, Robert H. and Lust, Nate B. and MacArthur, Lauren A. and Mawatari, Ken and Mineo, Sogo and Miyatake, Hironao and Miyazaki, Satoshi and More, Surhud and Morishima, Takahiro and Murayama, Hitoshi and Nakajima, Kimihiko and Nakata, Fumiaki and Nishizawa, Atsushi J. and Oguri, Masamune and Okabe, Nobuhiro and Okura, Yuki and Ono, Yoshiaki and Osato, Ken and Ouchi, Masami and Pan, Yen-Chen and Plazas Malagón, Andrés A. and Price, Paul A. and Reed, Sophie L. and Rykoff, Eli S. and Shibuya, Takatoshi and Simunovic, Mirko and Strauss, Michael A. and Sugimori, Kanako and Suto, Yasushi and Suzuki, Nao and Takada, Masahiro and Takagi, Yuhei and Takata, Tadafumi and Takita, Satoshi and Tanaka, Masayuki and Tang, Shenli and Taranu, Dan S. and Terai, Tsuyoshi and Toba, Yoshiki and Turner, Edwin L. and Uchiyama, Hisakazu and Vijarnwannaluk, Bovornpratch and Waters, Christopher Z. and Yamada, Yoshihiko and Yamamoto, Naoaki and Yamashita, Takuji},
	month = apr,
	year = {2022},
	pages = {247--272},
}

@article{masters_galaxy_2012,
	title = {Galaxy {Zoo} and {ALFALFA}: atomic gas and the regulation of star formation in barred disc galaxies},
	volume = {424},
	issn = {0035-8711},
	shorttitle = {Galaxy {Zoo} and {ALFALFA}},
	url = {https://ui.adsabs.harvard.edu/abs/2012MNRAS.424.2180M},
	doi = {10.1111/j.1365-2966.2012.21377.x},
	journal = {\mnras},
	author = {Masters, Karen L. and Nichol, Robert C. and Haynes, Martha P. and Keel, William C. and Lintott, Chris and Simmons, Brooke and Skibba, Ramin and Bamford, Steven and Giovanelli, Riccardo and Schawinski, Kevin},
	month = aug,
	year = {2012},
	pages = {2180--2192},
}

@article{nair_catalog_2010,
	title = {A {Catalog} of {Detailed} {Visual} {Morphological} {Classifications} for 14,034 {Galaxies} in the {Sloan} {Digital} {Sky} {Survey}},
	volume = {186},
	issn = {0067-0049},
	url = {https://ui.adsabs.harvard.edu/abs/2010ApJS..186..427N},
	doi = {10.1088/0067-0049/186/2/427},
	journal = {\apjs},
	author = {Nair, Preethi B. and Abraham, Roberto G.},
	month = feb,
	year = {2010},
	pages = {427--456},
}

@article{cheung_galaxy_2013,
	title = {Galaxy {Zoo}: {Observing} {Secular} {Evolution} through {Bars}},
	volume = {779},
	issn = {0004-637X},
	shorttitle = {Galaxy {Zoo}},
	url = {https://ui.adsabs.harvard.edu/abs/2013ApJ...779..162C},
	doi = {10.1088/0004-637X/779/2/162},
	journal = {\apj},
	author = {Cheung, Edmond and Athanassoula, E. and Masters, Karen L. and Nichol, Robert C. and Bosma, A. and Bell, Eric F. and Faber, S. M. and Koo, David C. and Lintott, Chris and Melvin, Thomas and Schawinski, Kevin and Skibba, Ramin A. and Willett, Kyle W.},
	month = dec,
	year = {2013},
	pages = {162},
}

@article{cheung_galaxy_2015,
	title = {Galaxy {Zoo}: {Are} bars responsible for the feeding of active galactic nuclei at 0.2 {\textless} z {\textless} 1.0?},
	volume = {447},
	issn = {0035-8711},
	shorttitle = {Galaxy {Zoo}},
	url = {https://ui.adsabs.harvard.edu/abs/2015MNRAS.447..506C},
	doi = {10.1093/mnras/stu2462},
	journal = {\mnras},
	author = {Cheung, Edmond and Trump, Jonathan R. and Athanassoula, E. and Bamford, Steven P. and Bell, Eric F. and Bosma, A. and Cardamone, Carolin N. and Casteels, Kevin R. V. and Faber, S. M. and Fang, Jerome J. and Fortson, Lucy F. and Kocevski, Dale D. and Koo, David C. and Laine, Seppo and Lintott, Chris and Masters, Karen L. and Melvin, Thomas and Nichol, Robert C. and Schawinski, Kevin and Simmons, Brooke and Smethurst, Rebecca and Willett, Kyle W.},
	month = feb,
	year = {2015},
	pages = {506--516},
}

@article{garland_most_2023,
	title = {The most luminous, merger-free {AGNs} show only marginal correlation with bar presence},
	volume = {522},
	issn = {0035-8711},
	url = {https://ui.adsabs.harvard.edu/abs/2023MNRAS.522..211G},
	doi = {10.1093/mnras/stad966},
	journal = {\mnras},
	author = {Garland, Izzy L. and Fahey, Matthew J. and Simmons, Brooke D. and Smethurst, Rebecca J. and Lintott, Chris J. and Shanahan, Jesse and Silcock, Maddie S. and Smith, Joshua and Keel, William C. and Coil, Alison and Géron, Tobias and Kruk, Sandor and Masters, Karen L. and O'Ryan, David and Thorne, Matthew R. and Wiersema, Klaas},
	month = jun,
	year = {2023},
	pages = {211--225},
}

@article{zee_unraveling_2023,
	title = {Unraveling {Joint} {Evolution} of {Bars}, {Star} {Formation}, and {Active} {Galactic} {Nuclei} of {Disk} {Galaxies}},
	volume = {949},
	issn = {0004-637X},
	url = {https://ui.adsabs.harvard.edu/abs/2023ApJ...949...91Z},
	doi = {10.3847/1538-4357/acc79a},
	urldate = {2024-03-19},
	journal = {\apj},
	author = {Zee, Woong-Bae G. and Paudel, Sanjaya and Moon, Jun-Sung and Yoon, Suk-Jin},
	month = jun,
	year = {2023},
	pages = {91},
}

@article{yu_edge-califa_2022,
	title = {The {EDGE}-{CALIFA} survey: {The} role of spiral arms and bars in driving central molecular gas concentrations},
	volume = {666},
	issn = {0004-6361},
	shorttitle = {The {EDGE}-{CALIFA} survey},
	url = {https://ui.adsabs.harvard.edu/abs/2022A&A...666A.175Y},
	doi = {10.1051/0004-6361/202244306},
	journal = {\aap},
	author = {Yu, Si-Yue and Kalinova, Veselina and Colombo, Dario and Bolatto, Alberto D. and Wong, Tony and Levy, Rebecca C. and Villanueva, Vicente and Sánchez, Sebastián F. and Ho, Luis C. and Vogel, Stuart N. and Teuben, Peter and Rubio, Mónica},
	month = oct,
	year = {2022},
	pages = {A175},
}

@article{alonso_impact_2018,
	title = {The impact of bars and interactions on optically selected {AGNs} in spiral galaxies},
	volume = {618},
	issn = {0004-6361},
	url = {https://ui.adsabs.harvard.edu/abs/2018A&A...618A.149A},
	doi = {10.1051/0004-6361/201832796},
	journal = {\aap},
	author = {Alonso, Sol and Coldwell, Georgina and Duplancic, Fernanda and Mesa, Valeria and Lambas, Diego G.},
	month = oct,
	year = {2018},
	pages = {A149},
}

@article{silva-lima_revisiting_2022,
	title = {Revisiting the role of bars in {AGN} fuelling with propensity score sample matching},
	volume = {661},
	issn = {0004-6361},
	url = {https://ui.adsabs.harvard.edu/abs/2022A&A...661A.105S},
	doi = {10.1051/0004-6361/202142432},
	journal = {\aap},
	author = {Silva-Lima, Luiz A. and Martins, Lucimara P. and Coelho, Paula R. T. and Gadotti, Dimitri A.},
	month = may,
	year = {2022},
	pages = {A105},
}

@article{cisternas_role_2015,
	title = {The {Role} of {Bars} in {AGN} {Fueling} in {Disk} {Galaxies} {Over} the {Last} {Seven} {Billion} {Years}},
	volume = {802},
	issn = {0004-637X},
	url = {https://ui.adsabs.harvard.edu/abs/2015ApJ...802..137C},
	doi = {10.1088/0004-637X/802/2/137},
	journal = {\apj},
	author = {Cisternas, Mauricio and Sheth, Kartik and Salvato, Mara and Knapen, Johan H. and Civano, Francesca and Santini, Paola},
	month = apr,
	year = {2015},
	pages = {137},
}

@article{driver_galaxy_2011,
	title = {Galaxy and {Mass} {Assembly} ({GAMA}): survey diagnostics and core data release},
	volume = {413},
	issn = {0035-8711},
	shorttitle = {Galaxy and {Mass} {Assembly} ({GAMA})},
	url = {https://ui.adsabs.harvard.edu/abs/2011MNRAS.413..971D},
	doi = {10.1111/j.1365-2966.2010.18188.x},
	journal = {\mnras},
	author = {Driver, S. P. and Hill, D. T. and Kelvin, L. S. and Robotham, A. S. G. and Liske, J. and Norberg, P. and Baldry, I. K. and Bamford, S. P. and Hopkins, A. M. and Loveday, J. and Peacock, J. A. and Andrae, E. and Bland-Hawthorn, J. and Brough, S. and Brown, M. J. I. and Cameron, E. and Ching, J. H. Y. and Colless, M. and Conselice, C. J. and Croom, S. M. and Cross, N. J. G. and de Propris, R. and Dye, S. and Drinkwater, M. J. and Ellis, S. and Graham, Alister W. and Grootes, M. W. and Gunawardhana, M. and Jones, D. H. and van Kampen, E. and Maraston, C. and Nichol, R. C. and Parkinson, H. R. and Phillipps, S. and Pimbblet, K. and Popescu, C. C. and Prescott, M. and Roseboom, I. G. and Sadler, E. M. and Sansom, A. E. and Sharp, R. G. and Smith, D. J. B. and Taylor, E. and Thomas, D. and Tuffs, R. J. and Wijesinghe, D. and Dunne, L. and Frenk, C. S. and Jarvis, M. J. and Madore, B. F. and Meyer, M. J. and Seibert, M. and Staveley-Smith, L. and Sutherland, W. J. and Warren, S. J.},
	month = may,
	year = {2011},
	pages = {971--995},
}

@article{hinshaw_nine-year_2013,
	title = {Nine-year {Wilkinson} {Microwave} {Anisotropy} {Probe} ({WMAP}) {Observations}: {Cosmological} {Parameter} {Results}},
	volume = {208},
	issn = {0067-0049},
	shorttitle = {Nine-year {Wilkinson} {Microwave} {Anisotropy} {Probe} ({WMAP}) {Observations}},
	url = {https://ui.adsabs.harvard.edu/abs/2013ApJS..208...19H},
	doi = {10.1088/0067-0049/208/2/19},
	journal = {\apjs},
	author = {Hinshaw, G. and Larson, D. and Komatsu, E. and Spergel, D. N. and Bennett, C. L. and Dunkley, J. and Nolta, M. R. and Halpern, M. and Hill, R. S. and Odegard, N. and Page, L. and Smith, K. M. and Weiland, J. L. and Gold, B. and Jarosik, N. and Kogut, A. and Limon, M. and Meyer, S. S. and Tucker, G. S. and Wollack, E. and Wright, E. L.},
	month = oct,
	year = {2013},
	pages = {19},
}

@article{kauffmann_host_2003,
	title = {The host galaxies of active galactic nuclei},
	volume = {346},
	issn = {0035-8711},
	url = {https://ui.adsabs.harvard.edu/abs/2003MNRAS.346.1055K},
	doi = {10.1111/j.1365-2966.2003.07154.x},
	journal = {\mnras},
	author = {Kauffmann, Guinevere and Heckman, Timothy M. and Tremonti, Christy and Brinchmann, Jarle and Charlot, Stéphane and White, Simon D. M. and Ridgway, Susan E. and Brinkmann, Jon and Fukugita, Masataka and Hall, Patrick B. and Ivezić, Željko and Richards, Gordon T. and Schneider, Donald P.},
	month = dec,
	year = {2003},
	pages = {1055--1077},
}

@article{garland_galaxy_2024,
       author = {{Garland}, Izzy L. and {Walmsley}, Mike and {Silcock}, Maddie S. and {Potts}, Leah M. and {Smith}, Josh and {Simmons}, Brooke D. and {Lintott}, Chris J. and {Smethurst}, Rebecca J. and {Dawson}, James M. and {Keel}, William C. and {Kruk}, Sandor and {Mantha}, Kameswara Bharadwaj and {Masters}, Karen L. and {O'Ryan}, David and {Popp}, J{\"u}rgen J. and {Thorne}, Matthew R.},
        title = "{Galaxy Zoo DESI: large-scale bars as a secular mechanism for triggering AGNs}",
      journal = {\mnras},
    year = 2024,
        month = aug,
       volume = {532},
       number = {2},
        pages = {2320-2330},
          doi = {10.1093/mnras/stae1620},
archivePrefix = {arXiv},
       eprint = {2406.20096},
 primaryClass = {astro-ph.GA},
       adsurl = {https://ui.adsabs.harvard.edu/abs/2024MNRAS.532.2320G}
}

@article{kruk_galaxy_2018,
	title = {Galaxy {Zoo}: secular evolution of barred galaxies from structural decomposition of multiband images},
	volume = {473},
	issn = {0035-8711},
	shorttitle = {Galaxy {Zoo}},
	url = {https://ui.adsabs.harvard.edu/abs/2018MNRAS.473.4731K},
	doi = {10.1093/mnras/stx2605},
	journal = {\mnras},
	author = {Kruk, Sandor J. and Lintott, Chris J. and Bamford, Steven P. and Masters, Karen L. and Simmons, Brooke D. and Häußler, Boris and Cardamone, Carolin N. and Hart, Ross E. and Kelvin, Lee and Schawinski, Kevin and Smethurst, Rebecca J. and Vika, Marina},
	month = feb,
	year = {2018},
	pages = {4731--4753},
}

@article{masters_galaxy_2011,
	title = {Galaxy {Zoo}: bars in disc galaxies},
	volume = {411},
	issn = {0035-8711},
	shorttitle = {Galaxy {Zoo}},
	url = {https://ui.adsabs.harvard.edu/abs/2011MNRAS.411.2026M},
	doi = {10.1111/j.1365-2966.2010.17834.x},
	journal = {\mnras},
	author = {Masters, Karen L. and Nichol, Robert C. and Hoyle, Ben and Lintott, Chris and Bamford, Steven P. and Edmondson, Edward M. and Fortson, Lucy and Keel, William C. and Schawinski, Kevin and Smith, Arfon M. and Thomas, Daniel},
	month = mar,
	year = {2011},
	pages = {2026--2034},
}

@article{hodges_significance_1958,
	title = {The significance probability of the smirnov two-sample test},
	volume = {3},
	issn = {0004-2080},
	url = {https://ui.adsabs.harvard.edu/abs/1958ArM.....3..469H},
	doi = {10.1007/BF02589501},
	urldate = {2023-05-30},
	journal = {Arkiv for Matematik},
	author = {Hodges, J. L.},
	month = jan,
	year = {1958},
	pages = {469--486},
}

@article{heckman_coevolution_2014,
	title = {The {Coevolution} of {Galaxies} and {Supermassive} {Black} {Holes}: {Insights} from {Surveys} of the {Contemporary} {Universe}},
	volume = {52},
	issn = {0066-4146, 1545-4282},
	shorttitle = {The {Coevolution} of {Galaxies} and {Supermassive} {Black} {Holes}},
	url = {https://www.annualreviews.org/doi/10.1146/annurev-astro-081913-035722},
	doi = {10.1146/annurev-astro-081913-035722},
	journal = {\araa},
	author = {Heckman, Timothy M. and Best, Philip N.},
	month = aug,
	year = {2014},
	pages = {589--660},
}

@article{martin_normal_2018,
	title = {Normal black holes in bulge-less galaxies: the largely quiescent, merger-free growth of black holes over cosmic time},
	volume = {476},
	issn = {0035-8711},
	shorttitle = {Normal black holes in bulge-less galaxies},
	url = {https://ui.adsabs.harvard.edu/abs/2018MNRAS.476.2801M},
	doi = {10.1093/mnras/sty324},
	journal = {\mnras},
	author = {Martin, G. and Kaviraj, S. and Volonteri, M. and Simmons, B. D. and Devriendt, J. E. G. and Lintott, C. J. and Smethurst, R. J. and Dubois, Y. and Pichon, C.},
	month = may,
	year = {2018},
	pages = {2801--2812},
}

@article{smethurst_evidence_2023,
	title = {Evidence for non-merger co-evolution of galaxies and their supermassive black holes},
	issn = {0035-8711},
	url = {https://ui.adsabs.harvard.edu/abs/2023MNRAS.tmp.1825S},
	doi = {10.1093/mnras/stad1794},
	journal = {\mnras},
	author = {Smethurst, R. J. and Beckmann, R. S. and Simmons, B. D. and Coil, A. and Devriendt, J. and Dubois, Y. and Garland, I. L. and Lintott, C. J. and Martin, G. and Peirani, S.},
	month = jun,
	year = {2023},
}

@article{hopkins_cosmological_2008,
	title = {A {Cosmological} {Framework} for the {Co}-{Evolution} of {Quasars}, {Supermassive} {Black} {Holes}, and {Elliptical} {Galaxies}. {I}. {Galaxy} {Mergers} and {Quasar} {Activity}},
	volume = {175},
	issn = {0067-0049},
	url = {https://ui.adsabs.harvard.edu/abs/2008ApJS..175..356H},
	doi = {10.1086/524362},
	journal = {\apjs},
	author = {Hopkins, Philip F. and Hernquist, Lars and Cox, Thomas J. and Kereš, Dušan},
	month = apr,
	year = {2008},
	pages = {356--389},
}

@article{hopkins_unified_2006,
	title = {A {Unified}, {Merger}-driven {Model} of the {Origin} of {Starbursts}, {Quasars}, the {Cosmic} {X}-{Ray} {Background}, {Supermassive} {Black} {Holes}, and {Galaxy} {Spheroids}},
	volume = {163},
	issn = {0067-0049},
	url = {https://ui.adsabs.harvard.edu/abs/2006ApJS..163....1H},
	doi = {10.1086/499298},
	journal = {\apjs},
	author = {Hopkins, Philip F. and Hernquist, Lars and Cox, Thomas J. and Di Matteo, Tiziana and Robertson, Brant and Springel, Volker},
	month = mar,
	year = {2006},
	pages = {1--49},
}

@article{di_matteo_energy_2005,
	title = {Energy input from quasars regulates the growth and activity of black holes and their host galaxies},
	volume = {433},
	issn = {0028-0836},
	url = {https://ui.adsabs.harvard.edu/abs/2005Natur.433..604D},
	doi = {10.1038/nature03335},
	journal = {\nat},
	author = {Di Matteo, Tiziana and Springel, Volker and Hernquist, Lars},
	month = feb,
	year = {2005},
	pages = {604--607},
}

@article{somerville_physical_2015,
	title = {Physical {Models} of {Galaxy} {Formation} in a {Cosmological} {Framework}},
	volume = {53},
	issn = {0066-4146, 1545-4282},
	url = {https://www.annualreviews.org/doi/10.1146/annurev-astro-082812-140951},
	doi = {10.1146/annurev-astro-082812-140951},
	number = {1},
	journal = {\araa},
	author = {Somerville, Rachel S. and Davé, Romeel},
	month = aug,
	year = {2015},
	pages = {51--113},
}

@article{barazza_bars_2008,
	title = {Bars in {Disk}-dominated and {Bulge}-dominated {Galaxies} at z {\textasciitilde} 0: {New} {Insights} from {\textasciitilde}3600 {SDSS} {Galaxies}},
	volume = {675},
	issn = {0004-637X},
	shorttitle = {Bars in {Disk}-dominated and {Bulge}-dominated {Galaxies} at z {\textasciitilde} 0},
	url = {https://ui.adsabs.harvard.edu/abs/2008ApJ...675.1194B},
	doi = {10.1086/526510},
	journal = {\apj},
	author = {Barazza, Fabio D. and Jogee, Shardha and Marinova, Irina},
	month = mar,
	year = {2008},
	pages = {1194--1212},
}

@article{geron_galaxy_2021,
	title = {Galaxy zoo: stronger bars facilitate quenching in star-forming galaxies},
	volume = {507},
	issn = {0035-8711},
	shorttitle = {Galaxy zoo},
	url = {https://ui.adsabs.harvard.edu/abs/2021MNRAS.507.4389G},
	doi = {10.1093/mnras/stab2064},
	journal = {\mnras},
	author = {Géron, Tobias and Smethurst, R. J. and Lintott, Chris and Kruk, Sandor and Masters, Karen L. and Simmons, Brooke and Stark, David V.},
	month = nov,
	year = {2021},
	pages = {4389--4408},
}

@article{aguerri_population_2009,
	title = {The population of barred galaxies in the local universe. {I}. {Detection} and characterisation of bars},
	volume = {495},
	issn = {0004-6361},
	url = {https://ui.adsabs.harvard.edu/abs/2009A&A...495..491A},
	doi = {10.1051/0004-6361:200810931},
	journal = {\aap},
	author = {Aguerri, J. A. L. and Méndez-Abreu, J. and Corsini, E. M.},
	month = feb,
	year = {2009},
	pages = {491--504},
}

@article{athanassoula_what_2003,
	title = {What determines the strength and the slowdown rate of bars?},
	volume = {341},
	issn = {0035-8711},
	url = {https://ui.adsabs.harvard.edu/abs/2003MNRAS.341.1179A},
	doi = {10.1046/j.1365-8711.2003.06473.x},
	journal = {\mnras},
	author = {Athanassoula, E.},
	month = jun,
	year = {2003},
	pages = {1179--1198},
}

@article{sakamoto_bar-driven_1999,
	title = {Bar-driven {Transport} of {Molecular} {Gas} to {Galactic} {Centers} and {Its} {Consequences}},
	volume = {525},
	issn = {0004-637X},
	url = {https://ui.adsabs.harvard.edu/abs/1999ApJ...525..691S},
	doi = {10.1086/307910},
	journal = {\apj},
	author = {Sakamoto, K. and Okumura, S. K. and Ishizuki, S. and Scoville, N. Z.},
	month = nov,
	year = {1999},
	pages = {691--701},
}

@article{lin_hydrodynamical_2013,
	title = {Hydrodynamical {Simulations} of the {Barred} {Spiral} {Galaxy} {NGC} 1097},
	volume = {771},
	issn = {0004-637X},
	url = {https://ui.adsabs.harvard.edu/abs/2013ApJ...771....8L},
	doi = {10.1088/0004-637X/771/1/8},
	journal = {\apj},
	author = {Lin, Lien-Hsuan and Wang, Hsiang-Hsu and Hsieh, Pei-Ying and Taam, Ronald E. and Yang, Chao-Chin and Yen, David C. C.},
	month = jul,
	year = {2013},
	pages = {8},
}

@article{galloway_galaxy_2015,
	title = {Galaxy {Zoo}: the effect of bar-driven fuelling on the presence of an active galactic nucleus in disc galaxies},
	volume = {448},
	issn = {0035-8711},
	shorttitle = {Galaxy {Zoo}},
	url = {https://ui.adsabs.harvard.edu/abs/2015MNRAS.448.3442G},
	doi = {10.1093/mnras/stv235},
	journal = {\mnras},
	author = {Galloway, Melanie A. and Willett, Kyle W. and Fortson, Lucy F. and Cardamone, Carolin N. and Schawinski, Kevin and Cheung, Edmond and Lintott, Chris J. and Masters, Karen L. and Melvin, Thomas and Simmons, Brooke D.},
	month = apr,
	year = {2015},
	pages = {3442--3454},
}

@article{lee_bars_2012,
	title = {Do {Bars} {Trigger} {Activity} in {Galactic} {Nuclei}?},
	volume = {750},
	issn = {0004-637X},
	url = {https://ui.adsabs.harvard.edu/abs/2012ApJ...750..141L},
	doi = {10.1088/0004-637X/750/2/141},
	journal = {\apj},
	author = {Lee, Gwang-Ho and Woo, Jong-Hak and Lee, Myung Gyoon and Hwang, Ho Seong and Lee, Jong Chul and Sohn, Jubee and Lee, Jong Hwan},
	month = may,
	year = {2012},
	pages = {141},
}

@article{alexander_what_2012,
	title = {What drives the growth of black holes?},
	volume = {56},
	issn = {1387-6473},
	url = {https://ui.adsabs.harvard.edu/abs/2012NewAR..56...93A},
	doi = {10.1016/j.newar.2011.11.003},
	journal = {\nar},
	author = {Alexander, D. M. and Hickox, R. C.},
	month = jun,
	year = {2012},
	pages = {93--121},
}

@article{skibba_galaxy_2012,
	title = {Galaxy {Zoo}: the environmental dependence of bars and bulges in disc galaxies},
	volume = {423},
	issn = {0035-8711},
	shorttitle = {Galaxy {Zoo}},
	url = {https://ui.adsabs.harvard.edu/abs/2012MNRAS.423.1485S},
	doi = {10.1111/j.1365-2966.2012.20972.x},
	journal = {\mnras},
	author = {Skibba, Ramin A. and Masters, Karen L. and Nichol, Robert C. and Zehavi, Idit and Hoyle, Ben and Edmondson, Edward M. and Bamford, Steven P. and Cardamone, Carolin N. and Keel, William C. and Lintott, Chris and Schawinski, Kevin},
	month = jun,
	year = {2012},
	pages = {1485--1502},
}

@article{marconi_local_2004,
	title = {Local supermassive black holes, relics of active galactic nuclei and the {X}-ray background},
	volume = {351},
	issn = {0035-8711},
	url = {https://ui.adsabs.harvard.edu/abs/2004MNRAS.351..169M},
	doi = {10.1111/j.1365-2966.2004.07765.x},
	journal = {\mnras},
	author = {Marconi, A. and Risaliti, G. and Gilli, R. and Hunt, L. K. and Maiolino, R. and Salvati, M.},
	month = jun,
	year = {2004},
	pages = {169--185},
}

@article{sellwood_secular_2014,
	title = {Secular evolution in disk galaxies},
	volume = {86},
	issn = {0034-6861},
	url = {https://ui.adsabs.harvard.edu/abs/2014RvMP...86....1S},
	doi = {10.1103/RevModPhys.86.1},
	journal = {Reviews of Modern Physics},
	author = {Sellwood, J. A.},
	month = jan,
	year = {2014},
	pages = {1--46},
}

@article{walmsley_zoobot_2023,
	title = {Zoobot: {Adaptable} {Deep} {Learning} {Models} for {Galaxy} {Morphology}},
	volume = {8},
	shorttitle = {Zoobot},
	url = {https://ui.adsabs.harvard.edu/abs/2023JOSS....8.5312W},
	doi = {10.21105/joss.05312},
	journal = {The Journal of Open Source Software},
	author = {Walmsley, Mike and Allen, Campbell and Aussel, Ben and Bowles, Micah and Gregorowicz, Kasia and Slijepcevic, Inigo and Lintott, Chris and Scaife, Anna and Jabłońska, Maja and Karchev, Kosio and Lanzieri, Denise and Mohan, Devina and O'Ryan, David and Saiguhan, Bharath and Suárez, Crisel and Guerra-Varas, Nicolás and Velu, Renuka},
	month = may,
	year = {2023},
	pages = {5312},
}

@article{hart_galaxy_2016,
	title = {Galaxy {Zoo}: comparing the demographics of spiral arm number and a new method for correcting redshift bias},
	volume = {461},
	issn = {0035-8711},
	shorttitle = {Galaxy {Zoo}},
	url = {https://ui.adsabs.harvard.edu/abs/2016MNRAS.461.3663H},
	doi = {10.1093/mnras/stw1588},
	journal = {\mnras},
	author = {Hart, Ross E. and Bamford, Steven P. and Willett, Kyle W. and Masters, Karen L. and Cardamone, Carolin and Lintott, Chris J. and Mackay, Robert J. and Nichol, Robert C. and Rosslowe, Christopher K. and Simmons, Brooke D. and Smethurst, Rebecca J.},
	month = oct,
	year = {2016},
	pages = {3663--3682},
}

@article{lamarcaDustPowerUnravelling2024,
  title = {Dust and Power: {{Unravelling}} the Merger-Active Galactic Nucleus Connection in the Second Half of Cosmic History},
  shorttitle = {Dust and Power},
  author = {La Marca, A. and {Margalef-Bentabol}, B. and Wang, L. and Gao, F. and Goulding, A. D. and Martin, G. and {Rodriguez-Gomez}, V. and Trager, S. C. and Yang, G. and Dav{\'e}, R. and Dubois, Y.},
  year = {2024},
  month = oct,
  journal = {\aap},
  volume = {690},
  pages = {A326},
  publisher = {EDP},
  issn = {0004-6361},
  doi = {10.1051/0004-6361/202348188}
}

@article{treisterMajorGalaxyMergers2012,
  title = {Major {{Galaxy Mergers Only Trigger}} the {{Most Luminous Active Galactic Nuclei}}},
  author = {Treister, E. and Schawinski, K. and Urry, C. M. and Simmons, B. D.},
  year = {2012},
  month = oct,
  journal = {\apj},
  volume = {758},
  pages = {L39},
  issn = {0004-637X},
  doi = {10.1088/2041-8205/758/2/L39}
}

@article{glikmanMajorMergersHost2015,
  title = {Major {{Mergers Host}} the {{Most-luminous Red Quasars}} at z {\textasciitilde} 2: {{A Hubble Space Telescope WFC3}}/{{IR Study}}},
  shorttitle = {Major {{Mergers Host}} the {{Most-luminous Red Quasars}} at z {\textasciitilde} 2},
  author = {Glikman, Eilat and Simmons, Brooke and Mailly, Madeline and Schawinski, Kevin and Urry, C. M. and Lacy, M.},
  year = {2015},
  month = jun,
  journal = {\apj},
  volume = {806},
  pages = {218},
  issn = {0004-637X},
  doi = {10.1088/0004-637X/806/2/218}
}

@article{Chabrier2003,
       author = {{Chabrier}, Gilles},
        title = "{Galactic Stellar and Substellar Initial Mass Function}",
      journal = {\pasp},
         year = 2003,
        month = jul,
       volume = {115},
       number = {809},
        pages = {763-795},
          doi = {10.1086/376392},
archivePrefix = {arXiv},
       eprint = {astro-ph/0304382},
 primaryClass = {astro-ph},
       adsurl = {https://ui.adsabs.harvard.edu/abs/2003PASP..115..763C}
}

@article{Stalevski2016,
       author = {{Stalevski}, Marko and {Ricci}, Claudio and {Ueda}, Yoshihiro and {Lira}, Paulina and {Fritz}, Jacopo and {Baes}, Maarten},
        title = "{The dust covering factor in active galactic nuclei}",
      journal = {\mnras},
         year = 2016,
        month = may,
       volume = {458},
       number = {3},
        pages = {2288-2302},
          doi = {10.1093/mnras/stw444},
archivePrefix = {arXiv},
       eprint = {1602.06954},
 primaryClass = {astro-ph.GA},
       adsurl = {https://ui.adsabs.harvard.edu/abs/2016MNRAS.458.2288S}
}

@article{Holwerda2024GalaxyZoo,
       author = {{Holwerda}, Benne Willem and {Robertson}, Clayton and {Cook}, Kyle and {Pimbblet}, Kevin and {Casura}, Sarah and {Sansom}, Anne E. and {Patel}, Divya and {Butrum}, Trevor Alexander and {Glass}, David Henry William and {Kelvin}, Lee S. and {Baldry}, Ivan K. and {De Propris}, Roberto and {Bamford}, Steven and {Masters}, Karen and {Stone}, Maria Babakhanyan and {Hardin}, Tim and {Walmsley}, Mike and {Liske}, Jochen and {Adnan}, S.~M. Rafee},
        title = "{The Galaxy Zoo catalogues for Galaxy And Mass Assembly (GAMA) survey}",
      journal = {\pasa},
         year = 2024,
        month = dec,
       volume = {41},
          eid = {e115},
        pages = {e115},
          doi = {10.1017/pasa.2024.109},
archivePrefix = {arXiv},
       eprint = {2410.19985},
 primaryClass = {astro-ph.GA},
       adsurl = {https://ui.adsabs.harvard.edu/abs/2024PASA...41..115H}
}

@article{lecunGradientbasedLearningApplied1998,
  title = {Gradient-Based Learning Applied to Document Recognition},
  author = {Lecun, Y. and Bottou, L. and Bengio, Y. and Haffner, P.},
  year = {1998},
  month = nov,
  journal = {Proceedings of the IEEE},
  volume = {86},
  number = {11},
  pages = {2278--2324},
  issn = {1558-2256},
  doi = {10.1109/5.726791}
}

@article{krizhevsky2012imagenet,
  title={Imagenet classification with deep convolutional neural networks},
  author={Krizhevsky, Alex and Sutskever, Ilya and Hinton, Geoffrey E},
  journal={Advances in neural information processing systems},
  volume={25},
  year={2012}
}

@inproceedings{liu2022convnet,
  title={A convnet for the 2020s},
  author={Liu, Zhuang and Mao, Hanzi and Wu, Chao-Yuan and Feichtenhofer, Christoph and Darrell, Trevor and Xie, Saining},
  booktitle={Proceedings of the IEEE/CVF conference on computer vision and pattern recognition},
  pages={11976--11986},
  year={2022}
}

@article{KnapenSubarcsecond2000,
       author = {{Knapen}, Johan H. and {Shlosman}, Isaac and {Peletier}, Reynier F.},
        title = "{A Subarcsecond Resolution Near-Infrared Study of Seyfert and ``Normal'' Galaxies. II. Morphology}",
      journal = {\apj},
         year = 2000,
        month = jan,
       volume = {529},
       number = {1},
        pages = {93-100},
          doi = {10.1086/308266},
archivePrefix = {arXiv},
       eprint = {astro-ph/9907379},
 primaryClass = {astro-ph},
       adsurl = {https://ui.adsabs.harvard.edu/abs/2000ApJ...529...93K}
}

@article{KossHostGalaxy2011,
       author = {{Koss}, Michael and {Mushotzky}, Richard and {Veilleux}, Sylvain and {Winter}, Lisa M. and {Baumgartner}, Wayne and {Tueller}, Jack and {Gehrels}, Neil and {Valencic}, Lynne},
        title = "{Host Galaxy Properties of the Swift Bat Ultra Hard X-Ray Selected Active Galactic Nucleus}",
      journal = {\apj},
         year = 2011,
        month = oct,
       volume = {739},
       number = {2},
          eid = {57},
        pages = {57},
          doi = {10.1088/0004-637X/739/2/57},
archivePrefix = {arXiv},
       eprint = {1107.1237},
 primaryClass = {astro-ph.CO},
       adsurl = {https://ui.adsabs.harvard.edu/abs/2011ApJ...739...57K}
}

@ARTICLE{CisternasXray2013,
       author = {{Cisternas}, Mauricio and {Gadotti}, Dimitri A. and {Knapen}, Johan H. and {Kim}, Taehyun and {D{\'\i}az-Garc{\'\i}a}, Sim{\'o}n and {Laurikainen}, Eija and {Salo}, Heikki and {Gonz{\'a}lez-Mart{\'\i}n}, Omaira and {Ho}, Luis C. and {Elmegreen}, Bruce G. and {Zaritsky}, Dennis and {Sheth}, Kartik and {Athanassoula}, E. and {Bosma}, Albert and {Comer{\'o}n}, S{\'e}bastien and {Erroz-Ferrer}, Santiago and {Gil de Paz}, Armando and {Hinz}, Joannah L. and {Holwerda}, Benne W. and {Laine}, Jarkko and {Meidt}, Sharon and {Men{\'e}ndez-Delmestre}, Kar{\'\i}n and {Mizusawa}, Trisha and {Mu{\~n}oz-Mateos}, Juan Carlos and {Regan}, Michael W. and {Seibert}, Mark},
        title = "{X-Ray Nuclear Activity in S$^{4}$G Barred Galaxies: No Link between Bar Strength and Co-occurrent Supermassive Black Hole Fueling}",
      journal = {\apj},
         year = 2013,
        month = oct,
       volume = {776},
       number = {1},
          eid = {50},
        pages = {50},
          doi = {10.1088/0004-637X/776/1/50},
archivePrefix = {arXiv},
       eprint = {1307.7709},
 primaryClass = {astro-ph.CO},
       adsurl = {https://ui.adsabs.harvard.edu/abs/2013ApJ...776...50C}
}

@ARTICLE{LaineNested2002,
       author = {{Laine}, Seppo and {Shlosman}, Isaac and {Knapen}, Johan H. and {Peletier}, Reynier F.},
        title = "{Nested and Single Bars in Seyfert and Non-Seyfert Galaxies}",
      journal = {\apj},
         year = 2002,
        month = mar,
       volume = {567},
       number = {1},
        pages = {97-117},
          doi = {10.1086/323964},
archivePrefix = {arXiv},
       eprint = {astro-ph/0108029},
 primaryClass = {astro-ph},
       adsurl = {https://ui.adsabs.harvard.edu/abs/2002ApJ...567...97L}
}

@article{Euclid2025Euclid,
       author = {{Euclid Collaboration} and {Mellier}, Y. and {Abdurro'uf} and {Acevedo Barroso}, J.~A. and {Ach{\'u}carro}, A. and {Adamek}, J. and {Adam}, R. and {Addison}, G.~E. and {Aghanim}, N. and {Aguena}, M. and {Ajani}, V. and {Akrami}, Y. and {Al-Bahlawan}, A. and {Alavi}, A. and {Albuquerque}, I.~S. and {Alestas}, G. and {Alguero}, G. and {Allaoui}, A. and {Allen}, S.~W. and {Allevato}, V. and {Alonso-Tetilla}, A.~V. and {Altieri}, B. and {Alvarez-Candal}, A. and {Alvi}, S. and {Amara}, A. and {Amendola}, L. and {Amiaux}, J. and {Andika}, I.~T. and {Andreon}, S. and {Andrews}, A. and {Angora}, G. and {Angulo}, R.~E. and {Annibali}, F. and {Anselmi}, A. and {Anselmi}, S. and {Arcari}, S. and {Archidiacono}, M. and {Aric{\`o}}, G. and {Arnaud}, M. and {Arnouts}, S. and {Asgari}, M. and {Asorey}, J. and {Atayde}, L. and {Atek}, H. and {Atrio-Barandela}, F. and {Aubert}, M. and {Aubourg}, E. and {Auphan}, T. and {Auricchio}, N. and {Aussel}, B. and {Aussel}, H. and {Avelino}, P.~P. and {Avgoustidis}, A. and {Avila}, S. and {Awan}, S. and {Azzollini}, R. and {Baccigalupi}, C. and {Bachelet}, E. and {Bacon}, D. and {Baes}, M. and {Bagley}, M.~B. and {Bahr-Kalus}, B. and {Balaguera-Antolinez}, A. and {Balbinot}, E. and {Balcells}, M. and {Baldi}, M. and {Baldry}, I. and {Balestra}, A. and {Ballardini}, M. and {Ballester}, O. and {Balogh}, M. and {Ba{\~n}ados}, E. and {Barbier}, R. and {Bardelli}, S. and {Baron}, M. and {Barreiro}, T. and {Barrena}, R. and {Barriere}, J. -C. and {Barros}, B.~J. and {Barthelemy}, A. and {Bartolo}, N. and {Basset}, A. and {Battaglia}, P. and {Battisti}, A.~J. and {Baugh}, C.~M. and {Baumont}, L. and {Bazzanini}, L. and {Beaulieu}, J. -P. and {Beckmann}, V. and {Belikov}, A.~N. and {Bel}, J. and {Bellagamba}, F. and {Bella}, M. and {Bellini}, E. and {Benabed}, K. and {Bender}, R. and {Benevento}, G. and {Bennett}, C.~L. and {Benson}, K. and {Bergamini}, P. and {Bermejo-Climent}, J.~R. and {Bernardeau}, F. and {Bertacca}, D. and {Berthe}, M. and {Berthier}, J. and {Bethermin}, M. and {Beutler}, F. and {Bevillon}, C. and {Bhargava}, S. and {Bhatawdekar}, R. and {Bianchi}, D. and {Bisigello}, L. and {Biviano}, A. and {Blake}, R.~P. and {Blanchard}, A. and {Blazek}, J. and {Blot}, L. and {Bosco}, A. and {Bodendorf}, C. and {Boenke}, T. and {B{\"o}hringer}, H. and {Boldrini}, P. and {Bolzonella}, M. and {Bonchi}, A. and {Bonici}, M. and {Bonino}, D. and {Bonino}, L. and {Bonvin}, C. and {Bon}, W. and {Booth}, J.~T. and {Borgani}, S. and {Borlaff}, A.~S. and {Borsato}, E. and {Bose}, B. and {Botticella}, M.~T. and {Boucaud}, A. and {Bouche}, F. and {Boucher}, J.~S. and {Boutigny}, D. and {Bouvard}, T. and {Bouwens}, R. and {Bouy}, H. and {Bowler}, R.~A.~A. and {Bozza}, V. and {Bozzo}, E. and {Branchini}, E. and {Brando}, G. and {Brau-Nogue}, S. and {Brekke}, P. and {Bremer}, M.~N. and {Brescia}, M. and {Breton}, M. -A. and {Brinchmann}, J. and {Brinckmann}, T. and {Brockley-Blatt}, C. and {Brodwin}, M. and {Brouard}, L. and {Brown}, M.~L. and {Bruton}, S. and {Bucko}, J. and {Buddelmeijer}, H. and {Buenadicha}, G. and {Buitrago}, F. and {Burger}, P. and {Burigana}, C. and {Busillo}, V. and {Busonero}, D. and {Cabanac}, R. and {Cabayol-Garcia}, L. and {Cagliari}, M.~S. and {Caillat}, A. and {Caillat}, L. and {Calabrese}, M. and {Calabro}, A. and {Calderone}, G. and {Calura}, F. and {Camacho Quevedo}, B. and {Camera}, S. and {Campos}, L. and {Ca{\~n}as-Herrera}, G. and {Candini}, G.~P. and {Cantiello}, M. and {Capobianco}, V. and {Cappellaro}, E. and {Cappelluti}, N. and {Cappi}, A. and {Caputi}, K.~I. and {Cara}, C. and {Carbone}, C. and {Cardone}, V.~F. and {Carella}, E. and {Carlberg}, R.~G. and {Carle}, M. and {Carminati}, L. and {Caro}, F. and {Carrasco}, J.~M. and {Carretero}, J. and {Carrilho}, P. and {Carron Duque}, J. and {Carry}, B.},
        title = "{Euclid: I. Overview of the Euclid mission}",
      journal = {\aap},
         year = 2025,
        month = may,
       volume = {697},
          eid = {A1},
        pages = {A1},
          doi = {10.1051/0004-6361/202450810},
archivePrefix = {arXiv},
       eprint = {2405.13491},
 primaryClass = {astro-ph.CO},
       adsurl = {https://ui.adsabs.harvard.edu/abs/2025A&A...697A...1E}
}

@article{Euclid2025Q1BarFraction,
       author = {{Euclid Collaboration} and {Huertas-Company}, M. and {Walmsley}, M. and {Siudek}, M. and {Iglesias-Navarro}, P. and {Knapen}, J.~H. and {Serjeant}, S. and {Dickinson}, H.~J. and {Fortson}, L. and {Garland}, I. and {G{\'e}ron}, T. and {Keel}, W. and {Kruk}, S. and {Lintott}, C.~J. and {Mantha}, K. and {Masters}, K. and {O'Ryan}, D. and {Popp}, J.~J. and {Roberts}, H. and {Scarlata}, C. and {Makechemu}, J.~S. and {Simmons}, B. and {Smethurst}, R.~J. and {Spindler}, A. and {Baes}, M. and {Corsini}, E.~M. and {Dom{\'\i}nguez S{\'a}nchez}, H. and {Duran-Camacho}, E. and {Fu}, H. and {Junais}, J. and {Mendez-Abreu}, J. and {Nersesian}, A. and {Shankar}, F. and {Le}, M.~N. and {Vega-Ferrero}, J. and {Wang}, L. and {Aghanim}, N. and {Altieri}, B. and {Amara}, A. and {Andreon}, S. and {Auricchio}, N. and {Baccigalupi}, C. and {Baldi}, M. and {Balestra}, A. and {Bardelli}, S. and {Basset}, A. and {Battaglia}, P. and {Bernardeau}, F. and {Biviano}, A. and {Bonchi}, A. and {Branchini}, E. and {Brescia}, M. and {Brinchmann}, J. and {Camera}, S. and {Capobianco}, V. and {Carbone}, C. and {Carretero}, J. and {Casas}, S. and {Castellano}, M. and {Castignani}, G. and {Cavuoti}, S. and {Chambers}, K.~C. and {Cimatti}, A. and {Colodro-Conde}, C. and {Congedo}, G. and {Conselice}, C.~J. and {Conversi}, L. and {Copin}, Y. and {Courbin}, F. and {Courtois}, H.~M. and {Cropper}, M. and {Da Silva}, A. and {Degaudenzi}, H. and {De Lucia}, G. and {Di Giorgio}, A.~M. and {Dolding}, C. and {Dole}, H. and {Dubath}, F. and {Duncan}, C.~A.~J. and {Dupac}, X. and {Dusini}, S. and {Ealet}, A. and {Escoffier}, S. and {Fabricius}, M. and {Farina}, M. and {Farinelli}, R. and {Faustini}, F. and {Ferriol}, S. and {Finelli}, F. and {Fotopoulou}, S. and {Frailis}, M. and {Galeotta}, S. and {George}, K. and {Gillard}, W. and {Gillis}, B. and {Giocoli}, C. and {Gracia-Carpio}, J. and {Grazian}, A. and {Grupp}, F. and {Gwyn}, S. and {Haugan}, S.~V.~H. and {Hoekstra}, H. and {Holmes}, W. and {Hook}, I.~M. and {Hormuth}, F. and {Hornstrup}, A. and {Hudelot}, P. and {Jahnke}, K. and {Jhabvala}, M. and {Keih{\"a}nen}, E. and {Kermiche}, S. and {Kubik}, B. and {Kuijken}, K. and {K{\"u}mmel}, M. and {Kunz}, M. and {Kurki-Suonio}, H. and {Le Boulc'h}, Q. and {Le Brun}, A.~M.~C. and {Le Mignant}, D. and {Ligori}, S. and {Lilje}, P.~B. and {Lindholm}, V. and {Lloro}, I. and {Maino}, D. and {Maiorano}, E. and {Mansutti}, O. and {Marcin}, S. and {Marggraf}, O. and {Martinelli}, M. and {Martinet}, N. and {Marulli}, F. and {Massey}, R. and {McCracken}, H.~J. and {Medinaceli}, E. and {Melchior}, M. and {Mellier}, Y. and {Meneghetti}, M. and {Merlin}, E. and {Meylan}, G. and {Mora}, A. and {Moresco}, M. and {Moscardini}, L. and {Neissner}, C. and {Nichol}, R.~C. and {Niemi}, S. -M. and {Nightingale}, J.~W. and {Padilla}, C. and {Paltani}, S. and {Pasian}, F. and {Pedersen}, K. and {Percival}, W.~J. and {Pettorino}, V. and {Pires}, S. and {Polenta}, G. and {Poncet}, M. and {Popa}, L.~A. and {Pozzetti}, L. and {Raison}, F. and {Renzi}, A. and {Rhodes}, J. and {Riccio}, G. and {Romelli}, E. and {Roncarelli}, M. and {Saglia}, R. and {Sakr}, Z. and {Sapone}, D. and {Sartoris}, B. and {Schirmer}, M. and {Schneider}, P. and {Scodeggio}, M. and {Secroun}, A. and {Seidel}, G. and {Seiffert}, M. and {Serrano}, S. and {Simon}, P. and {Sirignano}, C. and {Sirri}, G. and {Stanco}, L. and {Steinwagner}, J. and {Tallada-Cresp{\'\i}}, P. and {Taylor}, A.~N. and {Tereno}, I. and {Toft}, S. and {Toledo-Moreo}, R. and {Torradeflot}, F. and {Tutusaus}, I. and {Valenziano}, L. and {Valiviita}, J. and {Vassallo}, T. and {Verdoes Kleijn}, G. and {Wang}, Y. and {Weller}, J. and {Zacchei}, A. and {Zamorani}, G. and {Zerbi}, F.~M. and {Zinchenko}, I.~A. and {Zucca}, E. and {Allevato}, V. and {Ballardini}, M. and {Bolzonella}, M.},
        title = "{Euclid Quick Data Release (Q1), A first look at the fraction of bars in massive galaxies at $z<1$}",
      journal = {\aap in press},
         year = 2025,
        month = mar,
          eid = {arXiv:2503.15311},
        pages = {arXiv:2503.15311},
          doi = {10.48550/arXiv.2503.15311},
archivePrefix = {arXiv},
       eprint = {2503.15311},
 primaryClass = {astro-ph.GA},
       adsurl = {https://ui.adsabs.harvard.edu/abs/2025arXiv250315311E}
}

@ARTICLE{Erwin2018Dependence,
       author = {{Erwin}, Peter},
        title = "{The dependence of bar frequency on galaxy mass, colour, and gas content - and angular resolution - in the local universe}",
      journal = {\mnras},
         year = 2018,
        month = mar,
       volume = {474},
       number = {4},
        pages = {5372-5392},
          doi = {10.1093/mnras/stx3117},
archivePrefix = {arXiv},
       eprint = {1711.04867},
 primaryClass = {astro-ph.GA}
}

@ARTICLE{Marels2025RoleBars,
       author = {{Marels}, V. and {Mesa}, V. and {Jaque Arancibia}, M. and {Alonso}, S. and {Coldwell}, G. and {Damke}, G. and {Contreras Rojas}, V.},
        title = "{The role of bars in triggering active galactic nucleus galaxies}",
      journal = {\aap},
         year = 2025,
        month = jul,
       volume = {699},
          eid = {A204},
        pages = {A204},
          doi = {10.1051/0004-6361/202554961},
archivePrefix = {arXiv},
       eprint = {2505.23958},
 primaryClass = {astro-ph.GA},
       adsurl = {https://ui.adsabs.harvard.edu/abs/2025A&A...699A.204M}
}

@ARTICLE{RamosAlmeida2012AreLuminous,
       author = {{Ramos Almeida}, C. and {Bessiere}, P.~S. and {Tadhunter}, C.~N. and {P{\'e}rez-Gonz{\'a}lez}, P.~G. and {Barro}, G. and {Inskip}, K.~J. and {Morganti}, R. and {Holt}, J. and {Dicken}, D.},
        title = "{Are luminous radio-loud active galactic nuclei triggered by galaxy interactions?}",
      journal = {\mnras},
         year = 2012,
        month = jan,
       volume = {419},
       number = {1},
        pages = {687-705},
          doi = {10.1111/j.1365-2966.2011.19731.x},
archivePrefix = {arXiv},
       eprint = {1109.0021},
 primaryClass = {astro-ph.CO},
       adsurl = {https://ui.adsabs.harvard.edu/abs/2012MNRAS.419..687R}
}

@ARTICLE{RamosAlmeida2011OpticalMorphologies,
       author = {{Ramos Almeida}, C. and {Tadhunter}, C.~N. and {Inskip}, K.~J. and {Morganti}, R. and {Holt}, J. and {Dicken}, D.},
        title = "{The optical morphologies of the 2 Jy sample of radio galaxies: evidence for galaxy interactions}",
      journal = {\mnras},
         year = 2011,
        month = jan,
       volume = {410},
       number = {3},
        pages = {1550-1576},
          doi = {10.1111/j.1365-2966.2010.17542.x},
archivePrefix = {arXiv},
       eprint = {1008.2683},
 primaryClass = {astro-ph.CO},
       adsurl = {https://ui.adsabs.harvard.edu/abs/2011MNRAS.410.1550R}
}

@ARTICLE{Hopkins2010HowDoMassive,
       author = {{Hopkins}, Philip F. and {Quataert}, Eliot},
        title = "{How do massive black holes get their gas?}",
      journal = {\mnras},
         year = 2010,
        month = sep,
       volume = {407},
       number = {3},
        pages = {1529-1564},
          doi = {10.1111/j.1365-2966.2010.17064.x},
archivePrefix = {arXiv},
       eprint = {0912.3257},
 primaryClass = {astro-ph.CO},
       adsurl = {https://ui.adsabs.harvard.edu/abs/2010MNRAS.407.1529H}
}

@ARTICLE{EuclidCollaboration2025Q1Mergers,
       author = {{Euclid Collaboration} and {La Marca}, A. and {Wang}, L. and {Margalef-Bentabol}, B. and {Gabarra}, L. and {Toba}, Y. and {Mezcua}, M. and {Rodriguez-Gomez}, V. and {Ricci}, F. and {Fotopoulou}, S. and {Matamoro Zatarain}, T. and {Allevato}, V. and {La Franca}, F. and {Shankar}, F. and {Bisigello}, L. and {Stevens}, G. and {Siudek}, M. and {Roster}, W. and {Salvato}, M. and {Tortora}, C. and {Spinoglio}, L. and {Man}, A.~W.~S. and {Knapen}, J.~H. and {Baes}, M. and {O'Ryan}, D. and {Aghanim}, N. and {Altieri}, B. and {Amara}, A. and {Andreon}, S. and {Auricchio}, N. and {Aussel}, H. and {Baccigalupi}, C. and {Baldi}, M. and {Bardelli}, S. and {Battaglia}, P. and {Biviano}, A. and {Bonchi}, A. and {Branchini}, E. and {Brescia}, M. and {Brinchmann}, J. and {Camera}, S. and {Ca{\~n}as-Herrera}, G. and {Capobianco}, V. and {Carbone}, C. and {Carretero}, J. and {Castellano}, M. and {Castignani}, G. and {Cavuoti}, S. and {Chambers}, K.~C. and {Cimatti}, A. and {Colodro-Conde}, C. and {Congedo}, G. and {Conselice}, C.~J. and {Conversi}, L. and {Copin}, Y. and {Costille}, A. and {Courbin}, F. and {Courtois}, H.~M. and {Cropper}, M. and {Da Silva}, A. and {Degaudenzi}, H. and {De Lucia}, G. and {Di Giorgio}, A.~M. and {Dolding}, C. and {Dole}, H. and {Dubath}, F. and {Duncan}, C.~A.~J. and {Dupac}, X. and {Ealet}, A. and {Escoffier}, S. and {Fabricius}, M. and {Farina}, M. and {Farinelli}, R. and {Faustini}, F. and {Ferriol}, S. and {Finelli}, F. and {Frailis}, M. and {Franceschi}, E. and {Galeotta}, S. and {George}, K. and {Gillis}, B. and {Giocoli}, C. and {G{\'o}mez-Alvarez}, P. and {Gracia-Carpio}, J. and {Granett}, B.~R. and {Grazian}, A. and {Grupp}, F. and {Guzzo}, L. and {Gwyn}, S. and {Haugan}, S.~V.~H. and {Holmes}, W. and {Hook}, I.~M. and {Hormuth}, F. and {Hornstrup}, A. and {Hudelot}, P. and {Jahnke}, K. and {Jhabvala}, M. and {Joachimi}, B. and {Keih{\"a}nen}, E. and {Kermiche}, S. and {Kiessling}, A. and {Kubik}, B. and {K{\"u}mmel}, M. and {Kunz}, M. and {Kurki-Suonio}, H. and {Le Boulc'h}, Q. and {Le Brun}, A.~M.~C. and {Le Mignant}, D. and {Ligori}, S. and {Lilje}, P.~B. and {Lindholm}, V. and {Lloro}, I. and {Mainetti}, G. and {Maino}, D. and {Maiorano}, E. and {Mansutti}, O. and {Marcin}, S. and {Marggraf}, O. and {Martinelli}, M. and {Martinet}, N. and {Marulli}, F. and {Massey}, R. and {Maurogordato}, S. and {Medinaceli}, E. and {Mei}, S. and {Melchior}, M. and {Mellier}, Y. and {Meneghetti}, M. and {Merlin}, E. and {Meylan}, G. and {Mora}, A. and {Moresco}, M. and {Moscardini}, L. and {Nakajima}, R. and {Neissner}, C. and {Niemi}, S. -M. and {Nightingale}, J.~W. and {Padilla}, C. and {Paltani}, S. and {Pasian}, F. and {Pedersen}, K. and {Percival}, W.~J. and {Pettorino}, V. and {Pires}, S. and {Polenta}, G. and {Poncet}, M. and {Popa}, L.~A. and {Pozzetti}, L. and {Raison}, F. and {Rebolo}, R. and {Renzi}, A. and {Rhodes}, J. and {Riccio}, G. and {Romelli}, E. and {Roncarelli}, M. and {Rusholme}, B. and {Saglia}, R. and {Sakr}, Z. and {Sapone}, D. and {Sartoris}, B. and {Schewtschenko}, J.~A. and {Schneider}, P. and {Schrabback}, T. and {Scodeggio}, M. and {Secroun}, A. and {Seidel}, G. and {Seiffert}, M. and {Serrano}, S. and {Simon}, P. and {Sirignano}, C. and {Sirri}, G. and {Stanco}, L. and {Steinwagner}, J. and {Tallada-Cresp{\'\i}}, P. and {Taylor}, A.~N. and {Teplitz}, H.~I. and {Tereno}, I. and {Tessore}, N. and {Toft}, S. and {Toledo-Moreo}, R. and {Torradeflot}, F. and {Tutusaus}, I. and {Valenziano}, L. and {Valiviita}, J. and {Vassallo}, T. and {Verdoes Kleijn}, G. and {Veropalumbo}, A. and {Wang}, Y. and {Weller}, J. and {Zacchei}, A. and {Zamorani}, G. and {Zerbi}, F.~M. and {Zinchenko}, I.~A. and {Zucca}, E. and {Ballardini}, M. and {Bolzonella}, M. and {Bozzo}, E. and {Burigana}, C. and {Cabanac}, R. and {Cappi}, A.},
        title = "{Euclid Quick Data Release (Q1). First Euclid statistical study of galaxy mergers and their connection to active galactic nuclei}",
      journal = {\aap in press},
         year = 2025,
        month = mar,
          eid = {arXiv:2503.15317},
        pages = {arXiv:2503.15317},
          doi = {10.48550/arXiv.2503.15317},
archivePrefix = {arXiv},
       eprint = {2503.15317},
 primaryClass = {astro-ph.GA}
}

@ARTICLE{Cisternas2011SecularEvolution,
       author = {{Cisternas}, Mauricio and {Jahnke}, Knud and {Bongiorno}, Angela and {Inskip}, Katherine J. and {Impey}, Chris D. and {Koekemoer}, Anton M. and {Merloni}, Andrea and {Salvato}, Mara and {Trump}, Jonathan R.},
        title = "{Secular Evolution and a Non-evolving Black-hole-to-galaxy Mass Ratio in the Last 7 Gyr}",
      journal = {\apjl},
         year = 2011,
        month = nov,
       volume = {741},
       number = {1},
          eid = {L11},
        pages = {L11},
          doi = {10.1088/2041-8205/741/1/L11},
archivePrefix = {arXiv},
       eprint = {1109.4633},
 primaryClass = {astro-ph.CO}
}

@ARTICLE{Shlosman1989BarsWithin,
       author = {{Shlosman}, Isaac and {Frank}, Juhan and {Begelman}, Mitchell C.},
        title = "{Bars within bars: a mechanism for fuelling active galactic nuclei}",
      journal = {\nat},
         year = 1989,
        month = mar,
       volume = {338},
       number = {6210},
        pages = {45-47},
          doi = {10.1038/338045a0}
}

@ARTICLE{Guo2025Abundance,
       author = {{Guo}, Yuchen and {Jogee}, Shardha and {Wise}, Eden and {Pritchett}, Keith and {McGrath}, Elizabeth J. and {Finkelstein}, Steven L. and {Iyer}, Kartheik G. and {Arrabal Haro}, Pablo and {Bagley}, Micaela B. and {Dickinson}, Mark and et al.},
        title = "{The Abundance and Properties of Barred Galaxies out to z {\ensuremath{\sim}} 4 Using JWST CEERS Data}",
      journal = {\apj},
         year = 2025,
        month = jun,
       volume = {985},
       number = {2},
          eid = {181},
        pages = {181},
          doi = {10.3847/1538-4357/adc8a7},
archivePrefix = {arXiv},
       eprint = {2409.06100},
 primaryClass = {astro-ph.GA}
}

@ARTICLE{LeConte2024JWSTInvestigation,
       author = {{Le Conte}, Zoe A. and {Gadotti}, Dimitri A. and {Ferreira}, Leonardo and {Conselice}, Christopher J. and {de S{\'a}-Freitas}, Camila and {Kim}, Taehyun and {Neumann}, Justus and {Fragkoudi}, Francesca and {Athanassoula}, E. and {Adams}, Nathan J.},
        title = "{A JWST investigation into the bar fraction at redshifts 1 {\ensuremath{\leq}} z {\ensuremath{\leq}} 3}",
      journal = {\mnras},
         year = 2024,
        month = may,
       volume = {530},
       number = {2},
        pages = {1984-2000},
          doi = {10.1093/mnras/stae921},
archivePrefix = {arXiv},
       eprint = {2309.10038},
 primaryClass = {astro-ph.GA}
}

@ARTICLE{MenendezDelmestre2007NearInfrared,
       author = {{Men{\'e}ndez-Delmestre}, Kar{\'\i}n and {Sheth}, Kartik and {Schinnerer}, Eva and {Jarrett}, Thomas H. and {Scoville}, Nick Z.},
        title = "{A Near-Infrared Study of 2MASS Bars in Local Galaxies: An Anchor for High-Redshift Studies}",
      journal = {\apj},
         year = 2007,
        month = mar,
       volume = {657},
       number = {2},
        pages = {790-804},
          doi = {10.1086/511025},
archivePrefix = {arXiv},
       eprint = {astro-ph/0611540},
 primaryClass = {astro-ph}
}

@ARTICLE{Eskridge2000FrequencyBarred,
       author = {{Eskridge}, Paul B. and {Frogel}, Jay A. and {Pogge}, Richard W. and {Quillen}, Alice C. and {Davies}, Roger L. and {DePoy}, D.~L. and {Houdashelt}, Mark L. and {Kuchinski}, Leslie E. and {Ram{\'\i}rez}, Solange V. and {Sellgren}, K. and {Terndrup}, Donald M. and {Tiede}, Glenn P.},
        title = "{The Frequency of Barred Spiral Galaxies in the Near-Infrared}",
      journal = {\aj},
         year = 2000,
        month = feb,
       volume = {119},
       number = {2},
        pages = {536-544},
          doi = {10.1086/301203},
archivePrefix = {arXiv},
       eprint = {astro-ph/9910479},
 primaryClass = {astro-ph}
}

@ARTICLE{MendezAbreu2017TwoDimensional,
       author = {{M{\'e}ndez-Abreu}, J. and {Ruiz-Lara}, T. and {S{\'a}nchez-Menguiano}, L. and {de Lorenzo-C{\'a}ceres}, A. and {Costantin}, L. and {Catal{\'a}n-Torrecilla}, C. and {Florido}, E. and {Aguerri}, J.~A.~L. and {Bland-Hawthorn}, J. and {Corsini}, E.~M. and {Dettmar}, R.~J. and {Galbany}, L. and {Garc{\'\i}a-Benito}, R. and {Marino}, R.~A. and {M{\'a}rquez}, I. and {Ortega-Minakata}, R.~A. and {Papaderos}, P. and {S{\'a}nchez}, S.~F. and {S{\'a}nchez-Blazquez}, P. and {Spekkens}, K. and {van de Ven}, G. and {Wild}, V. and {Ziegler}, B.},
        title = "{Two-dimensional multi-component photometric decomposition of CALIFA galaxies}",
      journal = {\aap},
         year = 2017,
        month = feb,
       volume = {598},
          eid = {A32},
        pages = {A32},
          doi = {10.1051/0004-6361/201629525},
archivePrefix = {arXiv},
       eprint = {1610.05324},
 primaryClass = {astro-ph.GA}
}

@ARTICLE{deSaFreitas2023NewMethod,
       author = {{de S{\'a}-Freitas}, Camila and {Fragkoudi}, Francesca and {Gadotti}, Dimitri A. and {Falc{\'o}n-Barroso}, Jes{\'u}s and {Bittner}, Adrian and {S{\'a}nchez-Bl{\'a}zquez}, Patricia and {van de Ven}, Glenn and {Bieri}, Rebekka and {Coccato}, Lodovico and {Coelho}, Paula and {Fahrion}, Katja and {Gon{\c{c}}alves}, Geraldo and {Kim}, Taehyun and {de Lorenzo-C{\'a}ceres}, Adriana and {Martig}, Marie and {Mart{\'\i}n-Navarro}, Ignacio and {Mendez-Abreu}, Jairo and {Neumann}, Justus and {Querejeta}, Miguel},
        title = "{A new method for age-dating the formation of bars in disc galaxies. The TIMER view on NGC1433's old bar and the inside-out growth of its nuclear disc}",
      journal = {\aap},
         year = 2023,
        month = mar,
       volume = {671},
          eid = {A8},
        pages = {A8},
          doi = {10.1051/0004-6361/202244667},
archivePrefix = {arXiv},
       eprint = {2211.07670},
 primaryClass = {astro-ph.GA}
}

@ARTICLE{deSaFreitas2025BarAges,
       author = {{de S{\'a}-Freitas}, Camila and {Gadotti}, Dimitri A. and {Fragkoudi}, Francesca and {Coelho}, Paula and {de Lorenzo-C{\'a}ceres}, Adriana and {Falc{\'o}n-Barroso}, Jes{\'u}s and {S{\'a}nchez-Bl{\'a}zquez}, Patricia and {Kim}, Taehyun and {Mendez-Abreu}, Jairo and {Neumann}, Justus and {Querejeta}, Miguel and {van de Ven}, Glenn},
        title = "{Bar ages derived for the first time in nearby galaxies: Insights into secular evolution from the TIMER sample}",
      journal = {\aap},
         year = 2025,
        month = jun,
       volume = {698},
          eid = {A5},
        pages = {A5},
          doi = {10.1051/0004-6361/202453367},
archivePrefix = {arXiv},
       eprint = {2503.20864},
 primaryClass = {astro-ph.GA}
}

@ARTICLE{Malek2018HELPModelling,
       author = {{Ma{\l}ek}, K. and {Buat}, V. and {Roehlly}, Y. and {Burgarella}, D. and {Hurley}, P.~D. and {Shirley}, R. and {Duncan}, K. and {Efstathiou}, A. and {Papadopoulos}, A. and {Vaccari}, M. and {Farrah}, D. and {Marchetti}, L. and {Oliver}, S.},
        title = "{HELP: modelling the spectral energy distributions of Herschel detected galaxies in the ELAIS N1 field}",
      journal = {\aap},
         year = 2018,
        month = nov,
       volume = {620},
          eid = {A50},
        pages = {A50},
          doi = {10.1051/0004-6361/201833131},
archivePrefix = {arXiv},
       eprint = {1809.00529},
 primaryClass = {astro-ph.GA}
}

@ARTICLE{Romeo2016WhatPowers,
       author = {{Romeo}, Alessandro B. and {Fathi}, Kambiz},
        title = "{What powers the starburst activity of NGC 1068? Star-driven gravitational instabilities caught in the act}",
      journal = {\mnras},
         year = 2016,
        month = aug,
       volume = {460},
       number = {3},
        pages = {2360-2367},
          doi = {10.1093/mnras/stw1147},
archivePrefix = {arXiv},
       eprint = {1602.03049},
 primaryClass = {astro-ph.GA}
}

@ARTICLE{Romeo2015DoubleMolecular,
       author = {{Romeo}, Alessandro B. and {Fathi}, Kambiz},
        title = "{A double molecular disc in the triple-barred starburst galaxy NGC 6946: structure and stability}",
      journal = {\mnras},
         year = 2015,
        month = aug,
       volume = {451},
       number = {3},
        pages = {3107-3116},
          doi = {10.1093/mnras/stv1220},
archivePrefix = {arXiv},
       eprint = {1503.01326},
 primaryClass = {astro-ph.GA}
}

\begin{appendix}

\section{Classification examples and additional confusion matrices}\label{app:example_img}

\begin{figure*}
    \centering
    \includegraphics[width=0.95\textwidth]{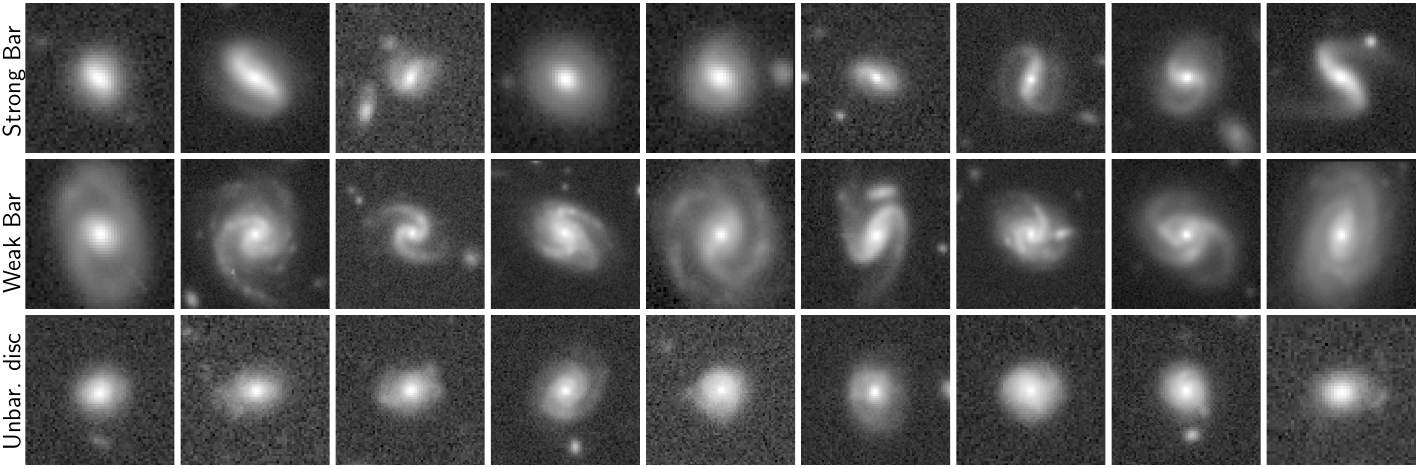}
    \includegraphics[width=0.95\textwidth]{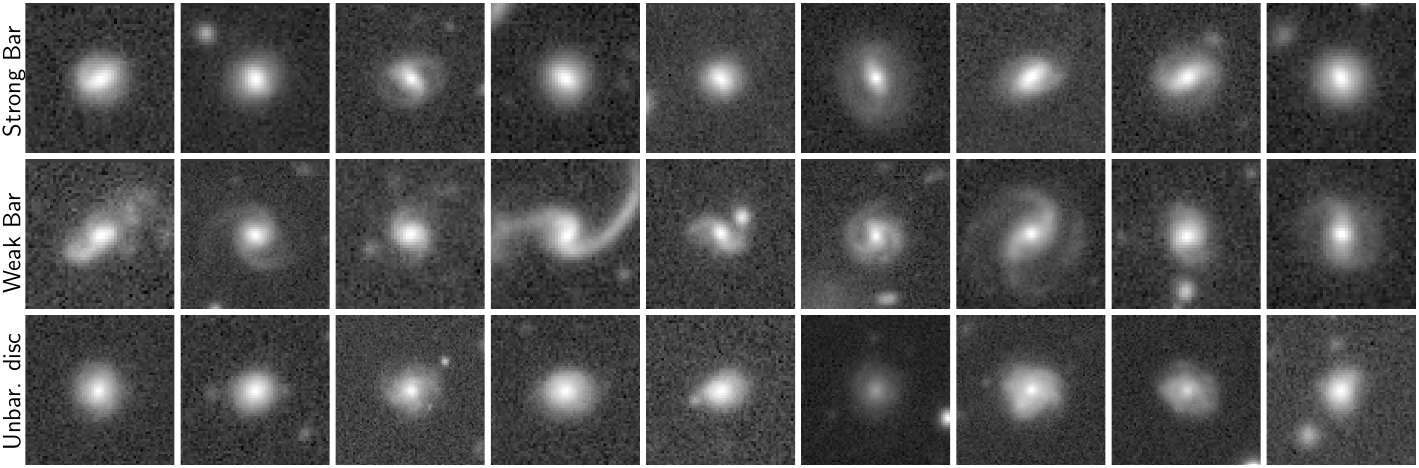}
    \caption{Random examples of barred and unbarred disc galaxies identified by the \texttt{Zoobot} model we trained. Top three rows show examples at $z<0.5$, while the lower three rows display examples at $z\geq0.5$. Same scaling as Fig.~\ref{fig:examples_zoobot}.}
    \label{fig:examples}
\end{figure*}

To visually illustrate the morphological characteristics of the galaxy classes defined by our fine-tuned \texttt{Zoobot} model and selection criteria (Sect.~\ref{sect:bar-selection}), Fig.~\ref{fig:examples} presents random examples of galaxies classified as strongly barred ($S_{\rm bar}$), weakly barred ($W_{\rm bar}$), and unbarred discs ($U_{\rm bar}$). The figure displays HSC-SSP $i$-band image cutouts for these three classes, divided into examples of galaxies at lower redshifts ($z < 0.5$, top three rows) and higher redshifts ($z \geq 0.5$, bottom three rows). 

Observing Fig.~\ref{fig:examples}, $S_{\rm bar}$ galaxies typically exhibit prominent, elongated bar structures at their centres, often connecting directly to spiral arms. $W_{\rm bar}$ galaxies display more subtle bar-like features. These can include less elongated or lower contrast central structures, sometimes appearing as oval distortions or thickened inner discs that are suggestive of a bar but lack the definitive, sharp features of strong bars. Distinguishing these from unbarred galaxies or galaxies with prominent bulges can be challenging, reflecting the lower purity of this class as discussed in Sect.~\ref{sect:bar-selection}. The $U_{\rm bar}$ galaxies generally lack any clear, centrally dominant linear bar structure. They often show smooth central light profiles or distinct spiral arm patterns originating closer to the galactic nucleus without an intervening bar. 

The examples at different redshifts demonstrate the challenge of morphological classification with increasing distance. Features become less distinct and more pixelated at higher z, which underscores the importance of carefully constructing the training sample for studies extending beyond the local Universe. Nevertheless, the \texttt{Zoobot} model, trained on GZ classifications which themselves span a range of redshifts, is capable of identifying these features across the redshift range of our study.

\begin{figure*}
    \centering
    \includegraphics[width=0.3\textwidth]{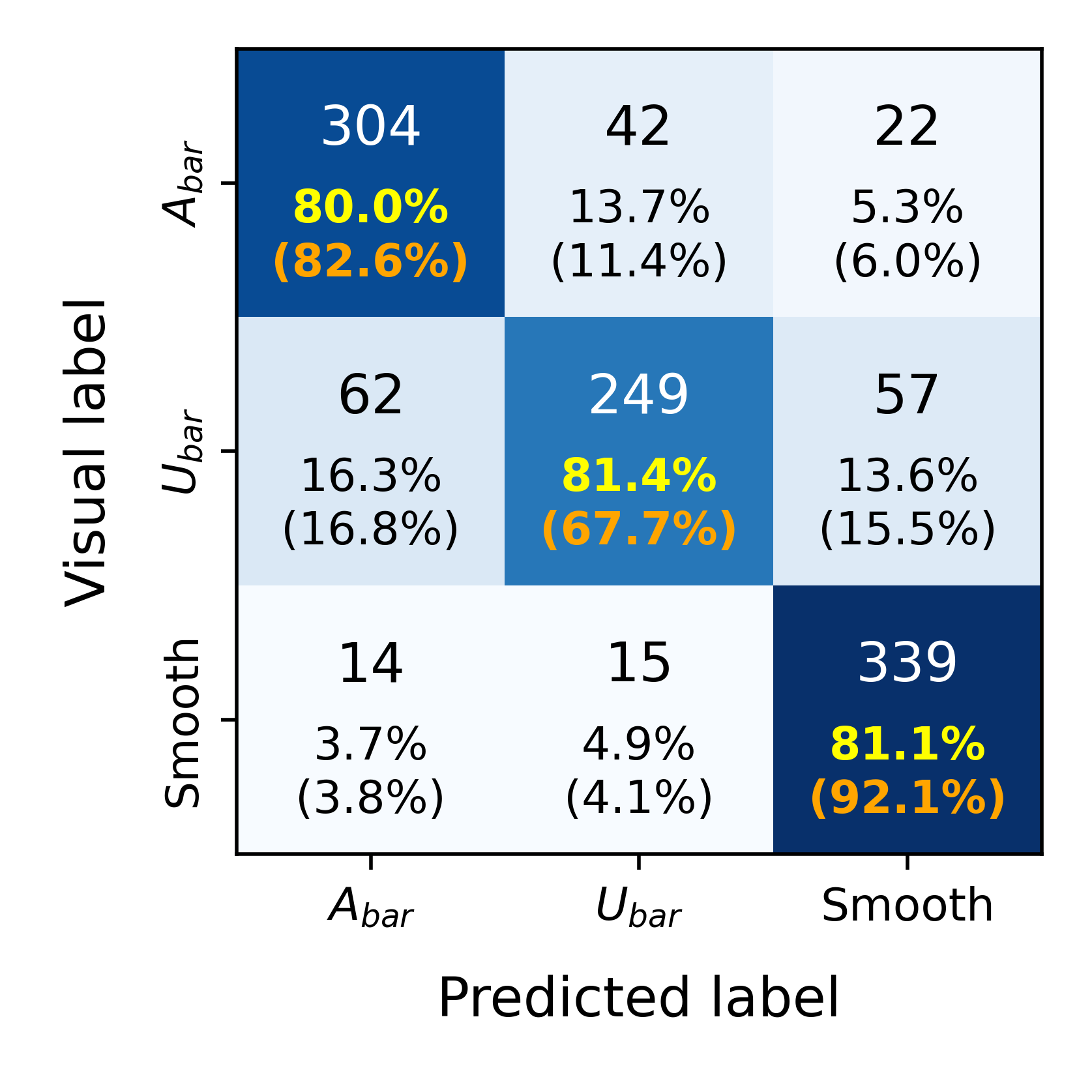}
    \includegraphics[width=0.39\textwidth]{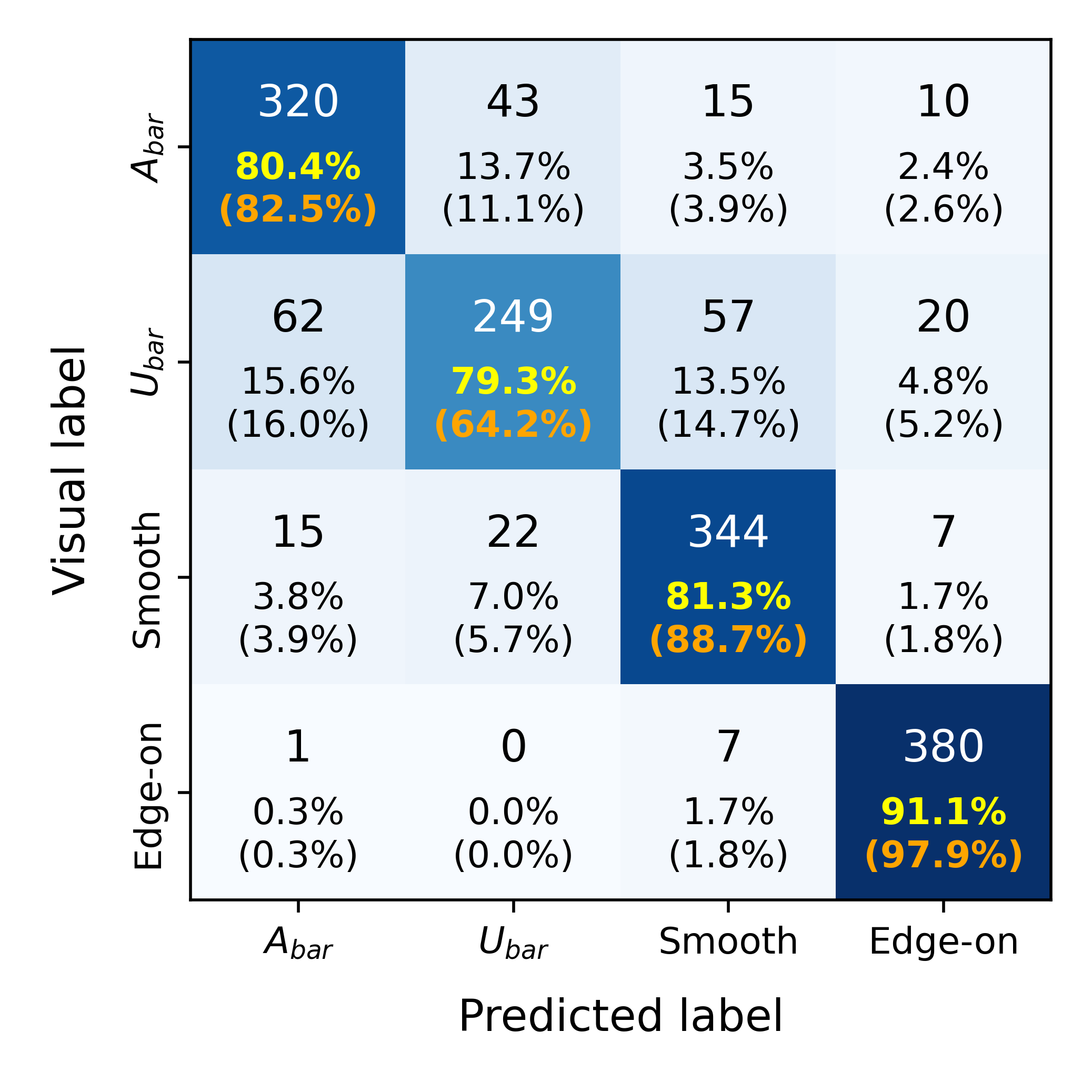}
    \caption{Confusion matrix for the \texttt{Zoobot} model we trained, colour-coded relative to the total number of galaxies in the test set. Each cell contains the raw galaxy counts, the ratio over the number of predictions per class as a percentage (second line), and the ratio over the number of labels per class (in brackets). Along the diagonal, we highlight the precision of each class (yellow, boldface text) and the recall (orange text). \textit{Left panel}, confusion matrix considering the $A_{\rm bar}$, $U_{\rm bar}$, and Smooth classes. The \textit{Right panel} also includes the edge-on class.}
    \label{fig:confusion_bar_nobar_smooth}
\end{figure*}

In this section, we also demonstrate how the definition of barred and unbarred galaxies we adopted affects the model's performance. Specifically, in Fig.~\ref{fig:confusion_bar_nobar_smooth}, we report the confusion matrices we get for the test set, using the $A_{\rm bar}$ and $U_{\rm bar}$ definitions given in Sect.~\ref{sect:bar-selection} and the smooth and edge-on definitions in Sect.~\ref{sect:CNN-perf}. As both matrices demonstrate, the confusion between the $U_{\rm bar}$ class and the smooth class considerably decreased, now being lower than $10\%$.

\section{SED AGN selection}\label{app:agn_thresh}

In our previous work \citepalias{lamarcaDustPowerUnravelling2024}, we defined SED AGN as galaxies with an AGN fraction $f_{\rm AGN} \geq 0.05$, based on the MIR (3--30$\,\mu$m) contribution derived from \texttt{CIGALE} SED fitting. While this threshold captures galaxies with non-negligible AGN contributions, we found that a more conservative selection improves the robustness of AGN identification.
To this end, we performed a test on the initial sample of SED AGN (i.e., those with $f_{\rm AGN} \geq 0.05$) by re-running \texttt{CIGALE} with the AGN module turned off. We then compared the resulting reduced chi-square values ($\chi^2_\nu$) between the original fit (with AGN) and the new fit (without AGN). We found that galaxies with $f_{\rm AGN} \geq 0.1$ typically showed a significant degradation in fit quality when the AGN component was excluded, with a median increase in $\chi^2_\nu$ exceeding 15\%. In contrast, galaxies with $0.05 \leq f_{\rm AGN} < 0.1$ could be adequately fit without an AGN component, showing only marginal or negligible increases in $\chi^2_\nu$.

This result indicates that $f_{\rm AGN} \geq 0.1$ provides a more secure threshold for identifying galaxies where the AGN contributes meaningfully to the infrared emission. Therefore, we adopt this stricter criterion in the present analysis.

\section{Parametrisation of the merger fraction - AGN fraction and $L_{\rm disc}$ relations from \citetalias{lamarcaDustPowerUnravelling2024}}\label{app:fmerg_fagn}

In Sect.~\ref{sect:fbar_fAGN}, we compare the bar fraction--AGN fraction ($f_{\rm bar}$--$f_{\rm AGN}$) relation derived in this work with the merger fraction--AGN fraction ($f_{\rm merg}$--$f_{\rm AGN}$) relation from our previous study \citepalias{lamarcaDustPowerUnravelling2024}. The overlaid lines in Fig.~\ref{fig:fbar_fagn} representing the $f_{\rm merg}$--$f_{\rm AGN}$ trends were derived from \citetalias{lamarcaDustPowerUnravelling2024} as follows.

The $f_{\rm merg}$--$f_{\rm AGN}$ relation showed distinct behaviours for different AGN types and $f_{\rm AGN}$ regimes. For MIR and X-ray selected AGN, the data were best described by a simple power-law function of the form:
\begin{equation}
    y = x^{\alpha} + c,
    \label{eq:power_law}
\end{equation}
where $y$ represents $f_{\rm merg}$, x is $f_{\rm AGN}$, and $\alpha$ and $c$ are free parameters. For SED AGN, which exhibited a more complex trend, with a mild rising $f_{\rm merg}$ in the low $f_{\rm AGN}$ regime, a combination of two power laws was used:
\begin{equation}
    y = x^{\alpha} + \beta \cdot x^{\gamma},
    \label{eq:double_power_law}
\end{equation}
where $\alpha$, $\beta$, $\gamma$ are free parameters.

To obtain robust parameter estimates, we employed a sampling method. First, the number of $f_{\rm AGN}$ bins used to calculate $f_{\rm merg}$ was varied (between 6 and 20 for X-ray/SED AGN, and up to 15 for MIR AGN due to smaller statistics). The merger fraction and its uncertainty were calculated for each bin. To account for these uncertainties, assumed to follow a Gaussian distribution, a bootstrapping method was applied by varying the $f_{\rm merg}$ values within their errors. The best-fit parameters for Equations~\ref{eq:power_law} and \ref{eq:double_power_law} were then found for each bootstrapped dataset. This entire process was repeated 10\,000 times. The final best-fit parameters reported in Table~\ref{tab:best_fit} are the median values from these 10\,000 fits, with uncertainties corresponding to their 25th and 75th percentile ranges. These parametrised relations are overlaid in Fig.~\ref{fig:fbar_fagn}. 

\begin{table}[]
    \centering
    \caption{Best-fit parameters of Eq.~\ref{eq:power_law} for X-ray and MIR AGN, and of Eq.~\ref{eq:double_power_law} for SED AGN.}
    \begin{tabular}{lcc}
    \hline\hline\\[-7pt]
    AGN type & $\alpha$ & $c$ \\
    \hline\\[-7pt]
    \noalign{\vskip 2mm}
    X-ray & $16.2^{+4.87}_{-3.52}$ & $0.287^{+0.009}_{-0.009}$ \\
    \noalign{\vskip 2mm}
    MIR & $35.0^{+43.1}_{-9.46}$ & $0.478^{+0.022}_{-0.023}$ \\
    \noalign{\vskip 2mm}
    \hline
    \vspace{2mm}
    \end{tabular}
    
    \begin{tabular}{lccc}
    \hline\hline\\[-7pt]
    AGN type & $\alpha$ & $\beta$ & $\gamma$ \\
    \hline\\[-7pt]
    \noalign{\vskip 2mm}
    SED & $47.2^{+29.1}_{-14.0}$ & $0.432^{+0.019}_{-0.018}$ & $0.637^{+0.041}_{-0.030}$ \\
    \noalign{\vskip 2mm}
    \hline
    \end{tabular}
    \tablefoot{Each parameter is estimated using a resampling process. The values reported are the median values of the 10\,000 best-fit parameters and their $25^{th}$--$75^{th}$ percentile ranges. }
    \label{tab:best_fit}
\end{table}

\begin{table}[h]
    \centering
    \caption{Best-fit parameters of Eq.~\ref{eq:exponential} for MIR, X-ray, and SED AGN.}
    \begin{tabular}{lcc}
    \hline\hline\\[-7pt]
    AGN type & $a$ & $c$ \\
    \hline\\[-7pt]
    \noalign{\vskip 2mm}
    MIR & $47.7^{+44.1}_{-8.3}\times10^3$ & $0.47^{+0.03}_{-0.03}$ \\
    \noalign{\vskip 2mm}
    X-ray & $23.2^{+2.3}_{-2.9}\times10^3$ & $0.26^{+0.01}_{-0.01}$ \\
    \noalign{\vskip 2mm}
    SED & $21.9^{+2.4}_{-6.0}\times 10^3$ & $0.119^{+0.002}_{-0.003}$ \\
    \noalign{\vskip 2mm}
    \hline
    \vspace{2mm}
    \end{tabular}
    \tablefoot{Each parameter is estimated using a resampling process. The values reported are the median values of the 10\,000 best-fit parameters and their $25^{th}$--$75^{th}$ percentile ranges. }
    \label{tab:best_fit_exp}
\end{table}

For the parametrisation of the $f_{\rm merg}$--$L_{\rm disc}$ relation, we followed the same methodology, fitting the following linear function for all AGN types analysed:
\begin{equation}
    f_{\rm merg} = \frac{1}{a} \left( \frac{L_{\rm disc}}{10^{42}\,{\rm erg\,s^{-1}}} \right) + c,
    \label{eq:exponential}
\end{equation}
where $a$ and $c$ are free parameters. We calculated the best-fit parameters following the same methodology adopted for the $f_{\rm merg}$--$f_{\rm AGN}$ relation and reported the results in Table~\ref{tab:best_fit_exp}, for each AGN type. These parametrised relations are overlaid in Fig.~\ref{fig:fbar_Ldisc}.

\section{$f_{\rm bar}$--$f_{\rm AGN}$ and $f_{\rm bar}$--$L_{\rm disc}$ including major mergers}\label{app:revised_relations}

\begin{figure*}
    \centering
    \includegraphics[width=0.85\textwidth]{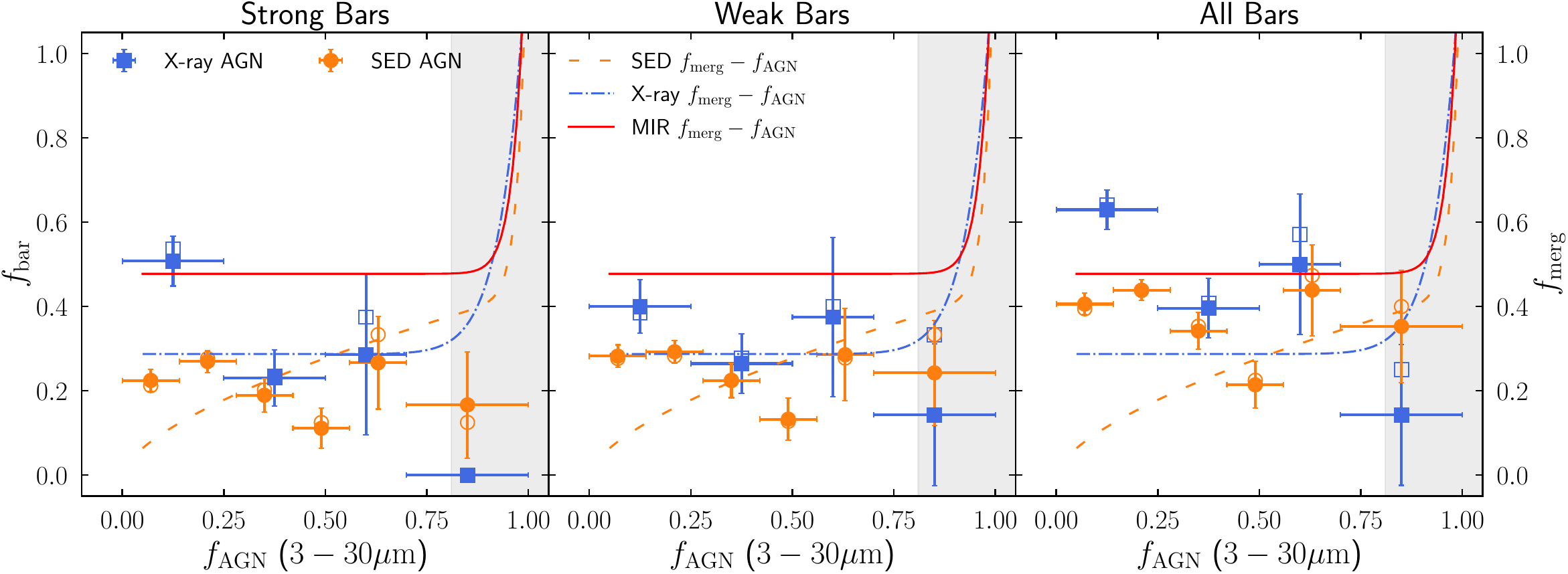}
    \includegraphics[width=0.85\textwidth]{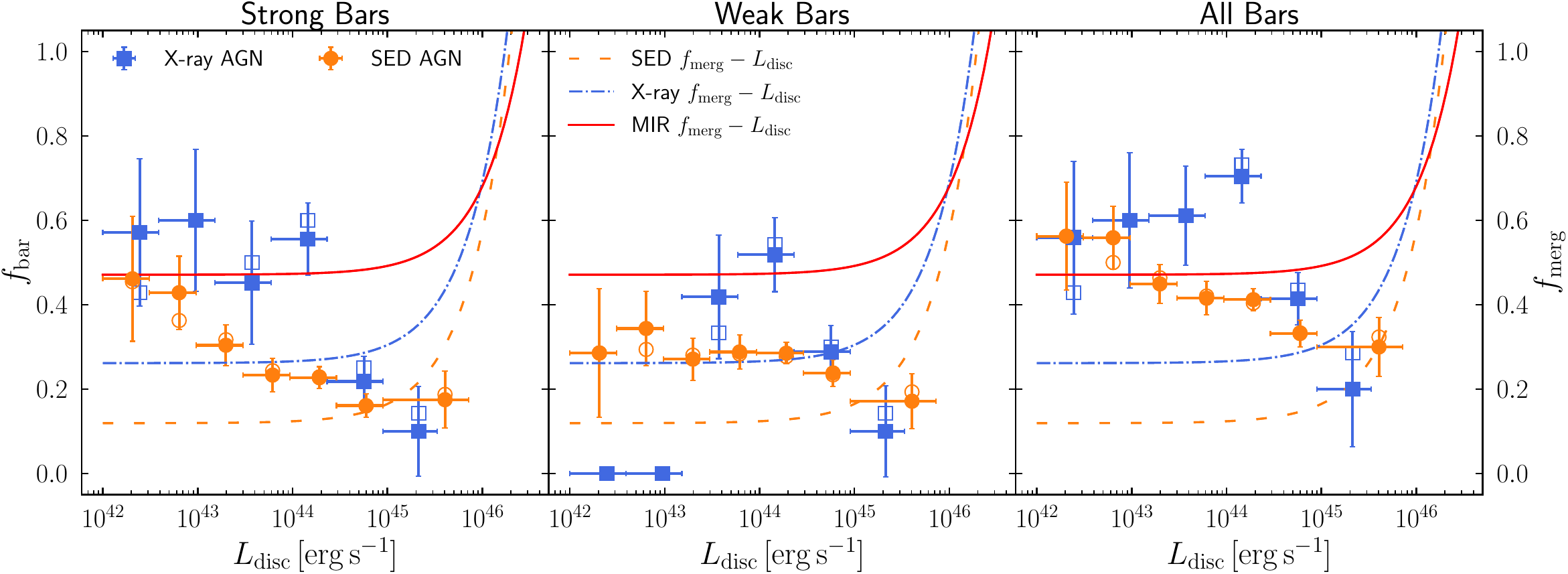}
    \caption{The bar fraction, $f_{\rm bar}$, as a function of $f_{\rm AGN}(3$--$30\,\mu{\rm m})$ (\emph{top}) and $L_{\rm disc}$ (\emph{bottom}) revised after including major mergers in the analysis. Same as Figures~\ref{fig:fbar_fagn} and \ref{fig:fbar_Ldisc}. Empty symbols show the results presented in Figures~\ref{fig:fbar_fagn} and \ref{fig:fbar_Ldisc}.}
    \label{fig:fbar_relations_merg}
\end{figure*}

Figure~\ref{fig:fbar_relations_merg} illustrates the $f_{\rm bar}$--$f_{\rm AGN}$ and $f_{\rm bar}$--$L_{\rm disc}$ relations presented in Sect.~\ref{sect:continuous_exp}, recomputed after adding back the major mergers excluded in the main analysis. Comparing Fig.~\ref{fig:fbar_relations_merg} with Figures~\ref{fig:fbar_fagn} and \ref{fig:fbar_Ldisc} (which excluded mergers entirely) reveals a nice agreement, with both trends being mostly unaffected. The bar fraction still remains relatively flat or mildly decreases with increasing $f_{\rm AGN}$ for both AGN types, regardless of the bar strength. Similarly, $f_{\rm bar}$ continues to show a peak at intermediate luminosities ($L_{\rm disc}\approx 10^{44}\,{\rm erg\,s^{-1}}$) for X-ray AGN, followed by a decrease at higher luminosities. On the other hand, SED AGN show a monotonic decline toward higher luminosities.


\end{appendix}


\end{document}